\documentclass{aa}  

\usepackage{subfig}
\usepackage{xspace}
\usepackage{graphicx}
\usepackage{lscape}
\usepackage{txfonts}
\usepackage{supertabular}
\usepackage{longtable}
\usepackage{amssymb}
\usepackage{caption}
\usepackage{color}
\usepackage{amsmath,amsfonts,amssymb}

\usepackage[titletoc,title]{appendix}
\usepackage[colorlinks=true,allcolors=blue]{hyperref}
\usepackage{natbib,twoopt}
\makeatletter
  \newcommandtwoopt{\citeads}[3][][]{\href{http://adsabs.harvard.edu/abs/#3}%
    {\def\hyper@linkstart##1##2{}%
     \let\hyper@linkend\@empty\citealp[#1][#2]{#3}}}
  \newcommandtwoopt{\citepads}[3][][]{\href{http://adsabs.harvard.edu/abs/#3}%
    {\def\hyper@linkstart##1##2{}%
     \let\hyper@linkend\@empty\citep[#1][#2]{#3}}}
  \newcommandtwoopt{\citetads}[3][][]{\href{http://adsabs.harvard.edu/abs/#3}%
    {\def\hyper@linkstart##1##2{}%
     \let\hyper@linkend\@empty\citet[#1][#2]{#3}}}
  \newcommandtwoopt{\citeyearads}[3][][]%
    {\href{http://adsabs.harvard.edu/abs/#3}
    {\def\hyper@linkstart##1##2{}%
     \let\hyper@linkend\@empty\citeyear[#1][#2]{#3}}}
\makeatother

\begin{document} 

\def\HII{H\,{\sc{ii}}\,}
\def\mm{\,$\mu$m\,}
\def\spitzer{$\it{Spitzer}$\,}
\def\herschel{$\it{Herschel}$\,}
\defcitealias{zav07}{ZAV07}
\defcitealias{deh09}{DEH09}
   \title{Star formation towards the Galactic H\,{\sc{ii}} region RCW~120}

   \subtitle{\herschel\thanks{\emph{Herschel} is an ESA space observatory with science instruments provided by European-led Principal Investigator consortia and with important participation from NASA.} observations of compact sources}

   \author{ M. Figueira\inst{1} 
           \and A. Zavagno\inst{1}
           \and L. Deharveng \inst{1} 
            \and D. Russeil\inst{1}        
          \and L.D. Anderson\inst{2}  
          \and A. Men'shchikov\inst{3}
          \and N. Schneider\inst{4}
      \and T. Hill\inst{1}
         \and F. Motte \inst{5,3} 
         \and P. M\`ege\inst{1}
          \and G. LeLeu\inst{1}
          \and H. Roussel\inst{6}
          \and J.-P. Bernard\inst{7,8}
          \and A. Traficante\inst{9}
          \and D. Paradis\inst{7,8}
          \and J. Tig\'e\inst{1} 
          \and P. Andr\'e\inst{3}  
          \and S. Bontemps\inst{10}
          \and A. Abergel\inst{11}}
          \offprints{M. Figueira, miguel.figueira@lam.fr }
          
   \institute{Aix Marseille Univ, CNRS, LAM, Laboratoire d'Astrophysique de Marseille, Marseille, France
        \and West Virginia University, Department of Physics \& Astronomy, Morgantown, WV 26506, USA
    \and Laboratoire AIM Paris-Saclay, CEA/IRFU - CNRS/INSU - Universit\'e Paris Diderot, 
    Service d'Astrophysique, B\^at. 709, CEA-Saclay, 91191 Gif-sur-Yvette CEDEX, France 
    \and Physik. Institut, University of Cologne, Zülpicher Str. 77, 50937, Koeln, Germany 
    \and  Institut de Planétologie et d'Astrophysique de Grenoble (IPAG), Univ. Grenoble Alpes / CNRS-INSU, BP 53, 38041 Grenoble Cedex 9, France 
    \and Institut d'Astrophysique de Paris, UMR 7095 CNRS, Universit\'e Pierre \& Marie Curie, 
    98 bis Boulevard Arago, 75014 Paris, France 
    \and CNRS, IRAP, 9 Av. colonel Roche, BP 44346, F-31028 Toulouse cedex 4, France
    \and Universit\'e de Toulouse, UPS-OMP, IRAP, F-31028 Toulouse cedex 4, France
    \and Istituto Nazionale di Astrofisica - IAPS, Via Fosso del Cavaliere 100, I-00133 Roma, Italy
    \and CNRS/INSU, Laboratoire d'Astrophysique de Bordeaux, UMR 5804, BP 89, 
    33271, Floirac CEDEX, France    
    \and Institut d'Astrophysique Spatiale, UMR 8617, CNRS, Universit\'e Paris-Sud 11, 
    91405, Orsay, France
     }

   \date{Received July 22 2016; accepted December 1 2016 }
 
  \abstract
  {The expansion of H\,{\sc{ii}} regions can trigger the formation of stars. An overdensity of young stellar objects (YSOs) is observed at the edges of \HII regions but the mechanisms that give rise to this phenomenon are not clearly identified. Moreover, it is difficult to establish a causal link between  \HII -region expansion and the star formation observed at the edges of these regions. A clear age gradient observed in the spatial distribution of young sources in the surrounding might be a strong argument in favor of triggering. }
   {We aim to characterize the star formation observed at the edges of \HII regions by studying the properties of young stars that form there. We aim to detect young sources, derive their properties and their evolution stage in order to discuss the possible causal link between the first-generation massive stars that form the \HII region and the young sources observed at their edges. }
   {We have observed  the Galactic \HII region RCW~120 with \herschel PACS and SPIRE photometers at 70, 100, 160, 250, 350 and 500\,$\mu$m. We produced temperature and H$_2$ column density maps and use the {\it{getsources}} algorithm to detect compact sources and measure their fluxes at \herschel wavelengths. We have complemented these fluxes with existing infrared data. Fitting their spectral energy distributions (SEDs) with a modified blackbody model, we derived their envelope dust temperature and envelope mass. We computed their bolometric luminosities and discuss their evolutionary stages. }
 {The overall temperatures of the region (without background subtraction) range from 15~K to 24~K. The warmest regions are observed towards the ionized gas. The coldest regions are observed outside the ionized gas and follow the emission of the cold material previously detected at 870\mm and 1.3\,mm. The H$_2$ column density map reveals the distribution of the cold medium to be organized in filaments and highly structured. Column densities range from $7\times 10^{21}$ cm$^{-2}$ up to $9\times 10^{23}$ cm$^{-2}$ without background subtraction. The cold regions observed outside the ionized gas are the densest and host star formation when the column density exceeds  $2\times~10^{22}$ cm$^{-2}$. The most reliable 35 compact sources are discussed.  Using exisiting CO data and morphological arguments we show that these sources are likely to be associated with the RCW~120 region. These sources' volume densities range from $2\times~10^{5}$~cm$^{-3}$ to $10^{8}$~cm$^{-3}$. Five sources have envelope masses larger than 50~$M_{\sun}$ and are all observed in high column density regions (>$7\times~10^{22}$~cm$^{-2}$). We find that the evolutionary stage of the sources primarily depends on the density of their hosting condensation and is not correlated with the distance to the ionizing star.  }
{ The \herschel data, with their unique sampling of the far infrared domain, have allowed us to characterize the properties of compact sources observed towards RCW~120 for the first time. We have also been able to determine the envelope temperature, envelope mass and evolutionary stage of these sources.  Using these properties we have shown that the density of the condensations that host star formation is a key parameter of the star-formation history, irrespective of their projected distance to the ionizing stars.}

\maketitle
\titlerunning{RCW~120 with \herschel}
\authorrunning{Figueira et al.}

%

\section{Introduction}
Massive stars (M>8 $M_{\sun}$) affect their surrounding medium due to the action of both their ionizing photons and stellar winds. They form ionized (\HII) regions that expand, bordered by a shell of swept-up neutral material \citep{dys97}. Star formation is observed at the edges of Galactic and extragalactic \HII regions \citep{ber16}. Young stars form there either spontaneously or through various mechanisms linked to the expansion of the ionized region \citep{deh10}.

Star formation observed at the edges of \HII regions  has been studied in detail during the past ten years. With the GLIMPSE \citep{ben03} and MIPSGAL \citep{car09} surveys, the \spitzer satellite has revealed that we live in a bubbling galactic disk where thousands of \HII regions have a clear impact on their environment. \citet{and11} have shown that half of all \HII regions have a bubble morphology. Studies of triggering have focused on bubble \HII regions. \citet{deh10} used \spitzer GLIMPSE and MIPSGAL data combined with ATLASGAL \citep{sch09} data on 102 bubbles. They showed that star formation observed at the edges of \HII regions is an important phenomenon in our Galaxy. Up to 25\% of the ionized regions show high-mass star formation  triggered on their edges. This result has been confirmed by \citet{tho12} and \citet{ken12,ken16} who found an overdensity of young stellar objects (YSOs), including massive objects, around \spitzer and ATLASGAL bubbles. \citet{sim12} have listed 5106 bubbles using these GLIMPSE and MIPSGAL surveys. Many studies of individual \HII regions, including numerical simulations, confirm that \HII regions impact on their surrounding, enhancing significantly the star formation there \citep{min13,sam14,liu15,lad15}. This impact is also observed at the waist of bipolar \HII regions as recently discovered by \citet{deh15}. 

However \citet{dal15} assessed the relevance of standard observational criteria used to decide whether the star-formation process is of spontaneous or triggered origin at the edges of \HII region. By comparing the observational criteria used to their own new numerical results they concluded that,  when interpreting observations of star formation in the vicinity of feedback-driven structures in terms of triggering, one should exercise
caution. 

While the large and rapidly increasing bulk of knowledge tends to offer empirical evidence in support of some impact of \HII regions on the local star formation, there are still many unanswered questions on the possible influence of these regions on star formation near their edges. One way to firmly establish the causal link existing between the ionized region and the star-formation process taking place on its surrounding could be to measure a clear difference between the age of the ionizing stars, located in the central \HII region and the ones formed at its edges \citep{mar10,bik10}. However, the determination of stellar ages is challenging \citep{mar10}.
 
We are left in a situation where we observe an overdensity of young stars at the edges of these \HII regions. These young stars are highly efficient (up to 25\%) at forming massive stars. \citep{bik10,ell13,cap14,tap14}. But we do not know how the material is assembled (uniformly distributed then collected versus pre-condensed in an inhomogeneous medium) and what are the mechanisms that control the formation of stars in these regions. For pre-existing clumps, star formation could occur spontaneously before encountering the ionization front or the ionizing radiation leaking from the \HII region. Dedicated observations can help in answering these questions. High resolution molecular spectroscopy reveals the distribution and velocity field of the material that surrounds \HII regions \citep{and15,liu15}. The spatial distribution, properties and evolutionary stage of YSOs are key points to address the triggering issue. We need to obtain an overview of all stages of star formation in a given region and access the distribution of the surrounding material on all spatial scales to discuss the history of star formation. The large scale distribution should help in understanding the initial distribution of the material (uniform versus clumpy, filamentary). A better knowledge of the distribution and properties (density, temperature) of the material that surrounds \HII regions could also help in better understanding how the material is assembled and how star formation occurs around ionized regions.   

The \herschel satellite offers a unique opportunity to study star formation around Galactic \HII regions and helps in answering some of the pending questions. Thanks to its sensitivity and its large wavelength coverage in the far-infrared, \herschel is perfectly suited to study the earliest phases of star formation. The six measured photometric points (70, 100, 160, 250, 350, 500\,$\mu$m) really help in constraining the young sources' properties (temperature, envelope mass, luminosity). Moreover, \herschel's wavelength range covers the peak of the YSOs' spectral energy distribution (SED) also helping to characterize the young source' evolutionary stage.  Combined with existing infrared and molecular data, \herschel observations allow us to obtain a global view of the star-formation history \citep{ngu15}. 
    
Here we present the results obtained for young compact sources observed towards the bubble \HII region RCW~120. 
Using \herschel photometric PACS and SPIRE data, we re-examine this region to better determine the nature and evolutionary stage of the YSOs observed there. We aim to  discuss the region's star-formation history there using sources' evolutionary stage. Section~\ref{reg} presents the current knowledge on RCW~120. The \textit{Herschel} observations are described in Sect.~\ref{obs}. The data reduction and sources' extraction are presented in Sect.~\ref{analysis}. The results are presented in Sect.~\ref{res} and discussed in Sect.~\ref{dis}. The main results and conclusions are given in Sect.~\ref{conc}.

\section{The RCW~120 region}\label{reg}
RCW~120 \citep{rod60} is an egg-shaped Galactic \HII region of 3.8~pc diameter, located 0.5$\degr$ above the Galactic plane. Due to its simple morphology and isolation, this region has been studied in detail during the past ten years. The main results are summarized below:\\ 

The region is ionized by an O8V star, CD$-38\degr11636$ (\citealt[hereafter ZAV07]{zav07}; \citealt{mar10}). An emission arc is observed at 24\mm below the star (\citealt[hereafter DEH09]{deh09}; \citealt[see their fig.~3]{mar10}) and is interpreted as representing the upstream boundary between the wind bubble and the photoionized interstellar medium \citep{mac15}.

The photometric distance of RCW~120 was computed by \citet{rus03} using UBV and H$\beta$ photometry. The uncertainty is estimated to be 0.6~kpc and comes from the uncertainty in the spectral type estimate (around 0.3 mag).

RCW~120 and its surrounding layer have been observed in the dust continuum at 870\mm (DEH09) and 1.3\,mm (ZAV07) and in CO molecular lines \citep{and15,tor15}. These observations show that RCW~120 is surrounded by a dense shell of gas and dust.

\citet{tor15} observed two molecular clouds towards RCW~120 and suggest that some collision between the clouds triggered the formation of the ionizing O star of RCW~120 in a short timescale of 0.2-0.4 Myr. An age of 0.4 Myr is also obtained by \citet{mac15}. Simulations from \citet{tre14b} lead to a similar age for the ionizing star of RCW~120.

\citet{and15} found no evidence for expansion of the molecular material associated with RCW~120 and therefore can make no claim about its geometry (2D or 3D). Dust emission simulations suggest that the \HII region RCW~120 is not spherical, but instead cylindrical, and that we observe the object along the axis of this cylinder \citep{pav13}. 

Using 1.3~mm continuum emission ZAV07 found eight condensations (five located at the edges of the ionized region, see their Fig.~4) and studied the young stellar content of these condensations, pointing out the possible importance of long-distance influence of the ionized region on its surrounding. This study has been completed by DEH09 who characterized the evolutionary stage by adding 24\mm data from MIPSGAL and confirmed the importance of long-distance interaction between the \HII region and its surroundings. Many YSOs, including Class~I and Class~II sources are observed at the edges of the ionized region. A noticeable massive Class~0 candidate is detected towards the highest density condensation (condensation~1), later confirmed with \herschel observations \citep{zav10}. A spectrophotometric study of the YSOs in the near-infrared confirm that these YSOs are associated with the RCW~120 region because they have the same velocity than that of the ionized gas \citep{mar10}.

DEH09 observed a series of eleven young sources aligned parallel to the ionization front towards the most massive condensation at 24\mm equally spaced by 0.1~pc and is thought to be the result of Jeans gravitational instabilities.

\citet{tre14} studied the probability density function (PDF) of a series of Galactic \HII regions, including RCW~120 (see their fig.~8 and fig.~9). They found evidence for compression, and the value of the exponent derived to fit the PDF towards condensation~one may indicate the role of ionization compression in the formation of this condensation and its collapse to form stars. According to numerical simulations lead by \citet{min13}, if the condensation had gravitationally collapsed prior to the passage of the ionization front, the condensation would be already sufficiently dense to resist the ionization front expansion. It would, for example, trigger the formation of a pillar rather than a condensation remaining in the shell.

\citet{wal15} performed three dimensional smoothed particle hydrodynamics (SPH) simulations of \HII regions expanding into fractal molecular clouds and then used RADMC-3D to compute the synthetic dust continuum emission at 870\mm from their simulations, applied to RCW~120. They found a hybrid form of triggering which combines elements of collect and collapse (C\&C) mechanism \citep{elm77} and radiation driven implosion (RDI) \citep{kes03}.

\begin{figure}
 \centering
 \includegraphics[angle=0,width=90mm]{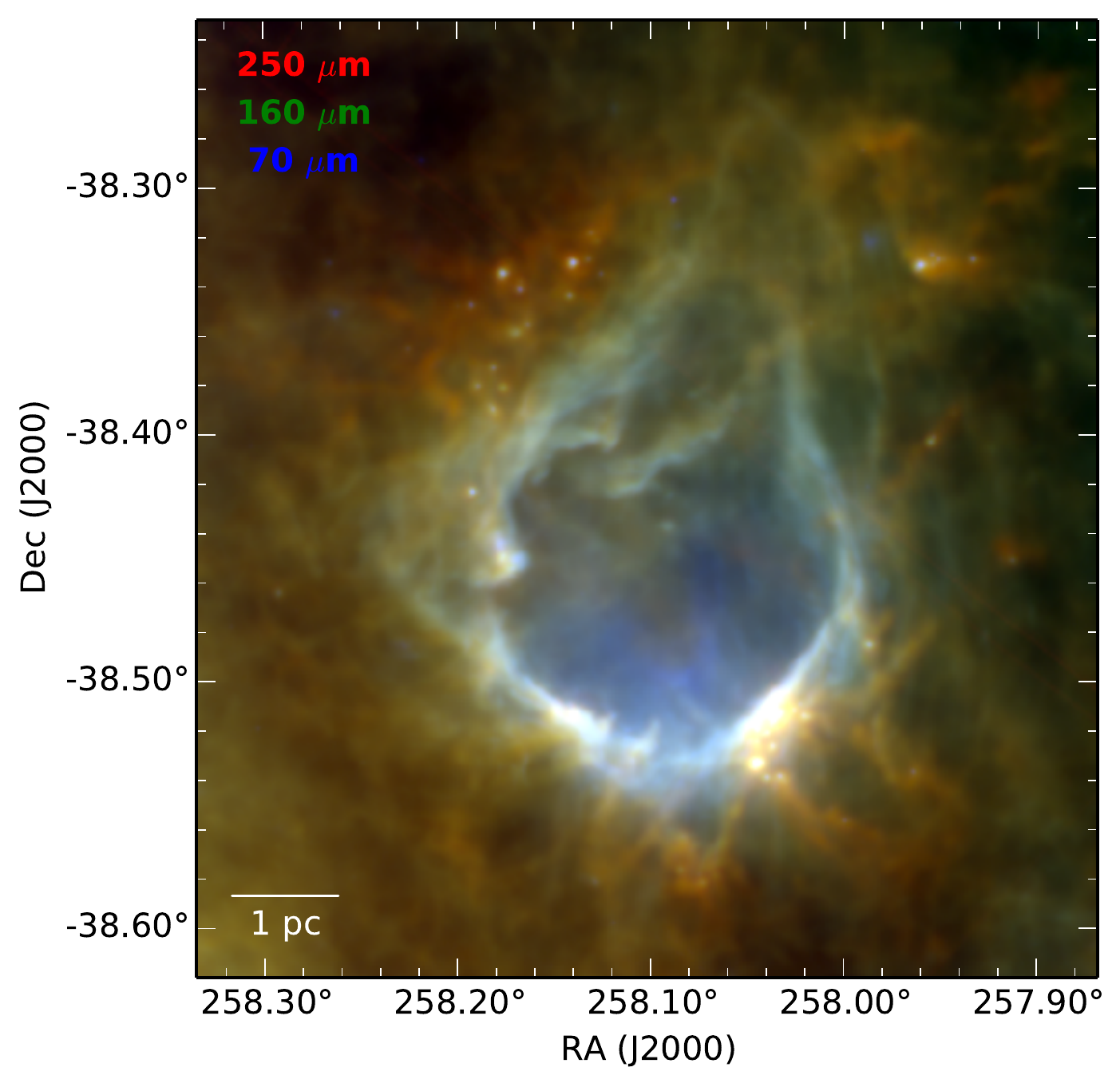}
  \caption{ RCW~120: \herschel-PACS 70\mm (blue), 160\mm (green) and \herschel-SPIRE 250\mm (red). The field size is 21.8\arcmin $\times$ 24.5\arcmin. North is up, east is left}
             \label{present}
\end{figure}

Figure~\ref{present} presents a three-color image of the RCW~120 region as seen by $\it{Herschel}$. The 70\mm emission (blue part) underlines the emission of the warm dust while the 250 \mm emission (red part) underlines the emission from the colder dust located in the dense material that surrounds the ionized region and that interacts with the ionizing radiation.  
\section{Observations and data reduction} \label{obs}

\subsection{Herschel observations}

RCW~120 was observed with the PACS and SPIRE photometers. Details of these observations (map size, observing time, observational identification (ObsID), observational date (Obs.) operational day (OD),  map center) are given in Table~\ref{tabobs}. The PACS photometer was used to make simultaneous photometric observations in two photometric bands as part of the HOBYS key program \citep{mot10}. Two cross-scan maps were done at angle 45$\degr$ and 135$\degr$ with a scanning speed of 20$\arcsec$/second. This observing mode is described in Section~5.2 of the PACS Observers' Manual{\footnote{\url{http:\\herschel.esac.esa.int/Docs/PACS/html/pacs_om.html}}}. The beam FWHM varies between 5$\farcs$9 at 70\mm, 6$\farcs$0 at 100\mm and 11$\farcs$4 at 160\mm. The total observing time is 2.6~hours.   

RCW~120 was observed with the SPIRE photometer as part of the Evolution of Interstellar Dust key program for the \herschel Science Demonstration Phase. The SPIRE photometer was used to make simultaneous photometric observations in the three photometer bands (250, 350 and 500~$\mu$m). The map is made by scanning the telescope at a given scan speed of 30$\arcsec$/second along lines. Cross-linked scanning is achieved by scanning at 42$\degr$ (Scan A angle) and then at $-42\degr$ (Scan B angle).
This ensures that the effect of 1/f noise on the map can be minimized and also leads to improved map coverage. This observing mode is described in details in the last version of the SPIRE Observers' Manual {\footnote{\url{http:\\herschel.esac.esa.int/Docs/SPIRE/html/spire_om.html\#x1-310003.2}, Version 2.2, November 29, 2010}} in Section~3.1.2. 
One map at each scanning angle was obtained. The beam FWHM varies between 18$\farcs$2 at 250\mm, 25$\farcs$2 at 350\mm and 36$\farcs$6 at 500\mm. The total observing time is 0.34 hour. 

\begin{table*}
\caption{Summary of \herschel observational parameters}\label{tabobs}      
\centering                          
\begin{tabular}{|ccclcl|}        
\hline          
 Map size & Time & ObsIDs & Date & OD & Map center \\
 (')  & (s)  &       & yyyy-mm-dd &  & J2000 \\
 \hline
 \multicolumn{5}{|c}{PACS 100 and 160\mm observations} & \\
$30\times 30$ & 3302 & 1342183978, 1342183979 & 2009-09-17 & 127 & 17$^{\mathrm{h}}$ 12$^{\mathrm{m}}$ 23$\fs$10 $-$38$\degr$ 27$\arcmin$ 43$\arcsec$  \\
$30\times 30$ & 3302 & 1342185553, 1342185554 & 2009-10-10 & 148 & 17$^{\mathrm{h}}$ 12$^{\mathrm{m}}$ 30$\fs$59 $-$38$\degr$ 27$\arcmin$ 25$\farcs$5   \\
  &    &    &    &   &  \\
  \hline
\multicolumn{5}{|c}{PACS 70 and 160\mm observations} & \\
$30\times 30$ & 2762 & 1342216585, 134216586 & 2011-03-22 & 677 &  17$^{\mathrm{h}}$ 12$^{\mathrm{m}}$ 23$\fs$21 $-$38$\degr$ 27$\arcmin$ 43$\farcs$21  \\
  &    &    &    &   & \\
  \hline
\multicolumn{5}{|c}{SPIRE observations} & \\
$22\times 22 $ & 1219 & 1342183678 & 2011-03-22 & 121 & 17$^{\mathrm{h}}$ 12$^{\mathrm{m}}$ 18$\fs$80 $-$38$\degr$ 27$\arcmin$ 58$\farcs$5 \\
\hline 
\end{tabular}                                 
\end{table*}
The PACS maps were produced using
the HIPE Level 1 data and then version 21 of the Scanamorphos software
package which performs baseline and drift removal before regridding \citep{rou12}.
The SPIRE images were reduced using modified pipeline
scripts of Version 10 of HIPE{\footnote{HIPE is a joint development software by the \herschel Science
Ground Segment Consortium, consisting of ESA, the NASA \herschel
Science Center, and the HIFI, PACS, and SPIRE consortia.}}, the \herschel Interactive Processing
Environment.  Each map direction (nominal and orthogonal) was first
reduced individually to Level 1 data, correcting for effects such as
temperature drifts and jumps, glitches and cooler burps.  The
individual maps were then combined to create one map (Level 2 data).  Map reconstruction was done using the SPIRE default 'naive' mapmaking
algorithm at the same time as a destriper module (including a median
correction and in bright source mode). The default gridding of 6$\arcsec$,
10$\arcsec$, 14$\arcsec$ for the SPIRE wavelengths 250, 350, 500\mm was
chosen. The fits output files for each SPIRE wavelength are in units
of Jy/beam. For an absolute calibration of the SPIRE maps, the
zeroPointCorrection task calculates the absolute offset for a SPIRE
map, based on cross-calibration with Planck HFI-545 and HFI-857 maps,
color-correcting HFI to SPIRE wavebands assuming a gray-body function
with fixed spectral index. The offsets determined in this way correspond well to
the ones provided by J.-P. Bernard (private communication). 

\subsection{Complementary data}
We complement the \herschel data with data from the Two Micron All-Sky Survey (2MASS) at 1.25\,$\mu$m ($J$), 1.65\,$\mu$m ($H$) and 2.17\,$\mu$m ($K_s$) with a resolution of 2$\arcsec$ \citep{skr06}, from the \textit{Spitzer}{\footnote{http://irsa.ipac.caltech.edu/data/SPITZER}} GLIMPSE and MIPSGAL surveys of the Galactic Plane at 3.6\,$\mu$m, 4.5\,$\mu$m, 5.8\,$\mu$m and 8\,$\mu$m and 24\,$\mu$m with a resolution of 17$\arcsec$, 1$\farcs$7, 1$\farcs$9, 2$\arcsec$ and 6$\arcsec$ \citep{ben03,car09}.

\section{Data analysis} \label{analysis}
\subsection{Compact sources' extraction method} \label{extract}
\herschel compact sources were extracted using the multi-wavelength, multi-scales {\it{getsources}} algorithm{\footnote{The \textit{getsources} algorithm is publicy available and can be downloaded at \url{http://www.herschel.fr/cea/gouldbelt/en/getsources/}}} (version 1.140127) \citep{{men12},{men13}}. The working method of {\it{getsources}} can be roughly decomposed into two steps : the detection and the measurement. While the latter is performed on all maps inserted into the algorithm, the detection can be made from a selected sample of maps depending on the aim of the study. In order to improve this step, an \herschel high-resolution density map \citep{hil12,pal13} was created (see Sect.~\ref{densitymap}) and added to better constrain the detection of compact sources. Moreover, some original maps were modified in order to enhance the contrast of the cooler and hence the densest regions since heated structures could be detected and misleading the final sample of sources. For this purpose and to provide valuable guidance to the detection algorithm, we use the 160\mm PACS and 250\mm SPIRE maps as they represent a good compromise between resolution and non-contamination by very small grains (VSG). The photometric offsets derived using IRAS-PLANCK model \citep{ber10} were added and the 160\mm map was convolved to the resolution of the 250\mm SPIRE observations (25\farcs2). We assumed a modified blackbody (hereafter, MBB) model with a spectral index of 2. This value is higher than the reference for the galaxy ($\sim$1.6, \citet{2016arXiv160509387P}) but for dense regions, in the inner regions of the galactic plane for instance, $\beta$ tends to increase and hence, a value of 2 should be more appropriate for compact regions \citep{par12}. Non-linear fitting of the SEDs was performed using the Levenberg-Marquardt's algorithm \citep{mar09}. From the SED, a color temperature can be found for each pixel by using the ratio of the two maps 

\begin{equation}
\frac{I_{250\mu m}}{I_{160\mu m}^{\theta=25\farcs2}}=\left(\frac{\nu_{250\mu m}}{\nu_{160\mu m}}\right)^5 \frac{e^{\frac{h\nu_{160\mu m}}{kT}}-1}{e^{\frac{h\nu_{250\mu m}}{kT}}-1}
,\end{equation}

where $I_{160\mu m}^{\theta=25\farcs2}$ is the 160\mm map convolved at the 250\mm resolution.

A weight-map is then created as the ratio between the map giving the MBB flux corresponding to the color temperature and a fiducial temperature of 20~K (median temperature). Multiplying the native 160\mm PACS map by the weight-map give the 160\mm corrected map where colder regions are enhance compared to warmer regions. The 250\mm corrected map is created in the same way and both are used in replacement of the native 160\mm PACS and 250\mm SPIRE maps for the detection step.

To summarize, for extraction of the sources, we use the original 70\,$\mu$m, 100\,$\mu$m, 350\,$\mu$m, 500\,$\mu$m maps and improved the detection by including the high-resolution density map and the 160\mm and 250\mm corrected maps.

\subsection{Pre-selection}
The final {\it{getsources}} catalog contains many useful informations about the detected sources. In the following subsections, we will keep only a small part of them : J2000 coordinates, detection significance at each wavelength, flux at peak and integrated flux with their corresponding errors and source ellipse parameters (major axis, minor axis, position angle). 

Before doing the analysis, the sources present in the catalog have to be filtered to select the well-defined ones. The selection criteria defined by the HOBYS consortium are listed below (see also Tig\'e et al. submitted) :\\
Each source must have a {\emph{deconvolved size}} (see Eq.~\ref{eq2}) smaller than 0.1~pc at the {\emph{reference wavelength}} (see below) and three reliable fluxes (including the reference wavelength). A flux is considered as reliable if the detection is reliable (the detection significance is higher than 7, see \citealt{men12}), the signal to noise of the peak and integrated flux is higher than 2 and the elongation (defined as the ratio of the major and minor axis of the ellipse footprint) lower than 2 in order to limit the sample to circular compact cores.

From these criteria, sources extracted by \textit{getsources} are expected to be dense and cold. Therefore, we consider as a good assumption that thermal emission from the cores is optically thin and not contaminated by VSGs for $\lambda \ge$100\,$\mu$m.
The {\emph{deconvolved size}} at wavelength $\lambda$, $\theta_{deconv}^{\lambda}$, is computed as
\begin{equation}\label{eq2}
\theta_{deconv}^{\lambda}=\sqrt{\theta^{a,\lambda}_{conv}\times \theta^{b,\lambda}_{conv}-HPBW^2_{\lambda}}
,\end{equation}
where $\theta^{a,\lambda}_{conv}$/$\theta^{b,\lambda}_{conv}$ stand for the major/minor convolved size estimate of the source at wavelength $\lambda$ (given in the \textit{getsources} catalog) and HPBW$_{\lambda}$ is the half-power beam width at wavelength $\lambda$. The reference wavelength was chosen to be 160\mm as a compromise between the resolution (11\farcs4) and the tracer of optically thin dust emission. This trade-off allows for both the correct identification of the peak of the SED and a good scaling of the flux \citep{mot10,ngu11}. Nevertheless, in some marginal cases, the 160\mm emission may be contaminated by small grains heated in the photo-dissociation region (PDR) leading to a deconvolved size larger than the one measured at 250\,$\mu$m. In such cases, the 250\mm is taken as the reference (resolution 18\farcs2). 

Detections complying with the above-mentionned criteria were kept for the analysis. However, the 70\mm data were systematically excluded from the SED fitting to avoid contamination from VSG's emission even though the criteria were satisfied.

Among the 359 detections of the {\it{getsources}} algorithm, 80 were kept at the end of the selection (put in the pre-selected sample). Rejected sources appear to be false detections, mainly filament pieces or sources with not enough flux measurements to fit the SED. Rejected sources were visually inspected and those which look like compact sources with at least one reliable wavelength (70\mm included) were kept in a tentative sample (80 sources). The physical properties of the tentative sources were derived throughout an indirect method (see Sect~\ref{tentative-prop}) since a SED fitting couldn't be done.
 
\subsection{Spectral energy distribution}\label{SED}
Before fitting the SED for each compact source, the fluxes must be scaled since we want them to be measured within the same aperture. 
A full treatment of this scaling can be found in \citet{mot10,ngu11} where the relation between flux and source's angular size is taken to be same as for protostellar cores. This aperture scaling is based on the assumptions that the source is optically thin for $\lambda$>100$\mu$m, $M(r)\propto r$ and the gradient of the temperature is weak within the region \citep{eli14}. The scaling was done when the size at the reference wavelength and the wavelength to be scaled could be deconvolved.
Following their procedure, we applied scaling factors to fluxes according to the formula 
\begin{equation}\label{eq:sizecor}
S_{\lambda}^{\mathrm{scaled}}=\zeta_{\lambda} \times S_{\lambda} = \frac{\theta_{deconv}^{Ref}}{\theta_{deconv}^{\lambda}} \times S_{\lambda}  
,\end{equation}
where $S^{\mathrm{scaled}}_{\lambda}$ represents the rescaled flux associated with scaling factor $\zeta_{\lambda}$ and $S_{\lambda}$ is the original flux.\\
The model to be fitted is a MBB (\ref{eq:Grey-body}) using the Hildebrand relation \citep{hil83} 
\begin{equation}\label{eq:Grey-body}
S_{\nu}(T)=\frac{M_{\rm{env}}\kappa_{300\mu m}}{RD^2\nu_0^2}\times v^{\beta}\times B_{\nu}(T)=C\times v^{\beta}\times B_{\nu}(T)
,\end{equation}
with a gas-to-dust ratio R=100, D the distance of RCW~120, 1.3 kpc, and C is introduced as a constant of the fit. The HOBYS consortium decided to use the dust opacity law $\kappa_{\nu}=\kappa _{300\mu m}(\nu/\nu_0)^2$ with $\kappa _{300\mu m}$=10 cm$^2$ g$^{-1}$, $\nu_0$=1000~GHz \citep{bec90,mot10}. As explained before, the spectral index has been fixed to two, reducing the model space of the fitting parameters to the C-T plane. The initial errors used to weight the data have been set to the quadratic sum of the {\it{getsources}} and the calibration errors (3\% at 100\,$\mu$m, 5\% at 160\mm and 7\% for SPIRE bands).

\begin{table}
\caption{Minimum, maximum and median values for the color correction factors at \textit{Herschel} wavelengths for the final sample of sources}\label{tab:colcor}
\centering
\begin{tabular}{|c|ccc|}
\hline 
$\lambda$ & Min & Max & Median \\   
($\mu$m) & & & \\
\hline
70 & 0.46 & 1.03 & 0.91 \\
100 & 0.79 & 1.03 & 1.02 \\
160 & 0.95 & 1.06 & 1.03 \\
250 & 0.94 & 0.94 & 0.94 \\
350 & 0.94 & 0.94 & 0.94 \\
500 & 0.94 & 0.94 & 0.95 \\
\hline
\end{tabular}
\end{table}

During the acquisition of the sources with the PACS and SPIRE instruments of \textit{Herschel}, the spectrum is assumed to be flat across the bands ($\nu S_{\nu}$=const) which is not true because we expect the sources to follow a MBB model. To correct for this assumption, we apply color correction factors given in the PACS and SPIRE observer's manual. The fitting algorithm reaches convergence when the absolute difference between two subsequent temperatures (obtained at two consecutive steps) is fainter than 0.1~K. Since the spectral index was fixed to two, these factors depends on the temperature for PACS and are constant for SPIRE. Table~\ref{tab:colcor} gives the minimum, maximum and median value for the color correction factors at \textit{Herschel} wavelengths. Color corrections are high for short wavelengths and low temperature. For 70$\mu$m and 100$\mu$m, they go up to 54\% and 21\% and corresponds to a source with T=11.2~K. However, considering the median value for all wavelengths, the color correction factors are low and do not change drastically the sources' fluxes of the final sample.

The final temperature is derived directly as one of the parameter of the MBB model and assuming optically thin emission for the dust, the envelope mass is derived as
\begin{equation} \label{Hildebrand}
{M}_{\mathrm{env}}=R\frac{S_{\nu}(T)D^2}{\kappa _{\nu}B_{\nu}(T)}=C\times 8.496 \times 10^{8} M_{\sun}
.\end{equation}

The uncertainties are derived from the fitting errors and are 3\% and 20\% in average for the dust temperature and the envelope mass, respectively.  Obviously, these error values on the physical parameters do not take into account the dependence of $\beta$ with the wavelength (from 1 to $\ge$2) and the uncertainty on the opacity factor which is at least a factor of two due to the unknown properties of dust grains \citep[see their Sect~4.1]{deh12}.

Each detected YSO's bolometric luminosity was computed by integrating the corresponding SED curve. For sources having infrared (IR) counterparts (from the 2MASS, GLIMPSE and MIPSGAL surveys) within a radius of 4$\arcsec$, the SED was bipartitioned and partial integrations are made over it. Below 70$\mu$m, a so-called IR luminosity is obtained using a trapezoidal integration scheme (numerical integration done by connecting the data points with straight lines) from the first IR counterpart found in the catalogs to the 70\mm flux. From 70\mm onwards a so-called "\herschel luminosity" is obtained by integrating over the \herschel SED. The bolometric luminosity is obtained by adding these two values. 
The SED fitting algorithm returned the error on \herschel luminosity while that affecting the IR one is obtained by computing the IR luminosity with the fluxes plus the uncertainties (higher-limit) and with the fluxes minus the uncertainties (lower-limit). On average this resulted in an overall uncertainty of 30\% on bolometric luminosity. The average volume density was computed by assuming a spherically symetric core with a diameter equal to its deconvolved size at the reference wavelength
\begin{equation}
<n_{H_{2}}>=\frac{M_{env}}{\frac{4}{3}\pi\times (\theta_{deconv}^{Ref})^3}
.\end{equation}
After the SED fitting, a second stage selection was applied to obtain the final sample of sources discussed in this paper (described below). 
\subsection{Final selection} \label{selproc}
At the end of the SED fitting procedure, the sample of sources obeying our first-stage selection scheme (pre-selected sample) was visually inspected at each \herschel wavelength, to ensure the detection of truly compact sources. The requirements for a source to pass successfully this second-stage selection scheme are two-fold : 
(1) The source had to be clearly seen by eye on one of the \herschel images, and 
(2) The source's SED had to be well-constrained. 

\noindent The first condition was checked by two different people to avoid subjective detections. The second condition allows us to eliminate dubious SED, for example SED with unconstrained peak or SED with increasing flux mostly at SPIRE wavelengths. From this second stage selection, 35 sources were kept for the study (final sample), seven sources having unconstrained SED were added to the tentative sample and 38 sources looking like small clumps or filamentary pieces were rejected.\\

\noindent To summarize the two-stage selection : from the 359 detections, 35 sources are included in the final sample and 87 sources are included in a tentative sample. The latter is composed of sources possessing at least one reliable wavelength (including 70\,$\mu$m) clearly seen (by eye) in the \textit{Herschel} images but not pre-selected due to the lack of flux measurements or pre-selected sources whose SED is unconstrained. These detections are thought to be real sources and kept in order to derive their physical properties with an indirect method since their SED cannot be used.

The second-stage selection is definitely highly non-conservative but ensures the reliability of the sources to be investigated.
As discussed before, part of the remaining detections could be real sources under filaments with well-constrained SED which are eliminated in order to have a reliable sample of sources. Finally, we are left with two samples : the final one which will be discussed in the paper and the tentative one whose physical parameters will be derived with an indirect method.
We assume that the selected sources are associated with RCW~120, that is, they are located at the same distance. This assumption is supported by the study by \citet{mar10} who showed using high resolution near-IR spectro-photometric observations that the YSOs with a IR counterpart observed towards RCW~120 are at the same velocity as that of the ionized gas. We further discuss this point in Sect.~\ref{molecular}.

\subsection{Dust temperature and column density maps} \label{densitymap}
\subsubsection{Method}
Following the procedure of \citet{hil11,hil12}, a map of dust temperature at 36$\farcs$6 can be obtained by fitting the flux at each pixel, using the MBB model 

\begin{equation}
F_{\nu}=\frac{B_{\nu}(T_{dust})\kappa_{\nu}\Sigma_{500\mu m}}{R}=\frac{B_{\nu}(T_{dust})\kappa_{\nu}\mu m_{H}N(H_2)}{R}
,\end{equation}
where $F_{\nu}$ is the brightness, $\mu$ the mean molecular weight ($\approx$ 2.8), $m_{H}$ the proton mass and $N$(H$_2$), the column density. The temperature and the gas surface density $\Sigma_{500\mu m}$ (the 500\mm subscript stands for the corresponding resolution) are the fitting parameters. Another direct byproduct of the fitting algorithm is a map of the H$_2$ column density at the same low-resolution assuming the dust opacity law of \citet{bec90} (See Sect.~\ref{SED}). Since most of the observed regions through the HOBYS project are not observed at 100~$\mu$m, the method considers only wavelengths higher than 160\mm for the SED fitting even if the data were available. This choice made the comparison of temperature and column density maps obtained for different regions easier.

Following the procedure described by \citet{pal13} based on a multi-scale decomposition method, a high-resolution column-density map at 18\farcs2 can be computed. The gas surface density smoothed at 250\mm resolution ($\Sigma_{250\mu m}$) can be written as a sum of gas surface density smoothed at 350\mm ($\Sigma_{350\mu m}$) and 500\mm resolution 
\begin{equation}
\label{hrcdm1}
\Sigma_{250\mu m}=\Sigma_{500\mu m}+(\Sigma_{350\mu m}-\Sigma_{500\mu m})+(\Sigma_{250\mu m}-\Sigma_{350\mu m})
.\end{equation}

The second term in parentheses of Eq.~\ref{hrcdm1} represents the spatial scale structure of the region seen at 350\mm without the largest structure corresponding to the 500~$\mu$m observations. An estimate of $\Sigma_{500\mu m}$ can be obtained by considering that these data are approximately equal to $\Sigma_{350\mu m}*G_{350-500}$ where $G_{350-500}$ is the full width at half maximum (FWHM) of the point spread function (PSF) needed to convolve the 350\mm map to the 500\mm resolution ($\sqrt{36\farcs6^2-25\farcs2^2}=$27\farcs2). The gas surface density $\Sigma_{350\mu m}$ is obtained in the same way than $\Sigma_{500\mu m}$ but excluding the 500\mm data from the SED fitting.

The third term of Eq.~\ref{hrcdm1} represents the structure seen at 250\mm without the largest scale structure seen at 350$\mu$m. As before, $\Sigma_{350\mu m}$ can be written as $\Sigma_{250\mu m}*G_{250-350}$ and $\sqrt{25\farcs2^2-18\farcs2^2}=$17\farcs4 is the FWHM of the PSF needed. The gas surface density $\Sigma_{250\mu m}$ is obtained using the ratio of the 160\mm and 250\mm maps as explained in Sect~\ref{extract}. 
Finally, Eq.~\ref{hrcdm1} can be rewritten as 

\begin{multline}
\label{hrcdm2}
\Sigma_{250\mu m}=\Sigma_{500\mu m}+(\Sigma_{350\mu m}-G_{350-500}*\Sigma_{350\mu m}) \\
+(\Sigma_{250\mu m}-G_{250-350}*\Sigma_{250\mu m}).
\end{multline}

Hence, the resulting high-resolution density map can be seen as a composite map representing the multi-scale structure of RCW~120 from 250\mm to 500\mm resolution.

\subsubsection{Comparison with \citet{and12} maps}

\begin{figure*}
\includegraphics[angle=0,width=180mm]{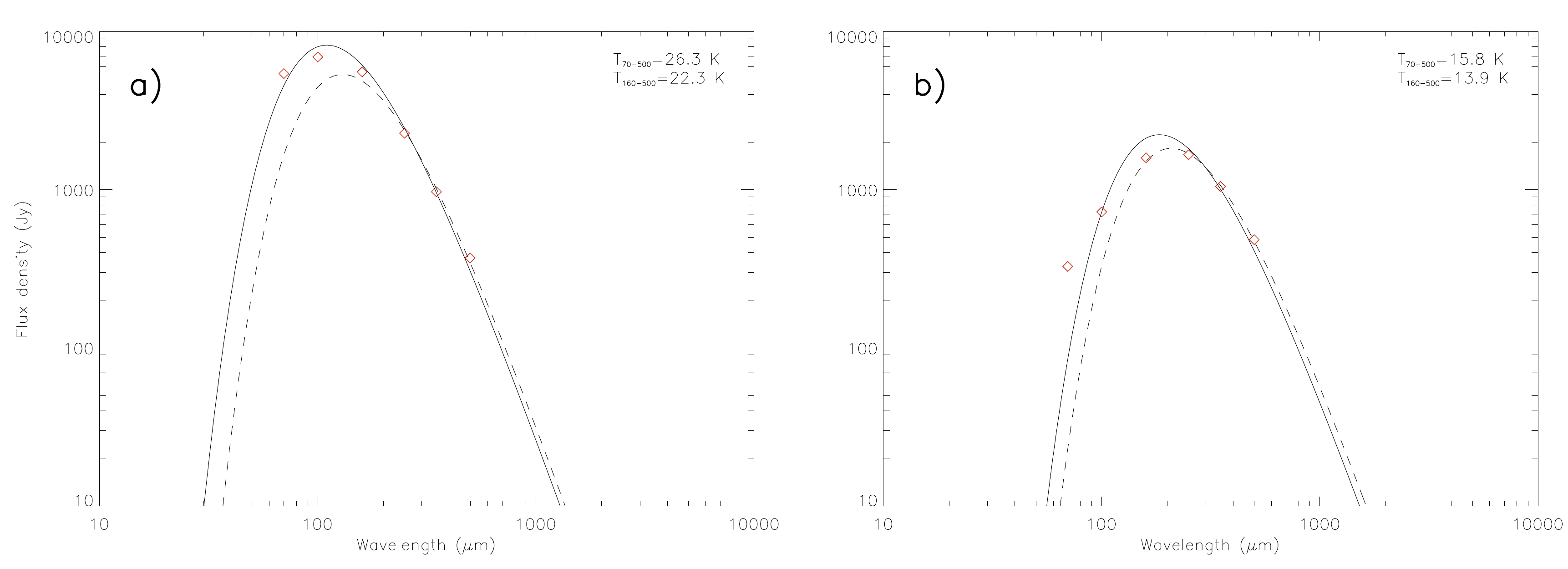}
\subfloat[\label{low_temp}]{\hspace{.5\linewidth}}
\subfloat[\label{high_temp}]{\hspace{.5\linewidth}}
\vspace{-1cm}
\caption{(a) SED fitting for the pixels giving the highest temperature. The continuous curve represents the fit made with all {\it{Herschel}} fluxes and dashed curved is the fit obtained by the method of \citet{hil12}. (b) Same for the pixel giving the lowest temperature. No background is subtracted in both cases}
\label{SEDpixel}
\end{figure*}

\begin{figure*}
\includegraphics[angle=0,width=180mm]{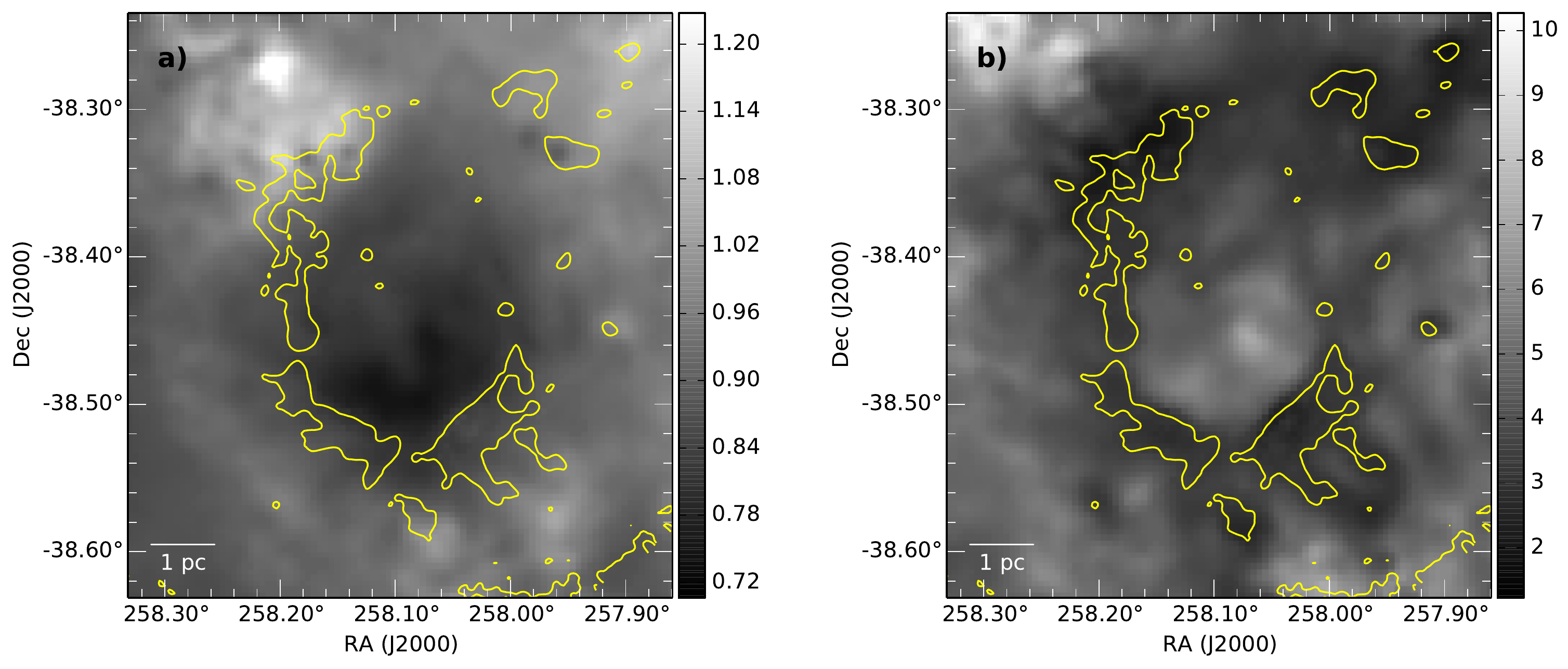}
\subfloat[\label{ratio_temp}]{\hspace{.5\linewidth}}
\subfloat[\label{ratio_dens}]{\hspace{.5\linewidth}}
\vspace{-1cm}
\caption{(a) Ratio of temperature between the maps obtained in this paper (no 70\mm and 100\mm data included and no background subtraction) over the ones obtained by \citet{and12} (see text). The yellow contours correspond to 870\mm emission at 0.1Jy/beam. (b) Same but with the column density maps}
\label{HillLoren}
\end{figure*}

\citet{and12} constructed temperature and column density maps for a sample of \HII regions (Sh~104, W5-E, Sh~241, RCW~71, RCW~79, RCW~82, G332.5-0.1 and RCW~120). Two differences exist between the method he used and the one we used (also described in \citet{hil12}). In the method used by \citet{and12}, the SED is fitted with all the data available (from 70\mm to 500\,$\mu$m) and a flat background is subtracted in each \textit{Herschel} map (see also \citet{bat11}). In warm regions (the ionized zone typically) where 70\mm and 100\mm fluxes are high (around 3$\times$10$^3$~MJy~sr$^{-1}$), the inclusion of these data in the fit induces a shift of the SED towards the high-frequency region and increases the temperature. The cold regions are less affected because the 70\mm and 100\mm fluxes are lower (around 3$\times$10$^2$~MJy~sr$^{-1}$). Spectral energy distributions representing both case for pixels giving a high and low temperature are shown in Figure~\ref{SEDpixel}. In hot regions, the difference in temperature reaches 4~K (18\%) while it is only 2~K (14\%) for cold regions.\\
\noindent To make a comparison between the temperature maps obtained using each method, we resample them at 14$\arcsec$pix$^{-1}$ to the same center and compute their ratio to see how the different methods lead to different temperature in specific regions (see Fig.~\ref{ratio_temp}). The structure seen on these images clearly reproduces the egg-shaped of RCW~120 and shows that the differences occur in specific regions. We define an area in the warmest (around the ionizing star) and coldest region (defined as condensation 5 hereafter) and compute the median and the standard deviation for them and the whole map. Results are shown in Table~\ref{tab:temp} where HR and CR stand for hot and cold regions. From the first and third line, we see, as expected, that the temperature found is higher when the 70\,$\mu$m, 100\mm and background subtraction are included particulary for the warmest region where the change is around 6~K. The colder region do not present significant difference ($\approx$0.2~K).\\
To estimate the change in temperature induced by the background subtraction, we created another temperature map using the method of \citet{and12} (including the 70\mm and 100\mm data points) but without removing any background. The range of temperature for this method is listed in the second line of Table~\ref{tab:temp}. We note, as expected, that hot regions are more affected by the inclusion of high-frequency maps ($\approx$2~K) and by the background subtraction ($\approx$5~K). The error on the fit for the temperature map has a mean value of 0.45~K hence the temperature for cold regions remains roughly the same. \\
The comparison between the two column density maps is less straighforward since \citet{and12} used only the 350\mm to obtain this map while our is the byproduct of the SED fitting. On Fig.~\ref{ratio_dens}, the ratio of the column density map shows that warm regions are more affected than colder ones. 
Due to the anticorrelation between column density and temperature, we expect warm regions to be more affected with the inclusion of the 70\mm and the 100\mm fluxes. Moreover, since the flux is linear with the column density to first order, we expect the background to be roughly equal for all the regions. Table~\ref{tab:cd} presents the values of column density for the whole map, the densest region (condensation 1 defined hereafter) and an area in the north-west of RCW~120 where the density is low (empty region) using the three different methods. Trends can be seen: as we include high frequency maps and background subtraction, the median column density decreases. The 70\mm and the 100\mm fluxes lead to a difference of 4$\times$10$^{21}$~cm$^{-2}$ for the warm region and does not change significantly the column density for cold ones. Removing the background causes a loss in column density of 1$\times$10$^{21}$~cm$^{-2}$. The method described in \citet{hil12} was the choice of the HOBYS consortium for the construction of temperature and column density maps and consequently, no background subtraction is made. This rule will allow an unbiased comparison between the different regions observed in the HOBYS project.
\begin{table}
\tiny
\caption{Range for the temperature map constructed following the method described in \citet{hil12} (first line), with all wavelengths and no background subtraction (second line) and following \citet{and12} (third line) for the whole map (first and fourth columns), the hottest region (second and fifth columns) and coldest region (third and sixth columns)}\label{tab:temp}
\centering
\begin{tabular}{|c|ccc|ccc|}
\hline 
 & \multicolumn{3}{c|}{Median} &\multicolumn{3}{c|}{$\sigma$} \\
 Method  & \multicolumn{3}{c|}{(K)} &\multicolumn{3}{c|}{(K)} \\
        & Map & HR & CR  & Map & HR & CR \\
\hline
160\mm-500\,$\mu$m, no BS &  18.5 & 23.1 & 16.4 & 1.2 & 0.9 & 0.4 \\
70\mm-500\,$\mu$m, no BS & 17.9 & 24.7 & 16.2 & 1.6 & 0.82 & 0.4 \\
70\mm-500\,$\mu$m, BS & 20.0 & 29.2 & 16.2 & 2.7 & 1.6 & 1.0 \\
\hline
\end{tabular}
\end{table}

\begin{table}
\tiny
\caption{Range for the column density map constructed following the method described in \citet{hil12} (first line), with all wavelengths and no background subtraction (second line) and following \citet{and12} (third line) for the whole map  (first and fourth columns), the densest region (second and fifth columns) and empty region (third and sixth columns)}\label{tab:cd}
\centering
\begin{tabular}{|c|ccc|ccc|}
\hline  
  & \multicolumn{3}{c|}{Median} &\multicolumn{3}{c|}{$\sigma$} \\
  Method  & \multicolumn{3}{c|}{(10$^{22}$~cm$^{-2}$)} &\multicolumn{3}{c|}{(10$^{22}$~cm$^{-2}$)} \\
 & Map & DR & ER & Map & DR & ER \\   
\hline
  160\mm-500\,$\mu$m, no BS   &  1.9 & 8.6 & 0.9 & 1.5 & 8.1 & 2.1 \\
  70\mm-500\,$\mu$m, no BS   & 1.5 & 4.4 & 1.1 & 0.8 & 3.5 & 0.2 \\
  70\mm-500\,$\mu$m, BS   & 0.4 & 3.0 & 0.3 & 0.6 & 0.3 & 0.2 \\
\hline
\end{tabular}
\end{table}

\section{Results} \label{res}
\subsection{Dust temperature and column density maps}
\begin{figure}
    \includegraphics[angle=0,width=90mm]{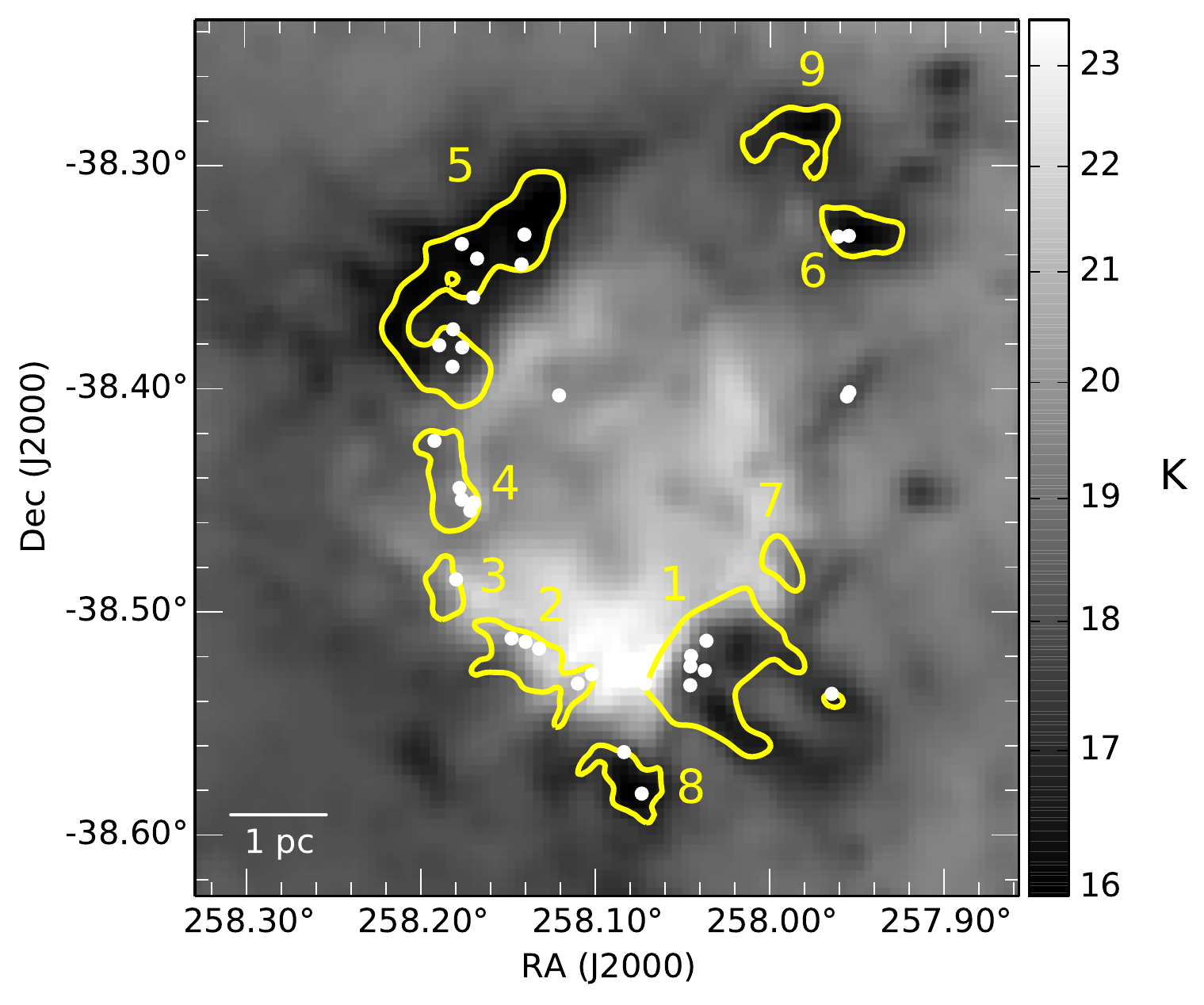}
   \caption{Temperature map of RCW~120 at 36\farcs6 resolution with 870\mm emission from LABOCA (in yellow countours) and the final sample of 35 compact sources (white dots) discussed in this paper. Condensations observed at 870\mm are identified following the labelling in DEH09. The temperature ranges from 15~K (dark) to 24~K (white). Warm regions are observed towards the ionized zone. Colder regions are located outside the ionized region and are distributed in cores, filaments and condensations}             
   \label{Figtemp}
    \end{figure}

Fig.~\ref{Figtemp} presents the temperature map obtained for RCW~120 with labelled condensations of DEH09 defined by yellow contours.  The temperature ranges from 15~K to 24~K. Temperatures between 19~K and 24~K are observed towards the ionized region, the highest temperature being observed to the south. A colder medium with a temperature around 15-18~K surrounds the \HII region. This colder medium is highly structured, organized in clumps, filaments and condensations that correspond to the condensations defined in DEH09 where the sources are located. A remarkable feature is the sharp edge seen on the temperature map at the south-western border of the ionized region. This drop in temperature (from 21~K to 16~K) corresponds to the presence of the (sub)millimeter condensation~1.

\begin{figure*}
\centering
 \includegraphics[angle=0,width=180mm]{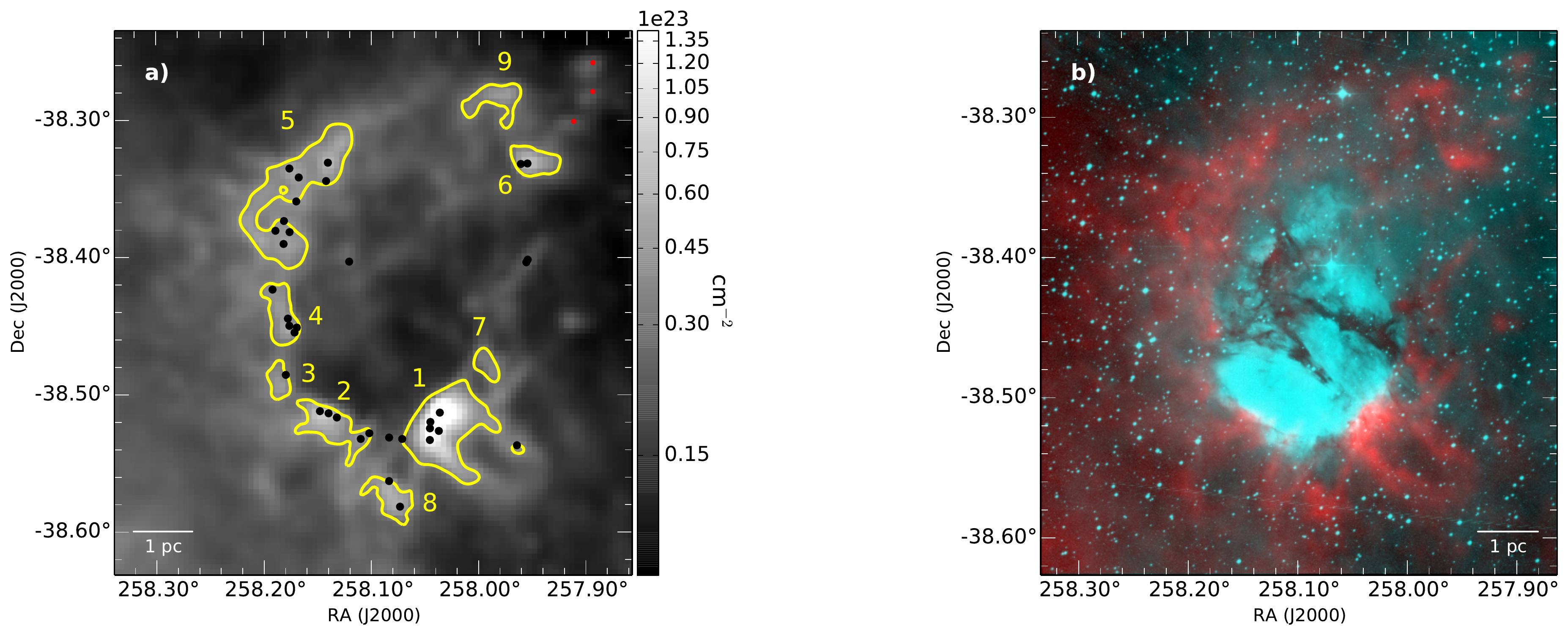}
 \subfloat[\label{low_dens}]{\hspace{.5\linewidth}}
\subfloat[\label{high_dens}]{\hspace{.5\linewidth}}
\vspace{-1cm}
   \caption{(a) On logarithmic scale, H$_2$ column density map of RCW~120 at 36$\farcs$6 resolution with 870\mm emission from LABOCA (in yellow countours), the final sample of sources (black dots) and the three prestellar clumps (red dots). Condensations observed at 870\mm are identified following the labelling in DEH09. The density values range from $7\times 10^{21}$~cm$^{-2}$ to $4\times 10^{23}$~cm$^{-2}$. (b) High resolution H$_2$ column density map of RCW~120 at 18$\farcs$2 resolution (in red) and H$\alpha$ emission (in blue) from the SuperCOSMOS H$\alpha$ Survey. The column density values range from $7\times 10^{21}$~cm$^{-2}$ to $9.4\times 10^{23}$~cm$^{-2}$}
              \label{Fig-dens}
\end{figure*}

\noindent Fig.~\ref{low_dens} presents the low resolution (36\farcs6) column density map with the condensations of DEH09 and Fig.~\ref{high_dens} the high-resolution column density map (18$\farcs$2) together with H$\alpha$ emission from the SuperCosmos H$\alpha$ survey \citep{par05}. The values range from $7\times 10^{21}$~cm$^{-2}$ to $4\times 10^{23}$~cm$^{-2}$ for the low-resolution map and goes up to $9\times 10^{23}$~cm$^{-2}$ for the high-resolution one. We checked that the convolution of the high resolution column density map to 36$\farcs$6 with the same grid agrees with the values found for the low resolution one. As expected, the ionized region with its egg-shaped corresponds to a drop in column density compared to the PDR and low column density filaments ($N$(H$_{2})=1.7\times 10^{22}$~cm$^{-2}$) are observed within it. These are seen in absorption in the optical (see fig.~1 in ZAV07) and show some compact structures that host sources (see Fig.~\ref{source_tentative} and Sect.~\ref{compsou}). Around the ionized region, a highly structured material is distributed in filaments and clumps where the nine condensations already observed at 1.3\,mm (ZAV07) and 870\mm (DEH09) are well seen. The leaking of the UV flux presented in DEH09 (see their fig. 16) is also seen on Fig.~\ref{high_dens}. It creates the extended elliptical structures observed on the southern part of the ionized region together with the structures observed on the north-eastern and north-western parts. 
Three pre-stellar clumps are seen on the temperature and density maps at ($\alpha,\delta$)=(257$\fdg$91,$-$38$\fdg$28), in absorption on the 70 and 100\mm images and in emission at 160\mm onward (see Fig.~\ref{low_dens}). In general, the size and elongation of these clumps are too large to be part of the studied sample but their detection at SPIRE wavelengths (and at 870\,$\mu$m, see Fig.~2 in DEH09) suggests that they are pre-stellar clumps.
The contrast between the high and low density regions is equal to 60. The highest density is observed in condensation 1 located at the south-western edge of the ionized region. This condensation could have been formed due to compression from the ionization region \citep{tre14}. Towards RCW~120, star formation is observed in column density region higher than $2.2\times 10^{22}$~cm$^{-2}$. 

\subsection{Compact sources} \label{compsou}
\subsubsection{Compact sources' spatial distribution} 
\begin{figure*}
 \centering
 \includegraphics[angle=0,width=170mm]{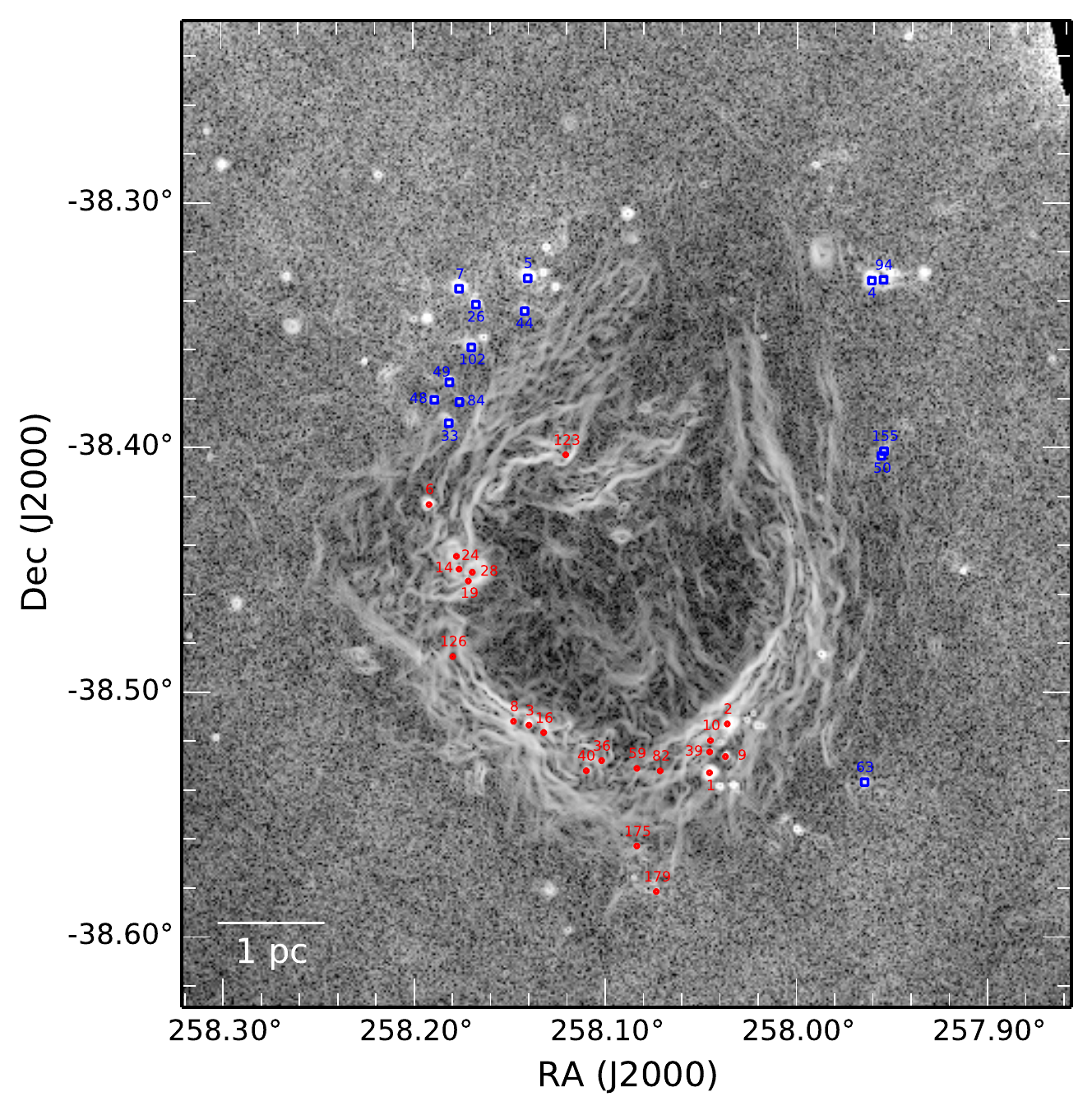}
  \caption{All 35 compact sources detected using {\it{getsources}} (and discussed in the text) superimposed on a 70\mm gradient image of RCW~120. The sources are color-coded depending on their location: red circles for sources observed towards the PDR, blue squares for sources outside (see text)}
  \label{Figsources}
\end{figure*}

\begin{figure*}
 \centering
 \includegraphics[angle=0,width=170mm]{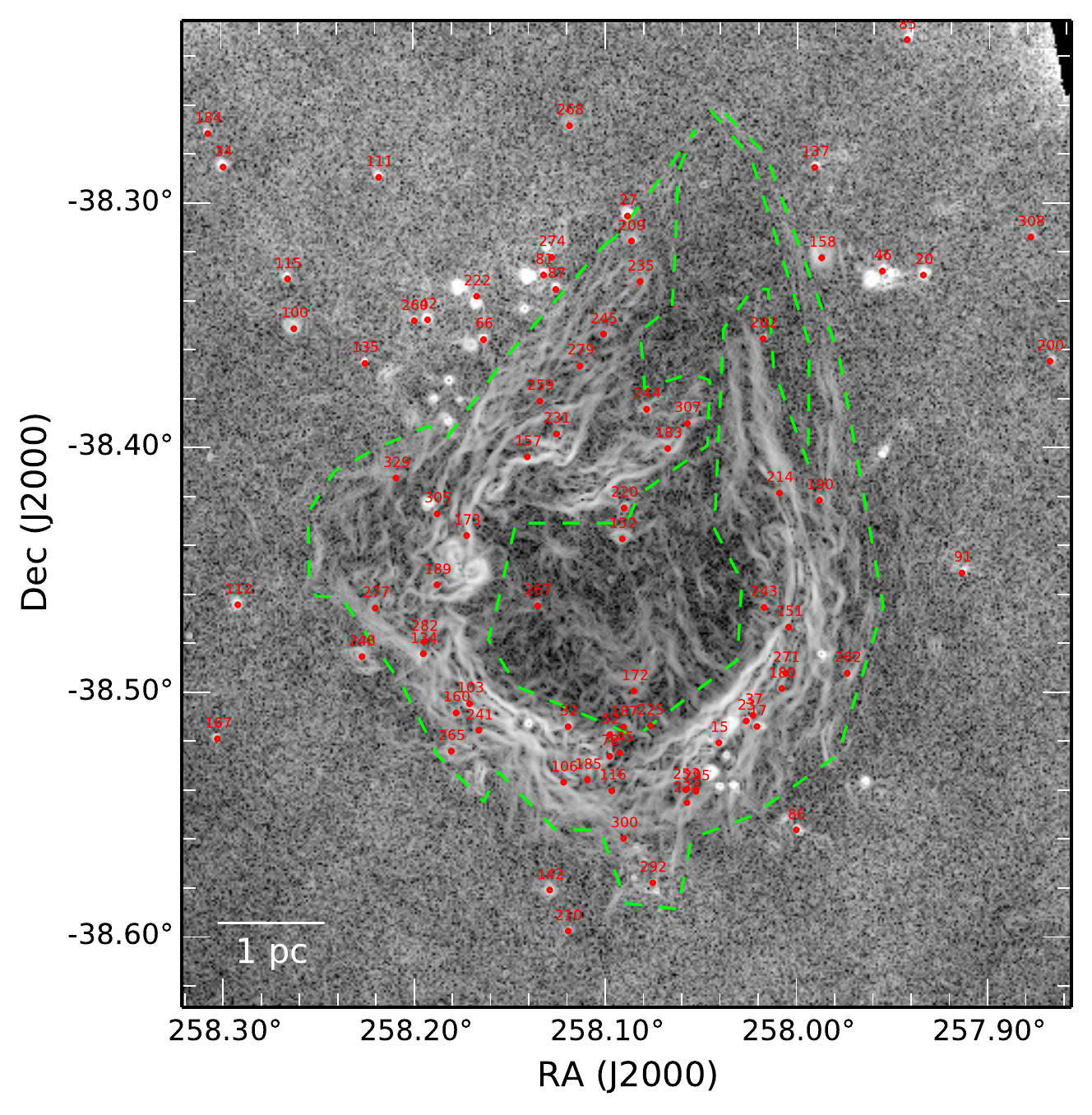}
  \caption{All 87 sources detected by {\it{getsources}} but not part of the final sample due to the lack of reliable flux measurements, mainly at SPIRE wavelengths. Physical parameters of these sources are derived in a secondhand way explained and presented in Sect.~\ref{tentative-prop}. The PDR region is enclosed in the green countours (see text)}
  \label{source_tentative}
\end{figure*}

Figure \ref{Figsources} shows the 35 selected sources superimposed on a PACS 70\mm gradient-filtered image of RCW~120. This image was produced with a standard 3$\times$3 bi-directional Sobel-Feldman convolution kernel applied to the original image. For each pixel, the derivatives along the horizontal $\left(\frac{\partial f}{\partial x}\right)$ and vertical $\left(\frac{\partial f}{\partial y}\right)$ directions are obtained and the final value for each pixel is computed as $\sqrt{\left(\frac{\partial f}{\partial x}\right)^2+\left(\frac{\partial f}{\partial y}\right)^2}$ giving an approximate value of the gradient norm. This gradient-filtering cuts-off the diffuse emission and enhances the contrast of steep emission regions. In the following we define the PDR as the filamentary emission region revealed by this gradient-filtering and shown by the green dashed contour seen in Fig.~\ref{source_tentative}.  
 
\noindent Using the selection criteria described in Sect.~\ref{analysis} and a visual inspection as a final check, we end up with 35 sources that are discussed (i.e., sources for which the temperature, envelope mass and bolometric luminosity can be derived). 87 additional detected sources are also shown in Fig.~\ref{source_tentative} but have less than two reliable fluxes (up to three if the 70\mm is included) or have unconstrained SEDs, meaning that their properties cannot be derived using SED fitting. Their original fluxes (given by {\it{getsources}} without any aperture scaling or color-corrections) are given in Table~\ref{comp}. 

As seen on Fig.~\ref{Figsources}, 14 sources are located outside the PDR and 21 are inside. 

\subsubsection{Compact sources' association with the region} \label{molecular}
Spectroscopic observations with SINFONI at the ESO-VLT showed the YSOs detected in the near IR towards RCW~120 have the same velocity as that of the ionized gas ($\approx $ $-$8 km s$^{-1}$) and are thus associated with RCW~120 \citep{mar10}. Even though most of the sources are thought to be part of RCW~120 because they are embedded in its filamentary region, sources located outside the PDR might not be associated with the region. 

\begin{figure*}
 \includegraphics[width=180mm]{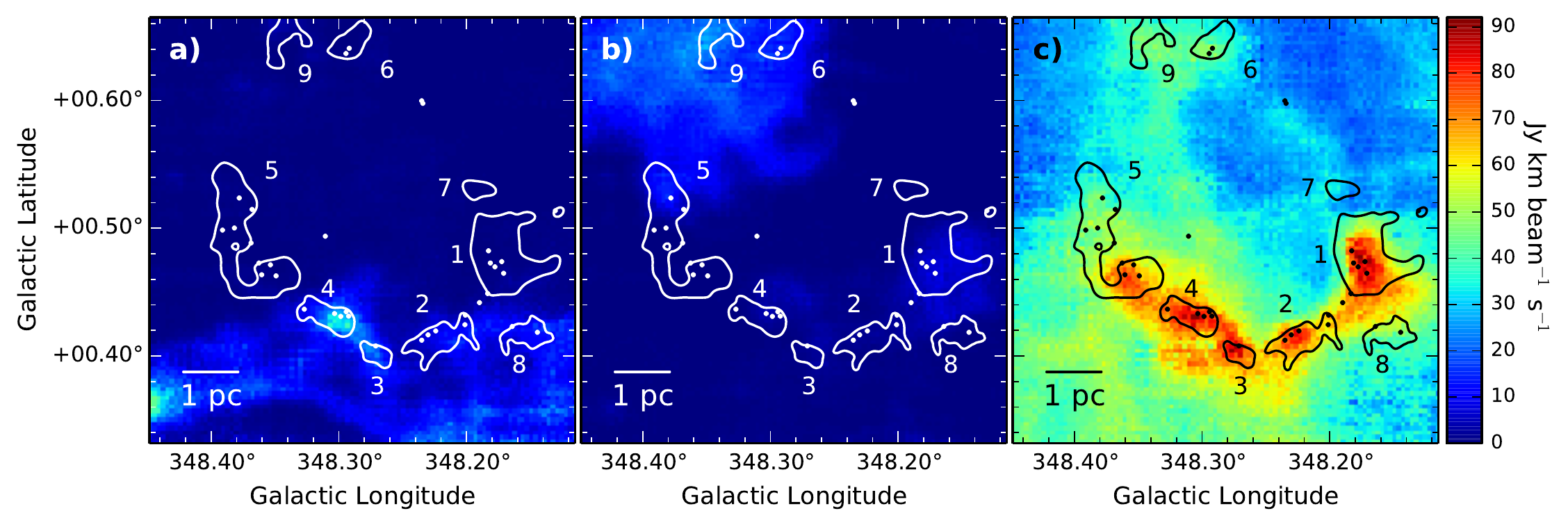}
  \caption{Integrated intensity of $^{12}$CO ($J=0 \to 1$) between (a) $-$75~km~s$^{-1}$ and $-$50~km~s$^{-1}$ (b) $-$35~km~s$^{-1}$ and $-$15~km~s$^{-1}$ (c) $-$15~km~s$^{-1}$ and 3~km~s$^{-1}$. The dots represent the 35 sources of the final sample and the contours stand for the 870\mm condensations of DEH09. The unit of the color image is in Jy~km~s$^{-1}$ beam$^{-1}$}
  \label{Map12CO}
\end{figure*}

\begin{figure*}
\includegraphics[width=180mm]{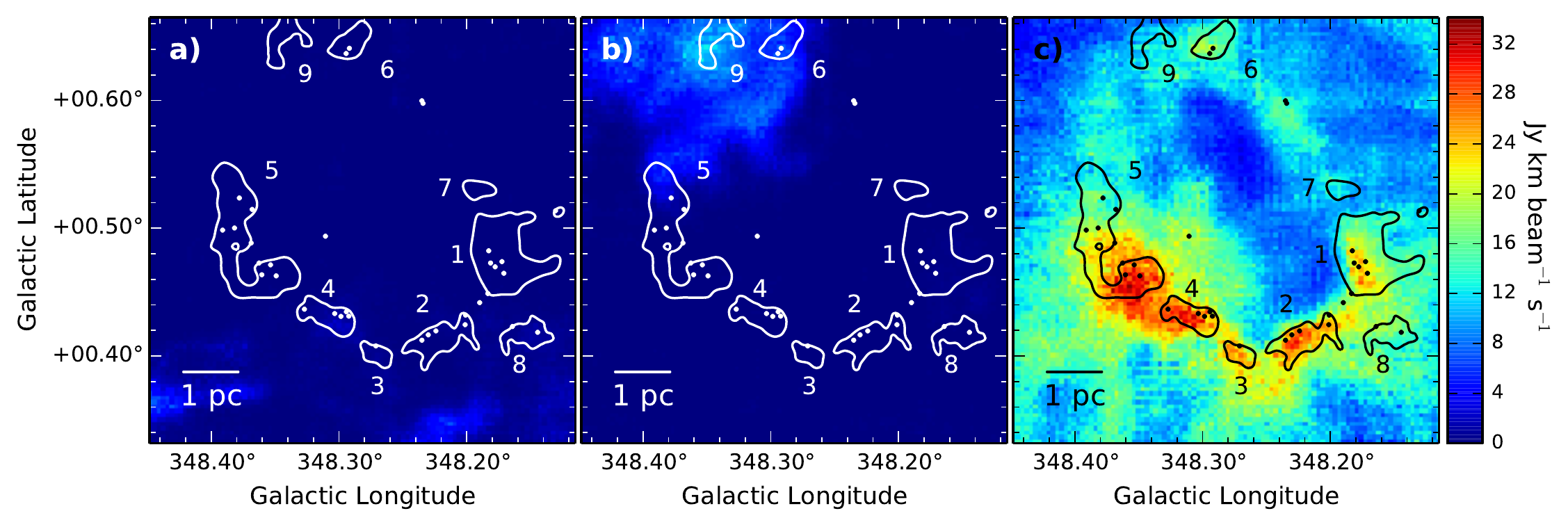}
  \caption{Integrated intensity of $^{13}$CO ($J=0 \to 1$) within the same velocity ranges. Dots and contours are the same as in Fig.~\ref{Map12CO}}
  \label{Map13CO}
\end{figure*}

\begin{figure*}
 \includegraphics[width=180mm]{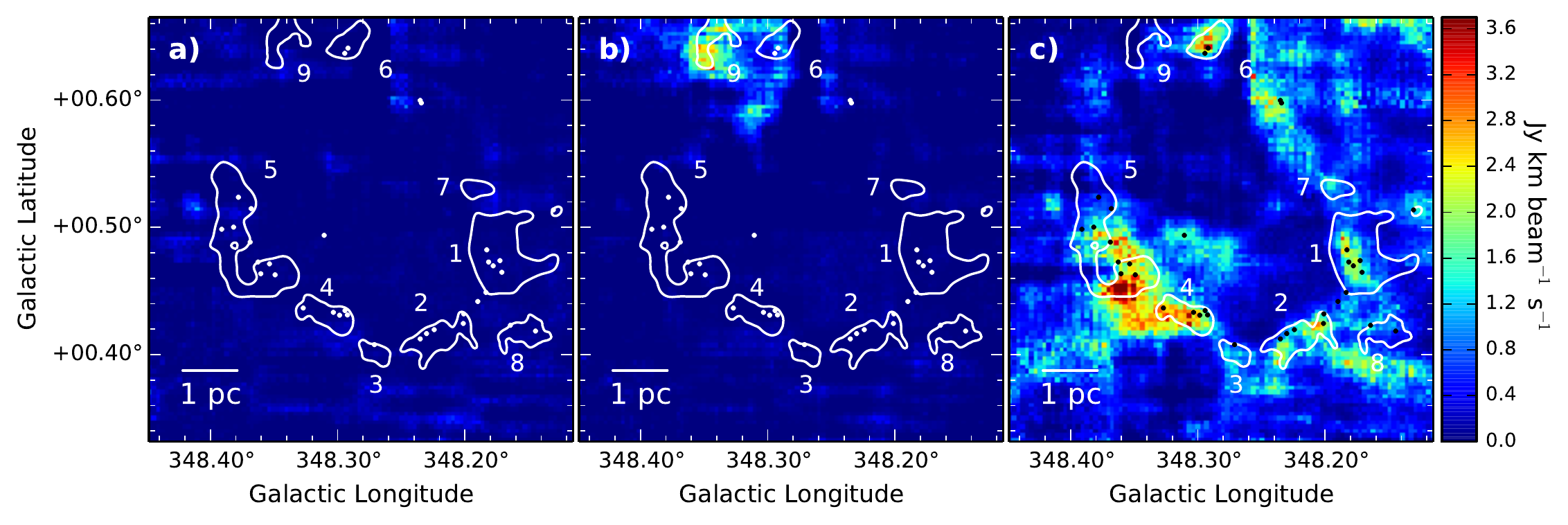}
  \caption{Integrated intensity of C$^{18}$O ($J=0 \to 1$) within the same velocity ranges. Dots and contours are the same as in Fig.~\ref{Map12CO}}
  \label{MapC18O}
\end{figure*}

Studying the $J=0\to1$ transition of $^{12}$CO, $^{13}$CO, C$^{18}$O and C$^{17}$O with the ATNF Mopra 22~m radio telescope, \citet{and15} identified three emission peaks at $-$7 km s$^{-1}$ (main temperature peak around the velocity of the ionized gas), $-$30~km~s$^{-1}$ and $-$60~km~s$^{-1}$. The $J=0\to1$ emission from the CO isotopologues \citep{and15} integrated between $-$75~km~s$^{-1}$ and $-$50~km~s$^{-1}$, $-$35~km~s$^{-1}$ and $-$15~km~s$^{-1}$ and $-$15~km~s$^{-1}$ and $+$3~km~s$^{-1}$ are presented in Fig.~\ref{Map12CO} for $^{12}$CO, Fig.~\ref{Map13CO} for $^{13}$CO and Fig.~\ref{MapC18O} for C$^{18}$O. We point out that, contrary to the other maps in this paper, these figures are given in galactic coordinates.

\noindent Condensation 6, the northern-part of condensation 5 and sources 55 and 150 (the western part, between condensation 6 and 7) are located outside the PDR but present an emission peak around $-$15~km~s$^{-1}$ and $+$3~km~s$^{-1}$ which indicates that they are part of RCW~120. Condensation 9 presents an emission peak in the same range but also between $-$35~km~s$^{-1}$ and $-$15~km~s$^{-1}$. Although we cannot rule out the fact that condensation 9 might in the foreground or background of the region, the emission peak is stronger in the main-peak velocity range and therefore, we consider this condensation to be part of RCW~120.
Between $-$15~km~s$^{-1}$ and $+$3~km~s$^{-1}$, the other condensations are distributed along the strong CO emission, following the PDR that surrounds the ionized region. 
This strongly suggests that the 35 sources of the final sample are indeed associated with RCW~120. 


 \subsubsection{Compact source properties}\label{source-prop}
We have shown that the detected sources are likely to be associated with RCW~120, that is, located at the same distance (see Sect.~\ref{molecular}). Table~\ref{tm} gives the physical properties derived for the 35 sources: the {\it{getsources}} identification number (identification number in DEH09 given in parenthesis if any), the envelope temperature T, envelope mass M$_{\rm{env}}$, bolometric luminosity L$_{\rm{bol}}$, ratio of the submillimetric luminosity (defined as the luminosity computed from 350\,$\mu$m onwards) over the bolometric luminosity, ratio of the envelope mass over the bolometric luminosity, associated condensation towards which the source is observed, evolutionary class derived from the study of DEH09 and from L$_{\lambda\ge 350\mu m}$/L$_{\rm{bol}}$, near- and mid-infrared counterparts, and volume density.

\noindent In the following, source ID refers to  IDs given in column 1 of Table~\ref{tm} and Table~\ref{comp}. 
\noindent Among the 35 sources of the final sample, 14 match the previous list discussed in DEH09. The sources previously identified on the basis on GLIMPSE and MIPSGAL data are now identified at \herschel wavelengths, based on their spatial correspondance in a radius of 4$\arcsec$. We discuss their evolutionary class in Sect.~\ref{dis}. Adopting $\beta=1.6$ from the latest $\it{Planck}$\, results, modifies the physical parameters (envelope temperature, envelope mass and bolometric luminosity) by 10\% in average throughout the MBB model and color corrections factors. A higher value of $\beta$ better represents denser regions \citep{par12}.

\clearpage
\begin{minipage}[t]{\textwidth}
\onecolumn
\centering
\tiny
\begin{longtable}{|ccrrrrccccc|}
\hline
ID\footnote{Identification number given by \textit{getsources} (corresponding identification number in DEH09 if present)} & T\footnote{Temperature derived using the SED fitting with a MBB model} & M$_{\rm{env}}$\footnote{Mass of the core using the Hildebrand model (See Sect~\ref{SED})} & L$_{\rm{bol}}$\footnote{Luminosity computed from the lowest IR counterpart (if present) to the end of the SED} & M$_{\rm{env}}$/L$_{\rm{bol}}$ & L$_{\lambda \geq 350\mu m}$/L$_{\rm{bol}}$ & Cond\footnote{Condensation to which belongs the source (numbering of DEH09)} & Class (DEH09)\footnote{Classification from DEH09 using color-color and magnitude-color diagrams} & L$_{\lambda \geq 350\mu m}$/L$_{\rm{bol}}$\footnote{Classification using the L$_{\lambda\ge 350\mu m}$/L$_{\rm{bol}}$ criteria from \citet{bon10}. If L$_{\lambda\ge 350\mu m}$/L$_{\rm{bol}}$ < 0.01, the class of the source is at least I} & IR\footnote{Infrared counterparts within a radius of 4\arcsec : * stands for IR counterparts seen but not measured. ** stands for source whose counterparts (if any) might be affected or misleaded by heated filament (not measured)} & <n$_{\rm{H_{2}}}$>\footnote{Average volume density computed from the radius at the reference wavelength. > symbol indicates a lower limit for non-resolved source} \\
      &   (K)      &   ($M_{\sun}$)  &  ($L_{\sun}$) &   & (\%)  & &  & & & (cm$^{-3}$) \\   
\hline
\hline
1(38)   &       19.8    &       36      &       281     &       0.13    &       1.78    &       1       &       I       &       0-I     &       3.6\mm  to      24\mm   &       >2$\times$10$^{7}$      \\      
2       &       20.3    &       174     &       1163    &       0.15    &       2.17    &       1       &       &       0-I     &       &       >1$\times$10$^{8}$      \\                                      
3(50)   &       20.9    &       14      &       167     &       0.08    &       1.26    &       2       &       I       &       0-I     &       2.17\mm to      24\mm   &       >8$\times$10$^{6}$      \\      
4(106)  &       19.1    &       9       &       92      &       0.1     &       1.22    &       6       &       I-flat? &       0-I     &       2.17\mm to      24\mm   &       >5$\times$10$^{6}$      \\      
5(107)  &       17.3    &       11      &       77      &       0.14    &       1.42    &       5       &       I-II?   &       0-I     &       1.65\mm to      24\mm   &       >6$\times$10$^{6}$      \\      
6(76)   &       23.2    &       1       &       76      &       0.02    &       0.36    &       4       &       I-II?   & I &     1.65\mm to      24\mm   &       >9$\times$10$^{5}$      \\      
7(105)  &       20.9    &       3       &       36      &       0.09    &       1.42    &       5       &       I       &       0-I     &       3.6\mm  4.5\mm  5.8\mm  24\mm   &       >2$\times$10$^{6}$      \\
8       &       16.4    &       30      &       58      &       0.53    &       4.86    &       2       &       &       0       &       &       >2$\times$10$^{7}$\\                                            9(40)   &       13.3    &       126     &       68      &       1.8     &       9.82    &       1       &       I       &       0       &       3.6\mm  to      24\mm   &       2$\times$10$^{6}$       \\      
10      &       13.8    &       86      &       69      &       1.24    &       7.44    &       1       &       &       0       &       4.5\mm  &       2$\times$10$^{7}$       \\                              
14(69)  &       19.2    &       16      &       97      &       0.16    &       2.13    &       4       &       I       &       0-I     &       1.25\mm to      24\mm   &       >1$\times$10$^{7}$      \\      
16      &       24.5    &       5       &       115     &       0.05    &       1       &       2       &       &       0-I     & ** &    1$\times$10$^{6}$       \\                                      
19(67)  &       21.6    &       4       &       85      &       0.05    &       0.85    &       4       &       II      &       I       &       1.25\mm to      24\mm   &       1$\times$10$^{6}$       \\      
24(ObjectA)     &       26.9    &       2       &       86      &       0.03    &       0.68    &       4       &       &       I       & 5.8\mm to 24\mm * &     >4$\times$10$^{5}$      \\                                      
26(103) &       21.1    &       1       &       17      &       0.03    &       0.53    &       5       &       I-II?   &       I       &       1.25\mm to      24\mm   &       >3$\times$10$^{5}$      \\      
28(ObjectB)     &       34.1    &       1       &       212     &       0.01    &       0.24    &       4       &       &       I & 1.25\mm       to      24\mm * &       >2$\times$10$^{5}$      \\                                      
33(88)  &       17.3    &       4       &       12      &       0.31    &       3.19    &       5       &       I       &       0       &       2.17\mm to      24\mm   &       >2$\times$10$^{6}$      \\      
36      &       27.1    &       1       &       28      &       0.03    &       0.66    &       2       bis     &       &       I       & ** &    >4$\times$10$^{5}$      \\                              
39      &       12.6    &       77      &       32      &       2.4     &       11.34   &       1       &       &       0       & ** &    >1$\times$10$^{7}$      \\              
40      &       15.7    &       11      &       16      &       0.69    &       5.72    &       2       bis     &       &       0       &       &       >7$\times$10$^{6}$      \\                              
44      &       18.3    &       1       &       6       &       0.23    &       2.66    &       5       &       &       0-I     &       All     except  8\mm    &       >9$\times$10$^{5}$      \\              
48      &       19.7    &       1       &       7       &       0.14    &       1.95    &       5       &       &       0-I     &       24\mm   &       >6$\times$10$^{5}$      \\                              
49(91)  &       19.6    &       1       &       7       &       0.11    &       1.46    &       5       &       I       &       0-I     &       1.65\mm to      24\mm   &       >5$\times$10$^{5}$      \\      
50      &       18.1    &       2       &       7       &       0.3     &       3.39    &       Fil.    &       &       0       &       &       >1$\times$10$^{6}$      \\                                      
59      &       20.8    &       2       &       16      &       0.13    &       1.97    &       &       &       0-I     & ** &    >1$\times$10$^{6}$      \\                                              
63(37)  &       12.7    &       16      &       13      &       1.17    &       5.61    &       Fil     &       I-II?   &       0       &       3.6\mm  to      24\mm   &       >1$\times$10$^{7}$      \\      
82      &       18.6    &       3       &       12      &       0.25    &       3.04    &       1       &       &       0-I     &       &       >1$\times$10$^{6}$      \\                                      
84      &       18.2    &       2       &       5       &       0.28    &       3.31    &       5       &       &       0-I     &       &       >9$\times$10$^{5}$      \\                                      
94      &       11.9    &       70      &       41      &       1.69    &       6.71    &       6       &       &       0       &       2.17\mm to      24\mm   &       >4$\times$10$^{7}$      \\              
102     &       18.7    &       3       &       15      &       0.18    &       2.22    &       5       &       &       0-I     &       1.65\mm to      24\mm   &       >2$\times$10$^{6}$      \\              
123     &       20      &       2       &       10      &       0.16    &       2.28    &       &       &       0-I     &       &       3$\times$10$^{5}$       \\                                              
126     &       23.4    &       2       &       31      &       0.06    &       1.22    &       3       &       &       0-I     &       &       >2$\times$10$^{5}$      \\                                      
155(86) &       15.1    &       2       &       5       &       0.42    &       3.13    &       Fil.    &       I       &       0       &       1.25\mm to      24\mm   &       >1$\times$10$^{6}$      \\      
175     &       13.3    &       9       &       5       &       1.78    &       9.6     &       8       &       &       0       &       1.25\mm to      24\mm   &       >5$\times$10$^{6}$      \\              
179     &       11.2    &       45      &       9       &       5.23    &       17.53   &       8       &       &       0       &       &       >3$\times$10$^{7}$      \\                                      
\hline
\caption{Properties of the 35 compact sources discussed in the text}\label{tm}
\end{longtable}
\end{minipage}
\twocolumn

\begin{figure}
 \centering
 \includegraphics[angle=0,width=90mm]{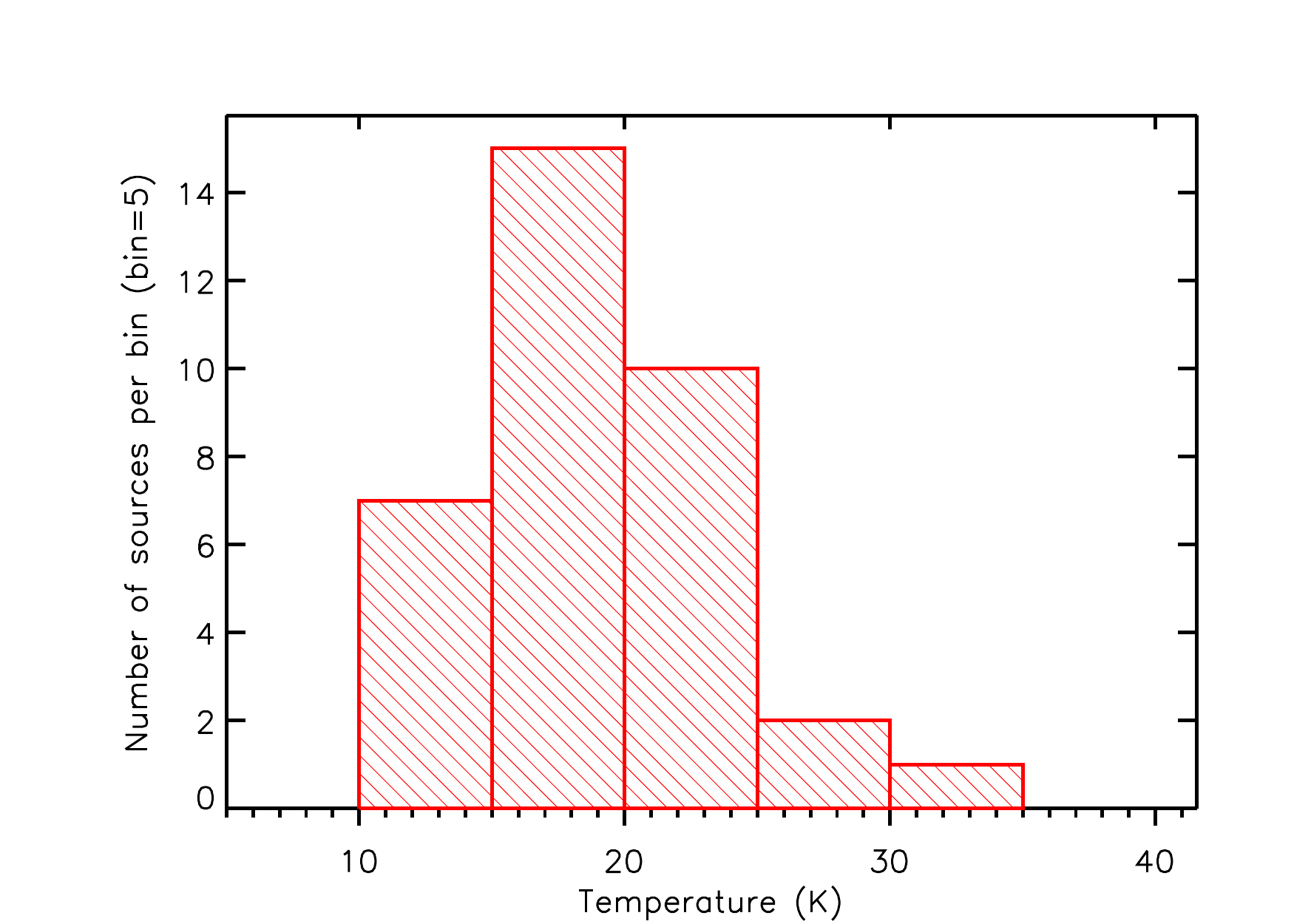}
  \caption{Envelope temperature distribution for the 35 sources observed towards RCW~120}
             \label{tempdis}
\end{figure}
Figure~\ref{tempdis} presents the distribution of dust envelope temperature for compact sources observed towards RCW~120. All but three sources (24, 28, 36) have envelope temperature lower than 25~K. As discussed in ZAV07, sources 24 and 28 are observed towards condensation 4. They are classified as Herbig Ae/Be objects and contain a central star of spectral type B4V for source 24 and B7V for source 28. Their extended nature is thought to be the result of local PDR due to radiation of the star but not massive enough to form \HII regions which is consistent with the envelope mass derived, 2~$M_{\sun}$ and 1~$M_{\sun}$ for source 24 and 28, respectively. Source 36 is located in a region of low density and high temperature (23~K).\\

\begin{figure}
 \centering
 \includegraphics[angle=0,width=90mm]{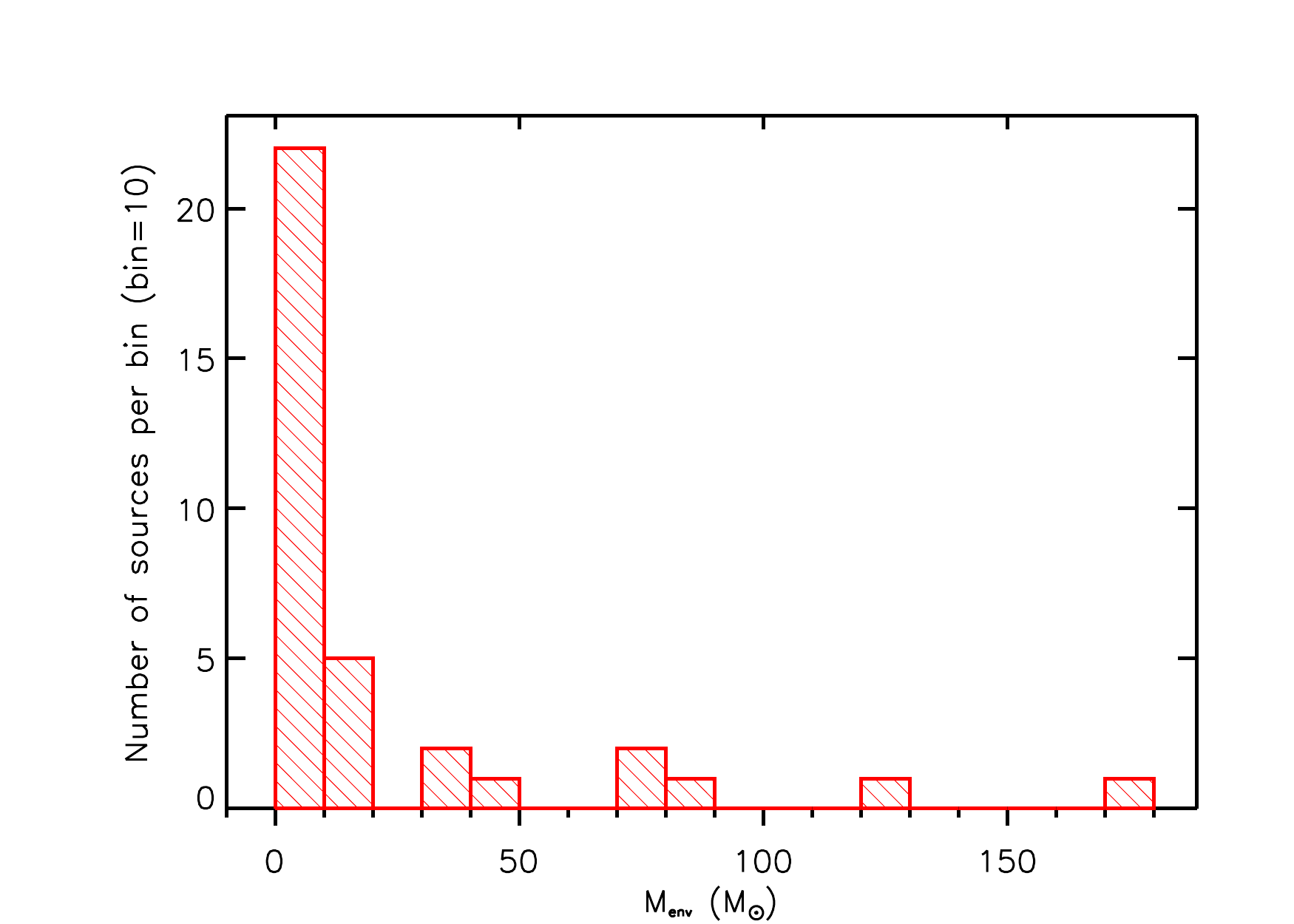}
  \caption{Histogram of envelope mass for sources observed towards RCW~120}
             \label{envmass}
\end{figure}

Figure~\ref{envmass} presents the distribution of envelope mass for sources observed towards RCW~120. Twenty-seven sources have a low mass (M$_{\rm{env}} \leq 20~M_{\sun}$) envelope. Sources with envelope mass up to 1~$M_{\sun}$ are detected here. From their \herschel study of dense cores in NGC~6334, Tigé et al. (2016, submitted) derived an envelope mass limit of 60~$M_{\sun}$ (lower limit at which they detect ongoing high-mass star activity) for a core to form a high-mass star.
Five sources (2, 9, 10, 39, 94) have envelope mass higher than this limit and four (2, 9, 10, 39) are located in condensation 1. Source 94 is located in condensation 6. The column density towards these condensations is higher than $1.7\times 10^{22}$ cm$^{-2}$. We point out the fact that the high-mass cores represent 15\% of the total number of sources in the final sample. Nevertheless, two biases could radically change the result. Firstly, it is possible that the cores are unresolved even at the best \herschel resolution (5\farcs9) and represents more than one YSO then decreasing the number of possible high-mass cores. We are confident that source 2 might represent this problematic case. Secondly, our final sample represent reliable sources but is incomplete. According to our selection criteria, the non-selected sources should represent low-mass objects. Consequently, this value of 15\% should represent a higher-limit to the number of high-mass cores.

\begin{table}
\caption{Physical properties of the final sample}\label{tab:prop_final}
\centering

\begin{tabular}{|c|cccc|}
\hline
Physical & T & M$_{\rm{env}}$ & L$_{\rm{bol}}$ & $<n_{H_{2}}>$ \\
parameter & (K) & ($M_{\sun}$) & ($L_{\sun}$) & (cm$^{-3}$) \\
\hline
Min & 11.2 & 1 & 5 & 2$\times$10$^5$ \\
Max & 34.1 & 174 & 1163 & 1$\times$10$^8$ \\
Median & 19.1 & 4 & 30 & 2$\times$10$^6$ \\
\hline
\end{tabular}
\end{table}

Table \ref{tab:prop_final} summarizes the physical properties of the final sample of sources.

\subsubsection{Properties of the tentative sources}\label{tentative-prop}
\begin{figure*}
 \centering
 \includegraphics[angle=0,width=180mm]{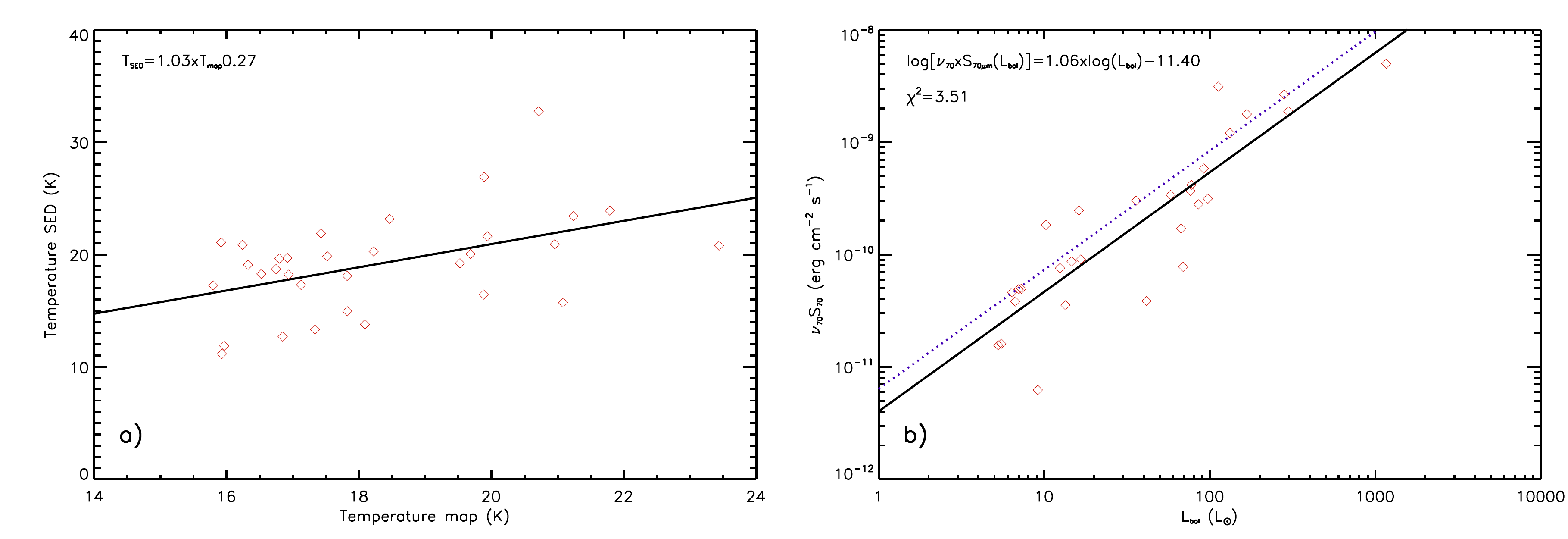}
 \subfloat[\label{Temp-Temp_fit}]{\hspace{.5\linewidth}}
  \subfloat[\label{Temp-Lbol}]{\hspace{.5\linewidth}}
\vspace{-1cm}
 \caption{(a) Temperature given by the SED fitting versus temperature obtained at the source location in the temperature map for the final sample of 35 sources (red diamonds). (b) $\nu_{70\mu m}\times$S$_{70\mu m}$ versus bolometric luminosity for the final sample of 35 sources following \citet{rag12} where the black continuous line represents the fit and the blue dotted one represented the relation from \citet{dun08}}
\label{Temp_map-Temp_fit}
\end{figure*}

Physical properties of the 87 tentative sources shown in Fig.~\ref{source_tentative} could not be obtained directly with the SED fitting due to a lack of \herschel measurements (see Sect.~\ref{analysis}). From the final sample of 35 sources, we fitted the envelope temperature obtained from the SED versus the temperature found at the source location on the temperature map (Fig.~\ref{Temp-Temp_fit}). We then used the fitted linear relation to assign a temperature to each of the tentative sources. The mass is computed using the Hildebrand formula \citep{hil83} with one of the \herschel fluxes. For the bolometric luminosity, we used the relation between the flux density and the bolometric luminosity (log$_{10}$(F$_{\nu}$)$\propto$log$_{10}$(L$_{bol}$)) using the final sample (Fig.~\ref{Temp-Lbol}) \citep{rag12}. The relation of \citet{dun08} for embedded protostars is normalized at 1.3~kpc 

\begin{equation}\label{Eq:dunham}
\begin{array}{ll}
Log_{10}(F_{\nu}^{1.3kpc})&=Log_{10}\left(\left(\frac{0.14}{1.3}\right)^2 F_{\nu}\right)\\
&=1.06Log_{10}(F_{\nu})-11.2
\end{array}
,\end{equation}

where $F_{\nu}^{1.3kpc}$ is the flux density at 1.3~kpc and $F_{\nu}$ is the flux density of \citet{dun08}. This relation is represented in Fig.~\ref{Temp_map-Temp_fit} by the blue-dotted line. Results are given in Appendix~\ref{apendix:prop_tent}.

\subsubsection{Mass of the condensations using the H$_2$ column density map}\label{condensation_mass}

Using the H$_2$ column density map, we derived the mass of each condensation defined by the same area as the one used by DEH09 to compute the mass from APEX 870\mm data. We used the following formula: 

\begin{equation} \label{eqmass}
M=A_{\rm{pixel}} \times \mu m_{\rm{H}} \times \sum_{ij} N^{ij}_{H_{2}}
,\end{equation}
where A$_{\rm{pixel}}$ is the area of a pixel in cm$^2$, $\mu$ is the mean molecular weight ($\approx$2.8), $m_{\rm{H}}$ is the hydrogen atom mass and N$^{ij}_{H_{2}}$ is the H$_2$ column density value at pixels (i,j). DEH09 computed the mass using the Hildebrand formula with T=20~K (and also with T=30~K but this value for the condensations is too high compared to the ones derived from the \herschel temperature map). The results are given in Table~\ref{colmass}. Column 1 gives the condensation number from DEH09. Column 2 gives the condensation's mass derived using the H$_2$ column density map and Eq.~\ref{eqmass}, and column 3 the mass derived by DEH09 using the 870~\mm data, assuming a dust temperature of 20~K. 

\noindent Compared to DEH09, we obtain higher masses for the condensations. At first sight, the absence of background subtraction could explain this difference but since they are massive, the background only accounts for a small amount of the total pixels value. The main difference between the two results could be explained by the extending emission filtering of the ground-based telescope at 870$\mu$m, leading to an underestimation of the mass \citep{cse16}. We point out that the condensation mass is critical for star-formation rate and star-formation efficiency estimates that will be further analyzed in a forthcoming paper (Liu et al. submitted).

\begin{table}
\caption{Condensations' mass using the low resolution density map (second column) and from DEH09 at 20~K (third column)}\label{colmass}
\centering
\begin{tabular}{|c|r|r|}
\hline %
Condensation  &  M$_{\rm{H_2}}$ & $M_{\rm{870\, \mu m}}$ \\
   & ($M_{\sun}$) &  ($M_{\sun}$) \\
\hline
  1 & 2530 & 800 \\
  2 & 540 & 192 \\
  3 & 140 & 63 \\
  4 & 350 & 88 \\
  5  & 1580 & 226 \\
  6  & 330 & 90 \\
  7 & 86 & 28.5 \\
  8  & 370 & 38  \\
  9 & 290 & 42 \\
\hline
\end{tabular}
\end{table}

\section{Discussion}  \label{dis}
\subsection{Compact sources' evolutionary stage}\label{evolution}


As the submillimetric luminosity depends on the envelope mass and the bolometric luminosity on the stellar mass, \citet{and93} proposed to use the submillimetric to bolometric luminosity ratio as an evolutionary indicator. \citet{bon10} used the same kind of criteria to distinguish Class~0 and Class~I objects in the Aquila Rift using \herschel data but the limits used were different. Class~0 are defined with  L$_{\lambda\ge 350\mu m}$/L$_{\rm{bol}}$>0.03 while it is higher than 0.01 for Class~I. The region between 0.01 and 0.03 contains sources with uncertain classification. 

\begin{figure}[ht!]
\includegraphics[angle=0,width=90mm]{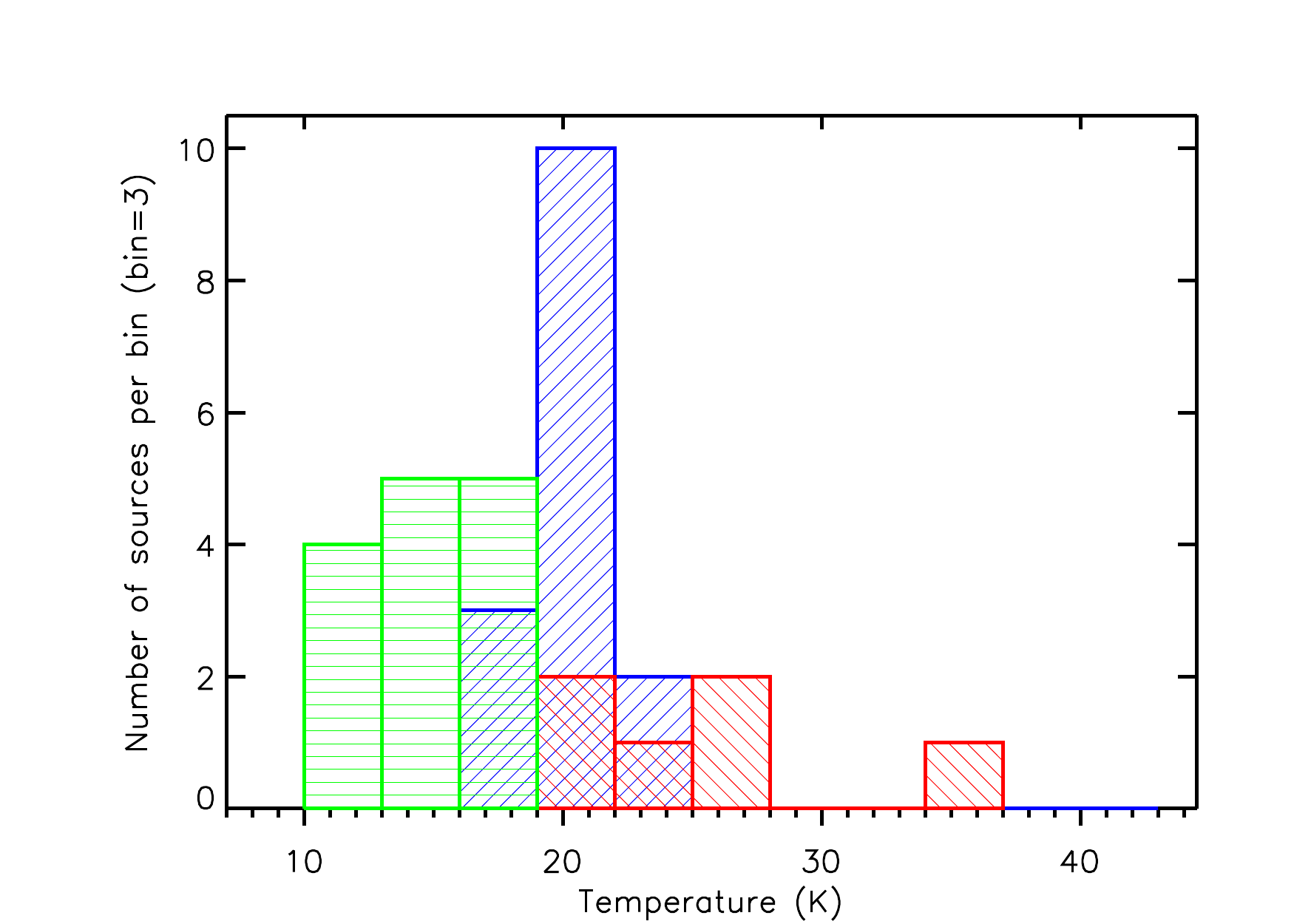}
   \caption{Histogram of the temperature color-coded according to L$_{\lambda\ge 350\mu m}$/L$_{\rm{bol}}$ in green for Class~0, red for Class~I and blue for uncertain cases}\label{Temp_class}
\end{figure}

\begin{figure*}
\includegraphics[angle=0,width=180mm]{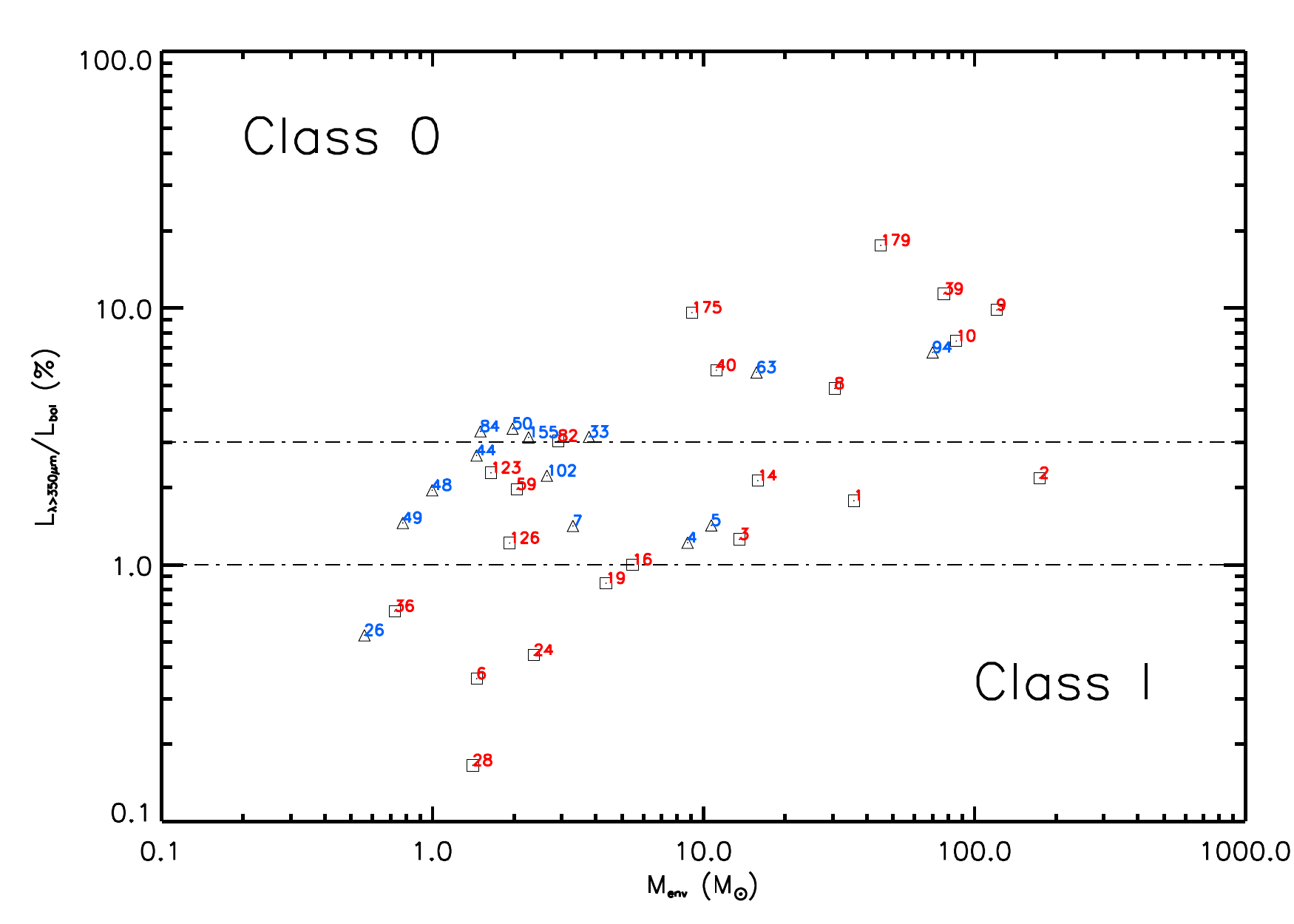}
   \caption{L$_{\lambda\ge 350\mu m}$/L$_{\rm{bol}}$ versus M$_{\rm{env}}$. The dotted-dashed lines represents the L$_{\lambda\ge 350\mu m}$/L$_{\rm{bol}}$ limits between Class~0 and Class~I from \citet{bon10} and sources are color-coded depending on their location: red squares for sources inside the PDR, blue triangles for outside}
              \label{ML-LL}
    \end{figure*}

Figure~\ref{Temp_class} shows the distribution of sources' envelope temperature, color-coded according to their L$_{\lambda\ge 350\mu m}$/L$_{\rm{bol}}$ value. As expected, there is a relation between the L$_{\lambda\ge 350\mu m}$/L$_{\rm{bol}}$ value and the envelope temperature of the source. Class~I objects (in red) have a higher temperature than Class~0 objects (in green) while the uncertain cases (in blue) are located in between.
    \newpage
Figure~\ref{ML-LL} shows the sample of the 35 compact sources on the L$_{\lambda\ge 350\mu m}$/L$_{\rm{bol}}$ versus M$_{\rm{env}}$ diagram coded depending on their location with respect to the PDR. 80\% of the sources are located in the Class~0 region with L$_{\lambda\ge 350\mu m}$/L$_{\rm{bol}}$>0.01. If an age gradient was at work in the region, sources towards the PDR would have been under the Class~I limit and sources outside the PDR would have been above the Class~0 limit. Depending on the Class~0 limit taken, 1\% for \citet{and93} or 3\% for \citet{bon10}, a weak trend in favor of very young objects out of the PDR can be seen. Towards the PDR, no trend is seen since these sources spread over the entire range of L$_{\lambda\ge 350\mu m}$/L$_{\rm{bol}}$ values.

\begin{figure*}
\includegraphics[angle=0,width=180mm]{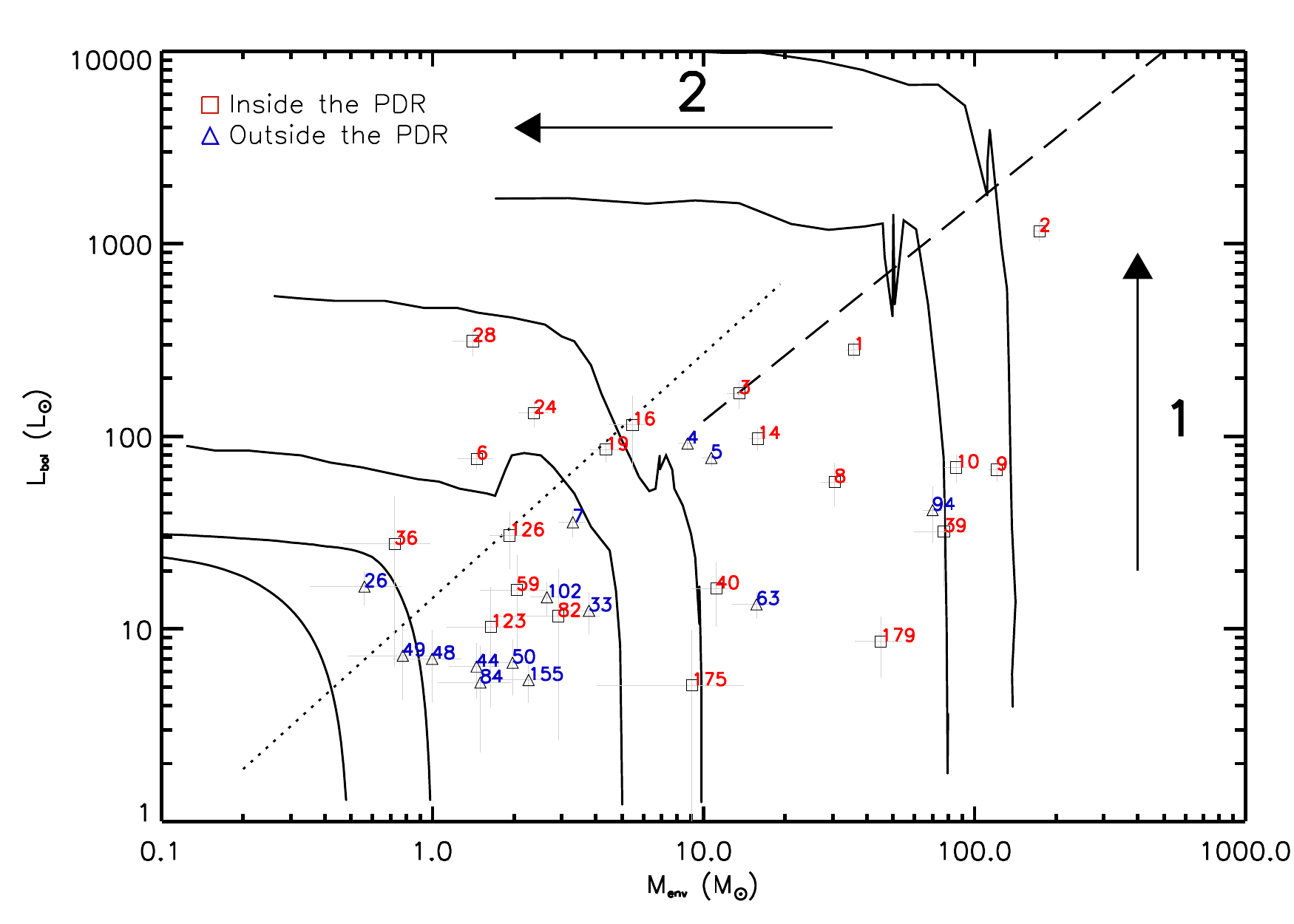}
   \caption{L$_{\rm{bol}}$ versus M$_{\rm{env}}$. Evolutionary tracks are adapted from \citet{sar96} and \citet{mol08}. Labeled arrows indicate (1) the accreation phase and (2) envelope cleaning phase. Sources are coded as a function of their location with respect to the PDR : red squares sources are for the ones observed towards the PDR and blue triangles for those outside. Error bars for L$_{bol}$ and M$_{env}$ are shown by gray lines}
              \label{evol}
    \end{figure*}

\begin{figure*}
\includegraphics[angle=0,width=180mm]{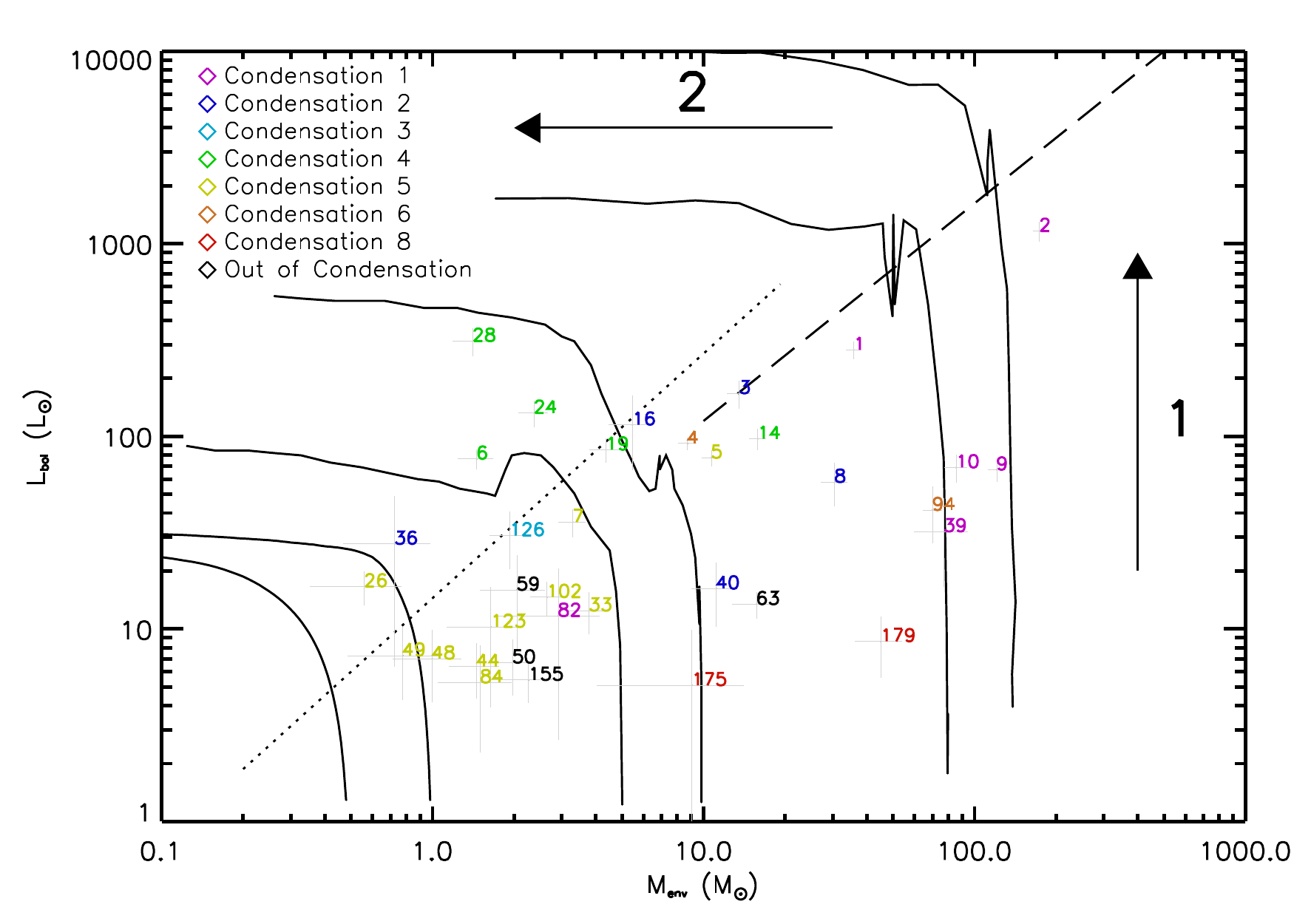}   \caption{Same as Fig.~\ref{evol} but sources are color-coded as a function of their hosting condensation previously identified using the 870~\mm and 1.3\,mm emission (ZAV07, DEH09). The condensation number refers as the one given in Fig.~\ref{Fig-dens} (\textit{left})}
              \label{evolC}
    \end{figure*}

\begin{figure*}
\includegraphics[angle=0,width=180mm]{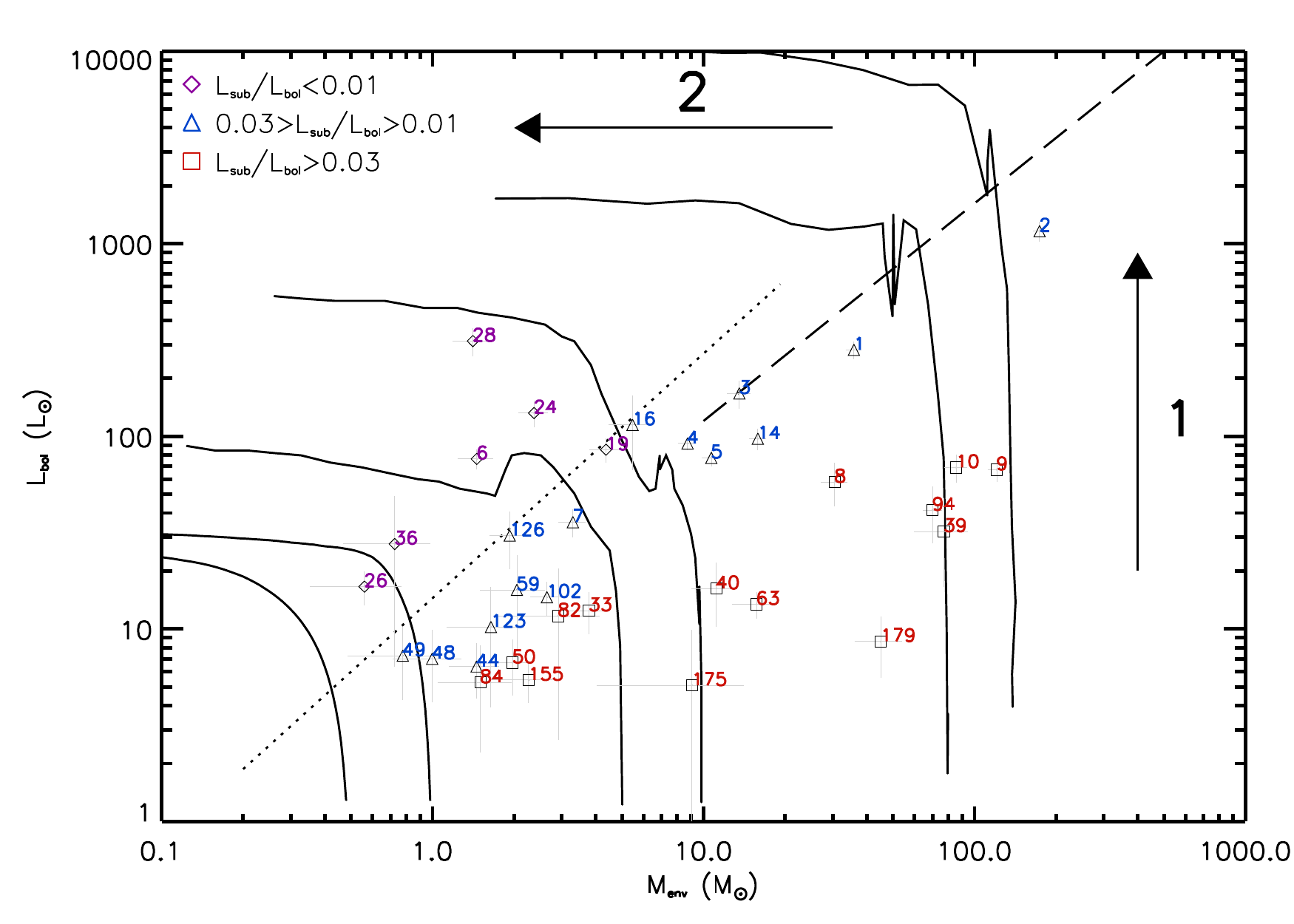}\caption{Same as Fig.~\ref{evol} but the sources are color-coded as a function of their L$_{\lambda\ge 350\mu m}$/L$_{\rm{bol}}$ ratio. Magenta diamonds for Class~I objects and red squares for Class~0, while blue-triangle sources are uncertain}
              \label{evolclass}
    \end{figure*}

\citet{sar96} presented the evolution of Class~0 to Class II via spherical accretion by a path in the L$_{\rm{bol}}$-M$_{\rm{env}}$ diagram. The unknown mechanism of massive star formation (scaled-up analogue of low-mass star or merging of low-mass stars) and the difficulties of establishing the evolution phase for individual YSOs (d$\ge$1~kpc, non-resolved clusters) make the construction of an evolutionary scenario for high-mass objects difficult. An attempt has been made by \citet{mol08} to reproduce the L$_{\rm{bol}}$-M$_{\rm{env}}$ evolutionary paths for massive objects. From a sample of 42 sources characterized by their [25-12] color value, they classified them as IR-sources if the SED could be fitted with a zero age main sequence (ZAMS) model or MM-sources if a MBB model was used. This difference in the SED translates into a different location in the L$_{\rm{bol}}$-M$_{\rm{env}}$ diagram well separated by a line representing IR-sources (see \citealt{mol08}) practically equivalent to the strip of low-mass Class~I objects from \citet{sar96}. Assuming a scaled-up analogue of the low-mass star regime with a turbulent core, a model of time dependent accretion rate \citep{mac03} with fixed final stellar masses and core surface densities, evolutionary paths in the L$_{\rm{bol}}$-M$_{\rm{env}}$ have been computed. The first sequence (indicated in Figs.~\ref{evol},~\ref{evolC} and \ref{evolclass}) represents the accretion phase where the luminosity is dominated by the accelerated accretion and the lost of envelope mass is due to accretion, outflows and possible draining by other YSOs. At the end of the first phase, the star reaches or is close to the ZAMS with a final stellar mass. During the second phase (also indicated in the figures), the envelope mass continues to decrease (the increase of stellar mass by residual accretion is neglected in this model) and the luminosity is now the sum of accretion and stellar luminosity. 
The final point of the paths corresponds to a lost of 90\% of the envelope mass for the four low-mass tracks and to a time of 2.1$\times$10$^{6}$~yrs and 2.7$\times$10$^{6}$~yrs when the star is optically visible for the two highest-mass tracks. We warn readers that this is a simple model which cannot be used to predict accurately the evolution of YSOs but rather to obtain indication about the evolutionary class of a source. 

\noindent In Fig.~\ref{evol}, we plot the evolutionary paths for low-mass \citep{sar96} and high-mass stars \citep{mol08} with their corresponding stripes for Class~I sources and include our sample. Sources with M$_{\rm{env}}>$10~$M_{\sun}$ are all located under the Class~I stripe and a qualitative analogy with Fig.~9 of \citet{mol08} permits a rough classification of them : sources 1,2,3,5,8,14 are Class~I and sources 9,10,39,40,63,94,179 are Class~0. On the contrary, the distribution of sources with M$_{\rm{env}}<$10~$M_{\sun}$ has a higher dispersion around the Class-I strip, also seen in fig.~9 of \citet{mol08}. As in Fig.~\ref{ML-LL}, sources located outside the PDR might tend to be younger but no evidence for a more evolved stage for sources located inside the PDR is seen, as it could have been expected if star formation progresses gradually in the surrounding medium, following the expansion of the ionization front and the leaking of the ionizing radiation.

In Figure~\ref{evolC}, the sources are color-coded depending on their location in the condensations. We see a clear trend for the sources' envelope mass and evolutionary stage to be determined by their hosting condensation: sources observed towards condensation 1 have the highest envelope mass and are in low evolutionary stage while sources in condensation 5 are low mass envelope sources, and possibly in a later evolutionary stage. Sources observed towards condensation 4 (pre-existing clump) tend to be evolved and of intermediate envelope mass. Condensation 8 is observed further away from the ionized front and hosts sources in a low-evolutionary stage. Sources 50 and 155 do not belong to any condensation according to DEH09 but are spatially close, outside the PDR and in a similar evolutionary state than the sources in condensation 5. Sources in condensation 2 show a higher dispersion in this diagram compared to the other condensations. The eastern part of condensation 2 contains Class~0-Class~I objects of intermediate mass and low-mass objects in the western part. The eastern-part of this condensation seems to be radiation-shielded thanks to the filament in front of sources 3, 8 and 16 while the western-part receives a significant part of the radiation through photons' leaking. This might explain the dispersion of sources' properties observed towards this condensation.

In Figure~\ref{evolclass}, the sources are color-coded depending on their L$_{\lambda\ge 350\mu m}$/L$_{\rm{bol}}$ value. We note that the magenta diamond sources (L$_{\lambda\ge 350\mu m}$/L$_{\rm{bol}}$<0.01) are above the Class~I stripe of \citet{sar96}, red square sources (L$_{\lambda\ge 350\mu m}$/L$_{\rm{bol}}$>0.03) are below the Class~I stripe of \citet{mol08} and blue triangle sources (0.03>L$_{\lambda\ge 350\mu m}$/L$_{\rm{bol}}$>0.01) are spread around these stripes. Hence, the two methods give consistent results to derive sources' evolutionary class.

We suggest that the main parameter that controls the star formation and the evolutionary stage of the YSOs is the column density of their hosting condensation. This means that a simple search for YSOs' age gradient around \HII region cannot be used as a simple indicator for establishing evidence for triggered star formation.  

\subsection{Evolutionary stage derived by DEH09}\label{deh09_comp}

\begin{figure}
 \centering
 \includegraphics[angle=0,width=90mm]{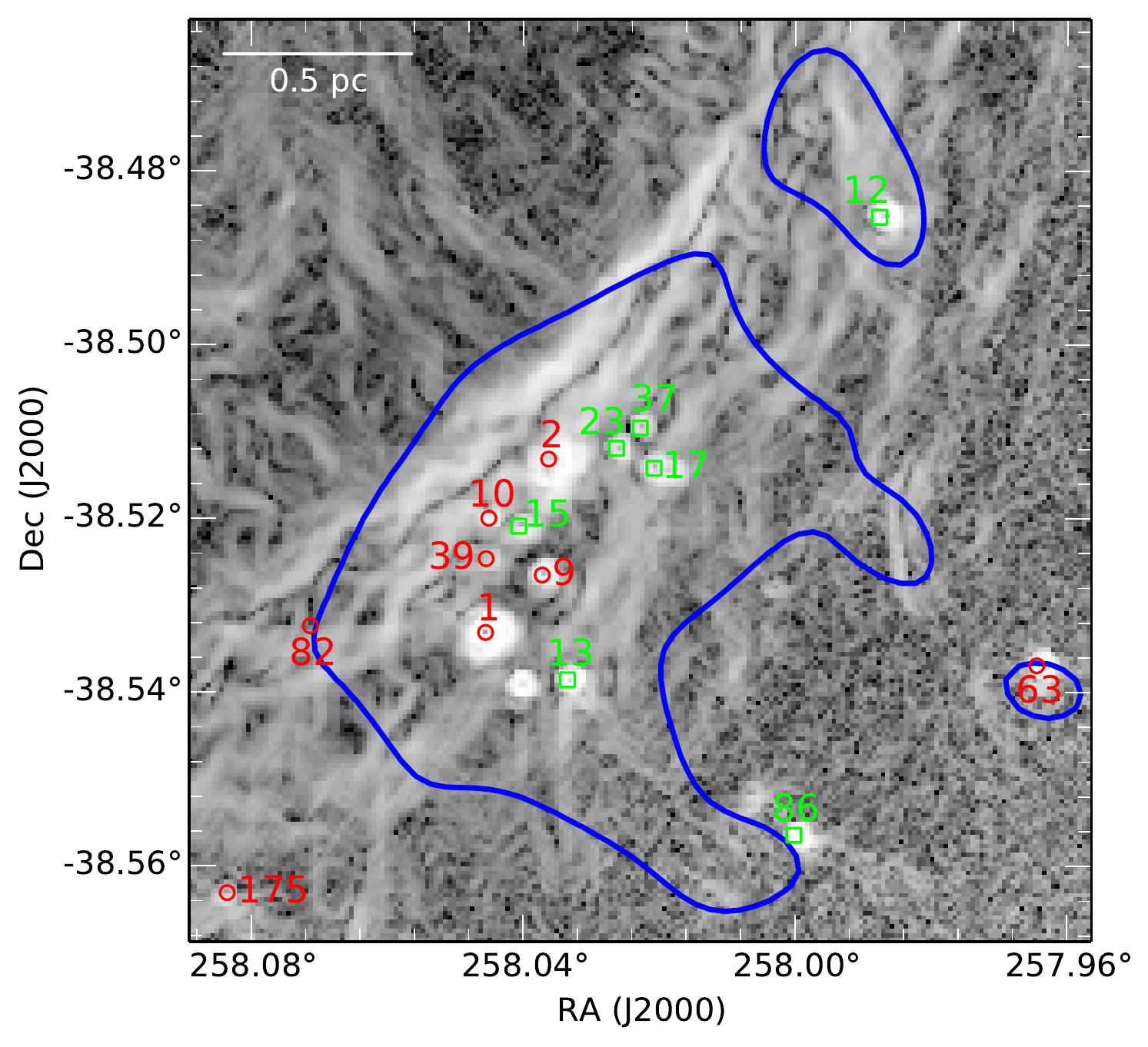}
  \caption{Condensation 1 and 7: 870\mm emission (countours) superimposed on the gradient image of the \herschel PACS 70\mm emission. Sources are identified with their {\it{getsources}} identification number. Sources coded with a red circle are those discussed (among the sample of 35 sources). The green square sources are detected but not discussed due to a lack of \herschel measurements (see text)}
             \label{cond1}
\end{figure}

\begin{figure}
 \centering
 \includegraphics[angle=0,width=90mm]{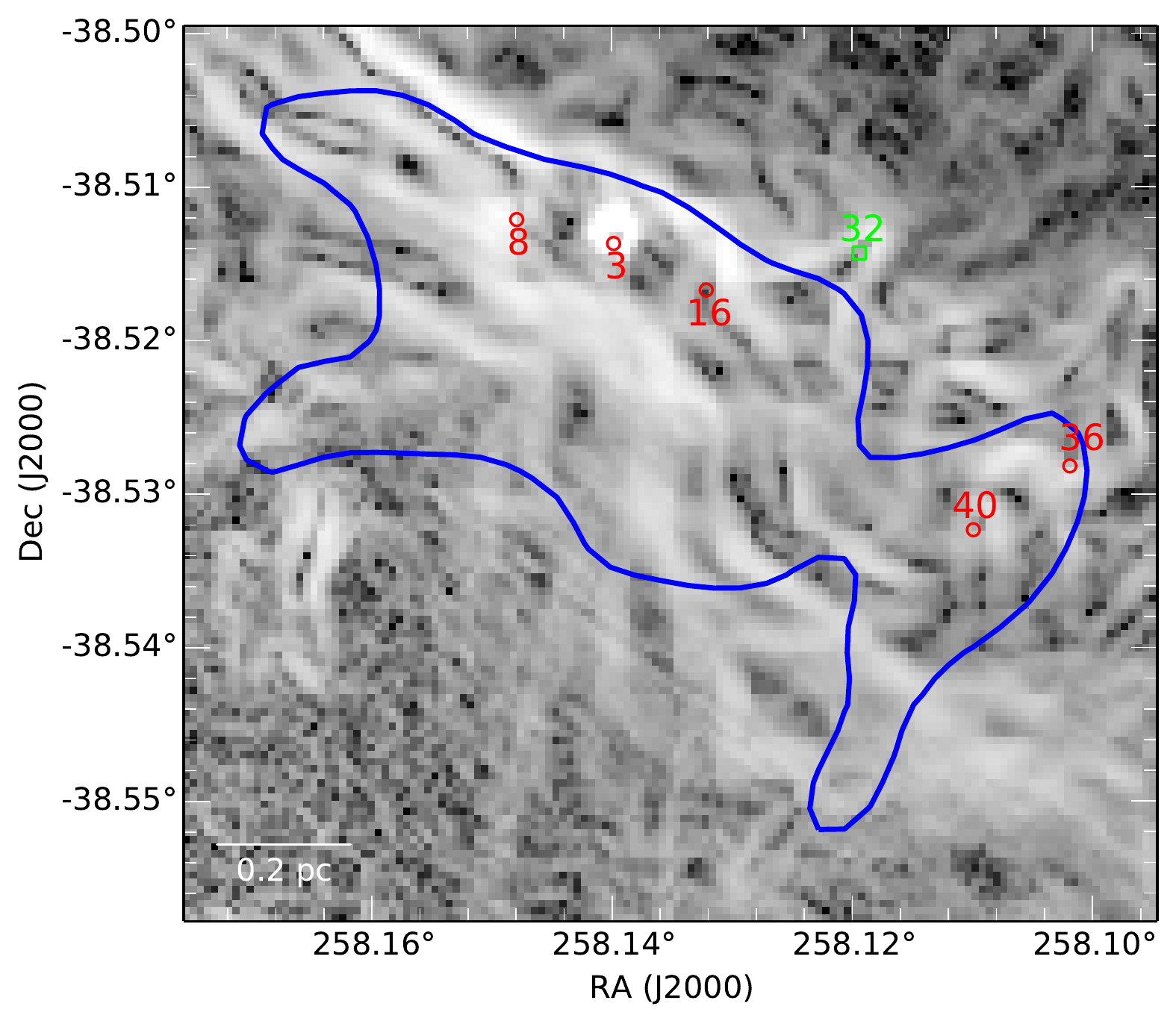}
  \caption{Condensation 2: same as for Figure~\ref{cond1}}
             \label{cond2}
\end{figure}

\begin{figure}
 \centering
 \includegraphics[angle=0,width=90mm]{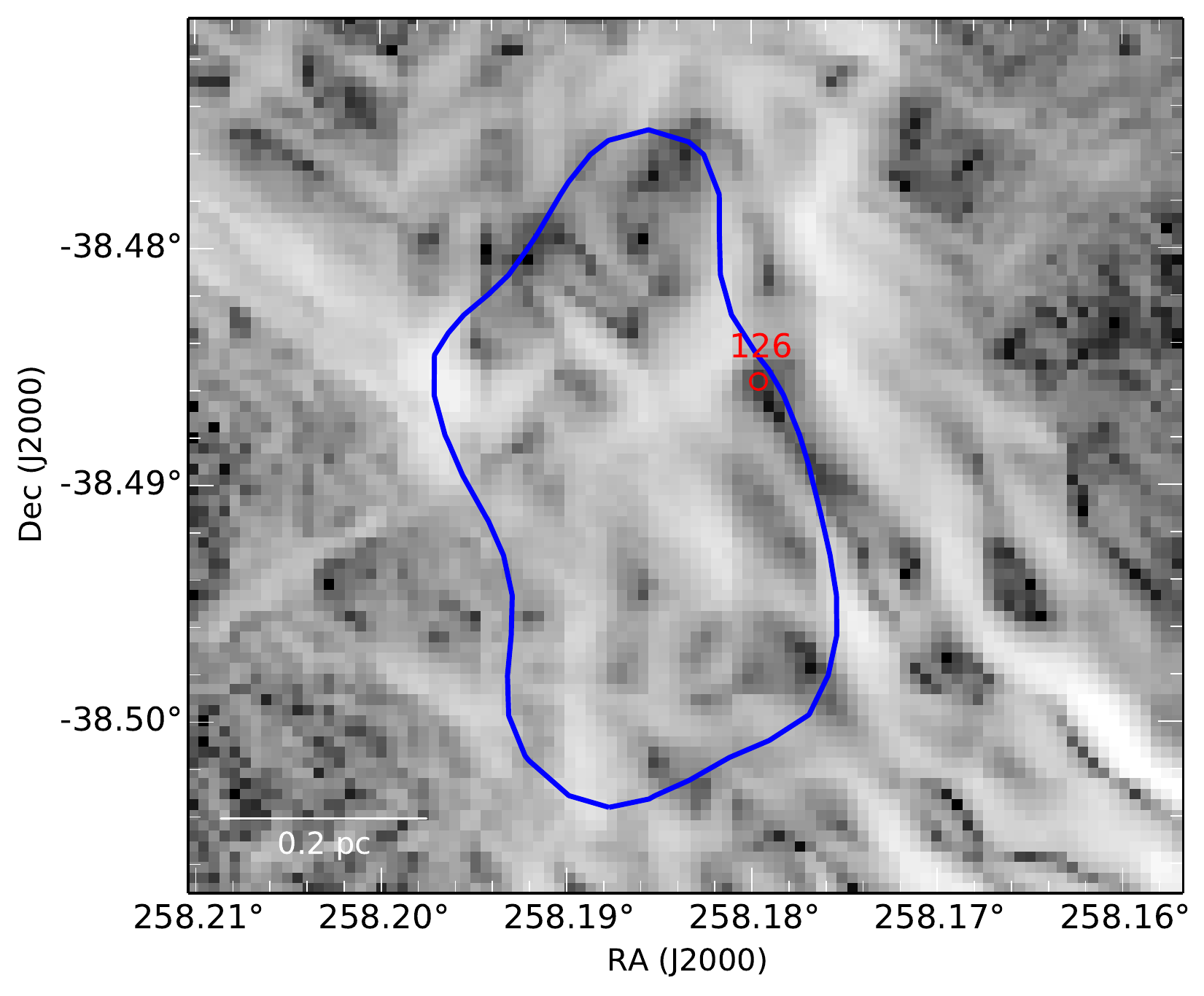}
  \caption{Condensation 3: same as for Figure~\ref{cond1}}
             \label{cond3}
\end{figure}

\begin{figure}
 \centering
 \includegraphics[angle=0,width=90mm]{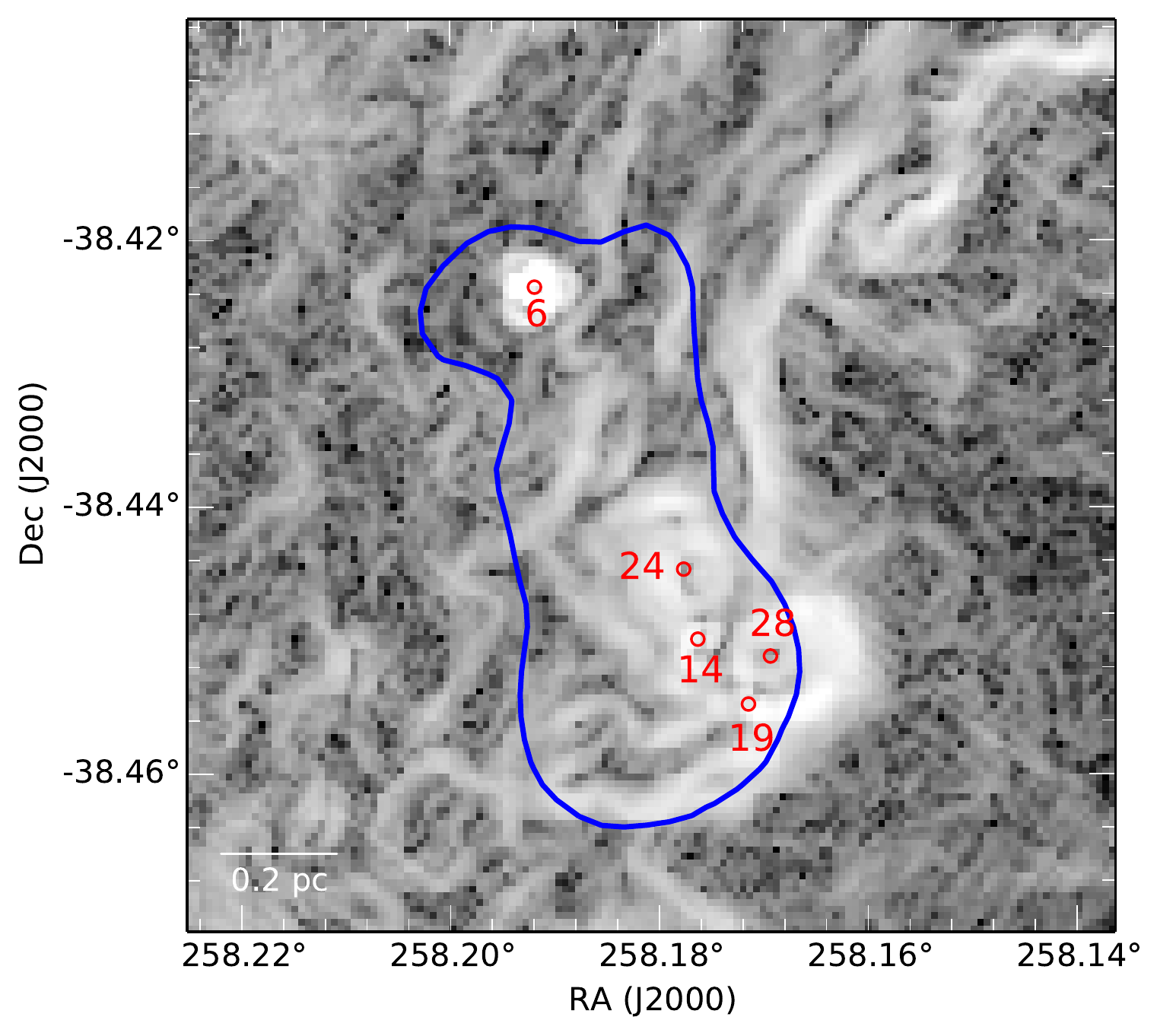}
  \caption{Condensation 4: same as for Figure~\ref{cond1} }
             \label{cond4}
\end{figure}

\begin{figure}
 \centering
 \includegraphics[angle=0,width=90mm]{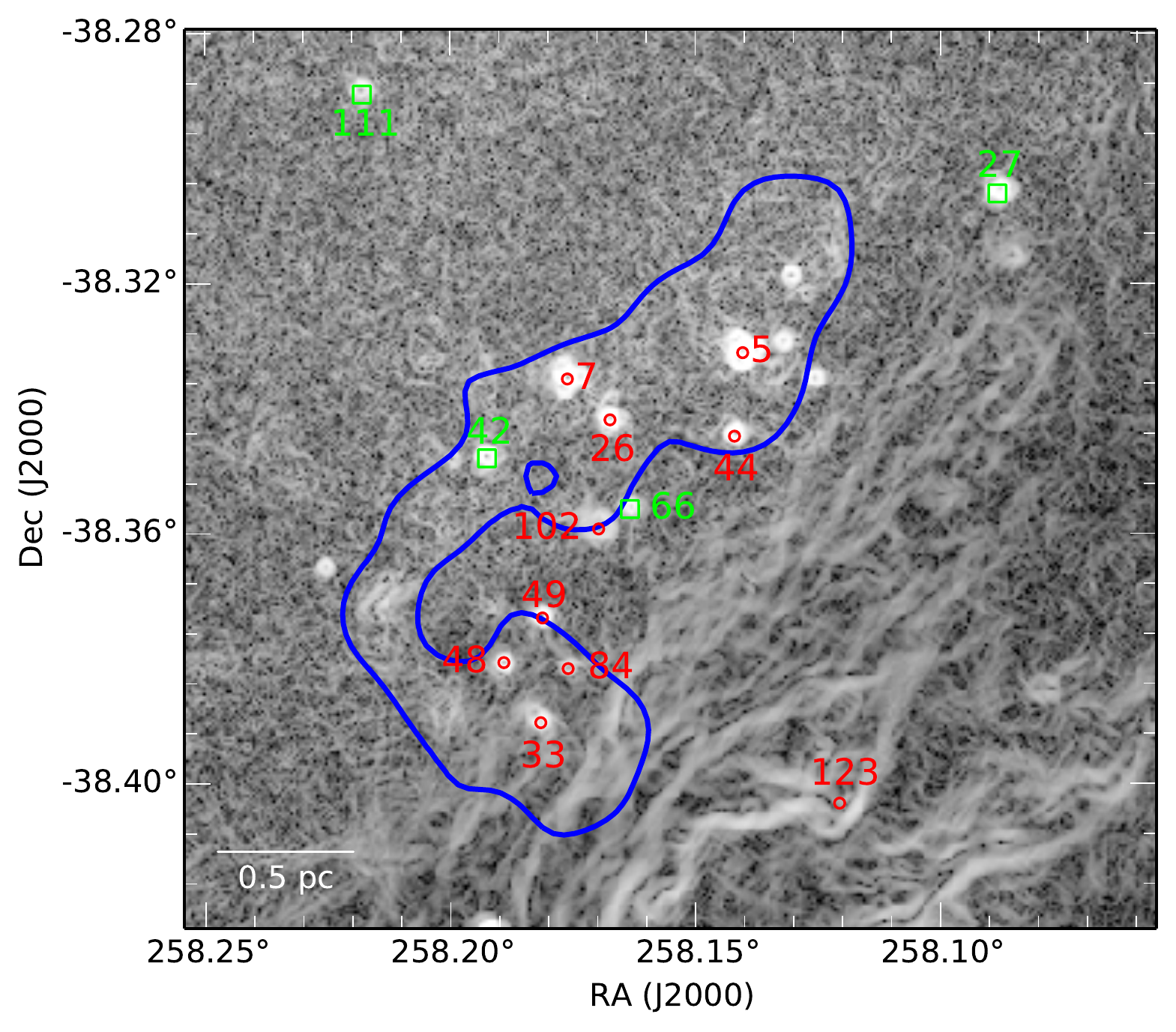}
  \caption{Condensation 5: same as for Figure~\ref{cond1} }
             \label{cond5}
\end{figure}

\begin{figure}
 \centering
 \includegraphics[angle=0,width=90mm]{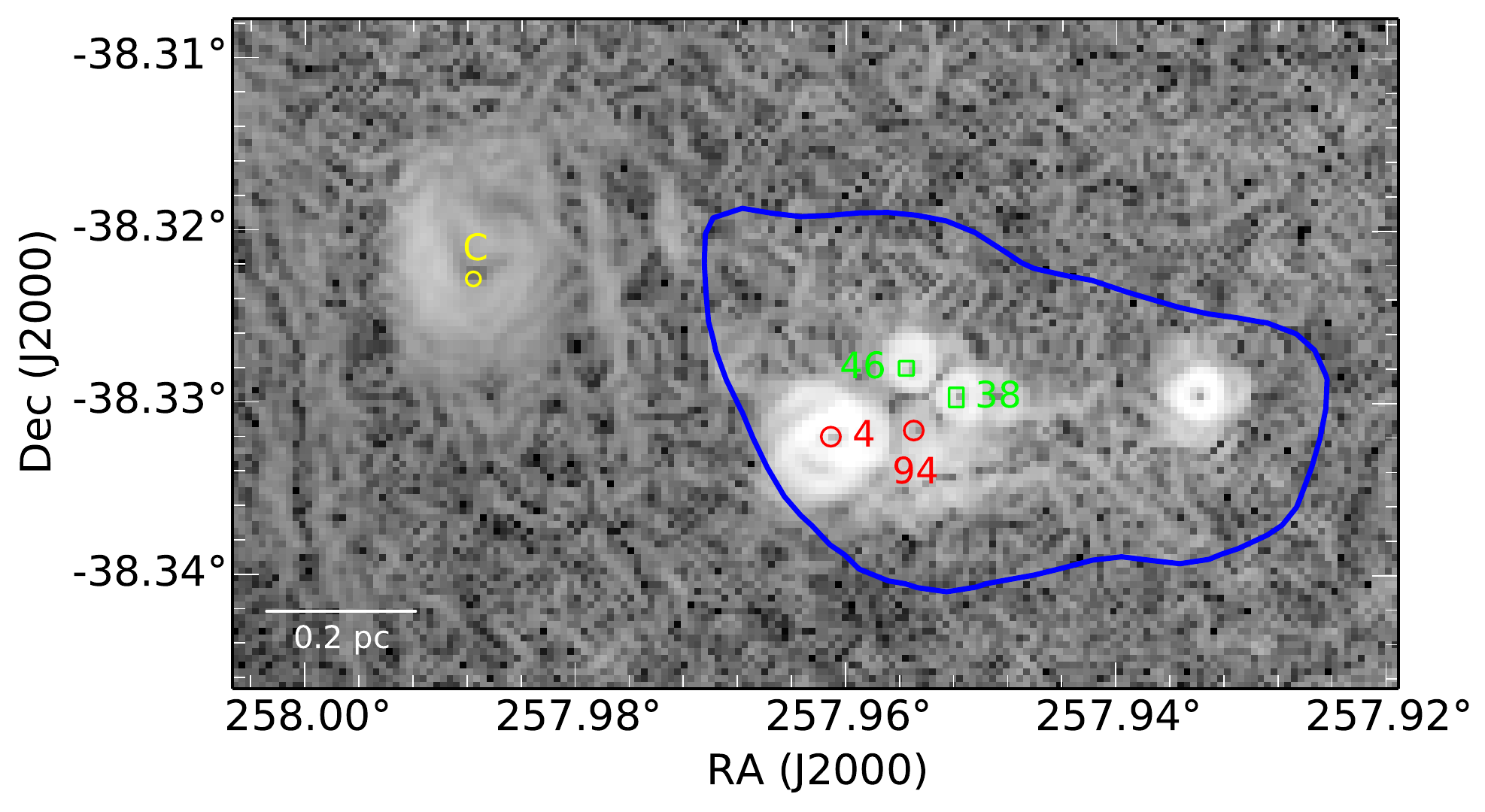}
  \caption{Condensation 6: same as for Figure~\ref{cond1} }
             \label{cond6}
\end{figure}

\begin{figure}
 \centering
 \includegraphics[angle=0,width=90mm]{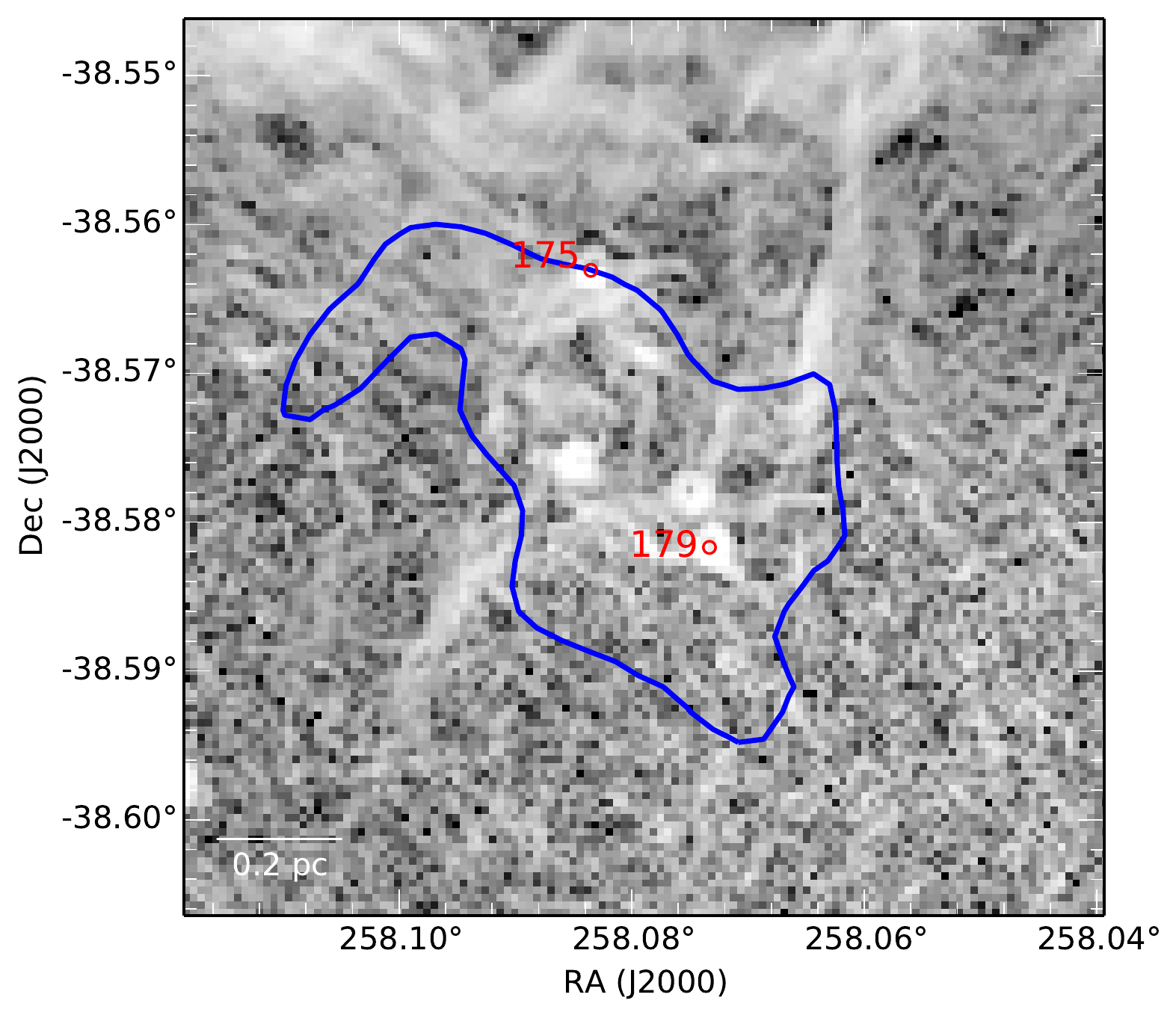}
  \caption{Condensation 8: same as for Figure~\ref{cond1} }

             \label{cond8}
\end{figure}

Color-color diagrams using near- and mid-IR data can also be used to infer the class of a source. DEH09 used \spitzer GLIMPSE and MIPSGAL colors to discuss the evolutionary stage of YSOs observed towards RCW~120. The results are given in columns 8 and 9 of Table~\ref{tm}. Fig.~\ref{cond1} to Fig.~\ref{cond8} present a zoom of the sources observed from condensation 1 to 8 on the gradient image of the \herschel PACS 70\mm emission and the 870\mm emission in countours. All the sources in these figures are detected by {\it{getsources}} and identified according to their {\it{getsources}} identification number in Table~\ref{tm} and Table~\ref{comp}. Final-sample sources, and those identified by DEH09 are indicated in the figures. Hence, non-labelled sources are either not part of the final sample and/or not detected by DEH09.
In the following we compare the evolutionary class of sources obtained from mid-IR color-color diagrams (DEH09) with the one obtained in this paper. 

\textbf{Condensation 1 (2530~$M_{\sun}$) : }This is the most massive and densest condensation observed. The classification of Source~9 (40) and 63 (37) do not agree with DEH09. Both objects do not have IR-counterpart in the J, H and K$_{s}$ and following Fig.~5 of DEH09, they have a $K-[24]$ higher than 10 mag. Hence, they are likely to be in an early evolutionary stage. Source~2 is the massive Class~0 object discussed in \citet{zav10} (see also DEH09). It is located at the peak of the column density distribution ($N(H_2)=4\times 10^{23}$~cm$^{-2}$) and has the highest envelope mass (M$_{\rm{env}}$=174~$M_{\sun}$) and bolometric luminosity (L$_{\rm{bol}}$=1163~$L_{\sun}$) of the sample. It is probably a Class~0 source,  since no IR-counterpart is detected. Sources 10, 39 and 82 are not detected by DEH09 and are classified as Class~0. This condensation hosts 80\% of the massive cores. Because condensation~1 is the densest and most massive in RCW~120, the core formation efficiency (CFE) is expected to be higher compared to the other condensations (\citealp{mot98,bon10b}; Liu et al. submitted).
\newpage

\textbf{Condensation 2 (540~$M_{\sun}$) : }Source 3 (50) has been classified as a Class I source by DEH09 in agreement with our classification. Sources 16 and 36 are not discussed by DEH09 maybe because of the high filamentary background around these compact sources. Therefore, no IR-counterpart could be reliably detected and the sources are classified as Class~0-I.

\textbf{Condensation 4 (350~$M_{\sun}$) : }Sources 6 (76), 14 (69) and 19 (67) are classified as at least Class~I objects and in agreement with DEH09. Source 24 (Object A) and 28 (Object B) are surrounded by local PDR revealed as shells on the gradient 70\mm image. Because their IR counterparts are diffuse, no attempt has been made by DEH09 to classify them but are likely Class~I or further. ZAV07 suggested that this condensation could be a pre-existing clump engulfed in the ionized region. A subsequent RDI process could have accelerated the collapse which might explain why the objects are in a higher evolutionary stage compared to the other condensations.

\textbf{Condensation 5 (1580~$M_{\sun}$) : }This region is highly structured and hosts nine YSOs among the 35 discussed. Among the sources of the final sample and discussed by DEH09, source 33 is the only one whose class does not agree. In the same way as sources 9 and 63 in condensation 1, DEH09 did not measured any near IR-counterpart for this source and its $K-[24]$ value should also be higher than ten. Therefore, this source is also in an early evolutionary stage. Sources 44 and 48 present IR-counterparts at all wavelengths except 8\mm and only at 24\mm respectively but are too weak (M$_{\rm{env}}$=1~$M_{\sun}$) to be discussed by DEH09. They probably are weak Class~I sources. Sources 84 and 123 do not present IR-counterparts and are classified as Class~0.

\textbf{Condensation 6 (330~$M_{\sun}$) : }We identify a massive YSO (source 94) of $M_{\rm{env}}$=70~$M_{\sun}$ with IR-counterparts but classified as Class~0. It is possible that the higher fluxes coming from source 4 contaminates source 94 at long wavelengths hence overestimating the  L$_{\lambda\ge 350\mu m}$/L$_{\rm{bol}}$ value and hence, the classification.

\textbf{Condensation 8 (370~$M_{\sun}$) : }Located south of the ionized region (see Fig.~\ref{high_dens}) the zone where the source is embedded, this source was probably formed by the leaking of UV photons passing through the low density medium seen on the high resolution density map (see Fig.~\ref{high_dens}) at (258$\fdg$07,$-$38$\fdg$52). Hence, sources 175 and 179 were probably formed later compared to the sources located in the PDR. This is confirmed by their low-temperature (between 11.2~K and 13.3~K), low evolutionary stage and the absence of IR counterparts.

\subsection{Comparison with the \citet{wal15} model}
\citet{wal12,wal13} show that clumpy, shell-like structures like that seen in RCW 120 are probably attributable to pre-existing density structures in the natal molecular cloud. During the expansion of the \HII region and the collection of the dense shell, the pre-existing density structures are enhanced and lead to a clumpy distribution within the shell. The masses and locations of the swept-up clumps depend on the fractal density structure of the molecular cloud, through the parameters $n$ and $\rho_0$, related to the fractal dimension of the cloud and the density, respectively \citep[see Sect. 2]{wal13}.

\citet{wal15} compared simulations and APEX-LABOCA 870\mm observations of RCW~120. They performed three-dimensional SPH simulations of \HII regions expanding into fractal molecular clouds in order to investigate whether the formation of massive clumps in the swept-up shell necessarily requires the C\&C mechanism \citep{elm77}. 
They show that a distribution of clumps similar to the one seen in RCW 120 can be explained by a non-uniform initial molecular cloud structure, implying that a shell-like configuration of massive clumps does not imply that the C\&C mechanism is at work. They find a hybrid form of triggering, which combines elements of C\&C mechanism and RDI. 

We discuss below how the \herschel results presented here compare with their findings. The temperature map obtained from \herschel images indicates that dust temperatures lower than 30~K, the temperature used by \citet{wal15}, are observed. This means that the mass they derived for the condensations represents a lower limit. 
The H$_2$ column density maps obtained from \herschel images show that the observations better correspond with a low value of $\rho_0$, where $\rho_0$ is the scaling constant for the density fluctuations field caracterizing the width of the density PDF \citep{wal12,wal13}. However pillars are not observed on the northern, lower density part of the ionized region as obtained in their simulations \citep[fig.~2]{wal15}. This suggests that the numerical treatment adopted better describe higher density regions while lower density regions seem to be better represented by the higher value of $\rho_0$.  

The distribution of sources observed towards the central part of the ionized region in the simulation is also not observed \citep[their Fig.~2 right]{wal15}. The distribution of sources in condensations is also not well reproduced by this model, as seen on Fig.~\ref{Figsources} and \ref{source_tentative}. The number of sources they found towards the three main condensations well corresponds with our findings - nine sources towards condensation 1 (their condensation 3), three sources towards condensation 2 (their condensation 1) and six sources towards condensation 4 (their condensation 2). For the two runs, the condensation 3 formed the highest number of high-mass protostars (12.7$M_{\sun}$ and 19$M_{\sun}$ in average). This is in agreement with the observations where high-mass cores are found towards our condensation 1.  We remind the reader that the mass computed in this paper is the envelope mass while \citet{wal15} use the protostars mass. Therefore, it is not surprising that the mass computed in condensation 1 are much higher compared to \citet{wal15}. Nonetheless, their condensation 2 host protostars of lower masses (9$M_{\sun}$ in average) while our corresponding condensation contains low-mass cores. A new modeling of this region using \herschel results would help for discussing the star formation history and its propagation in the ambient medium. It would be interesting to discuss the parameters (and mechanisms) that lead to the formation of the high number of high-mass cores observed towards condensation 1. 

\subsection{Comparison with the model of \citet{tor15}}
Using MOPRA observations of $^{12}$CO, $^{13}$CO and C$^{18}$O in the $J=1\rightarrow 0$ transition, \citet{lor14} did not detect any expansion of the \HII region which means that the expansion velocity is either too low to be observed or inexistant. Considering this fact, \citet{tor15} explained the formation of the O star and the corresponding ring-like structure following the cloud-cloud collision (CCC) scenario from \citet{hab92} which can be described in three stages. First, a small and a large clouds are heading towards each other. Secondly, a cavity is created in the large cloud due to the collision with the small cloud. The place where the two clumps collided is compressed, leading to massive star formation. Finally, the cavity in the large cloud is filled with the ionzing radiation coming from the recently formed massive star(s). A schematic explanation can be found in \citet{tor15} (see their Figs.~12 and 13). In the case of RCW~120, they suggest that the weak leaking of H$\alpha$ emission in the northern part of the ring indicates only the beginning of the erosion by the ionizing radiation. Hence, the triggering which is assumed to take place as a consequence of the C\&C mechanism cannot be seen yet. However, after the formation of the ionizing star, a triggering mechanism caused by the compression of the remaining small clump on the large clump is plausible. This could be an alternative explanation which should only affect the formation of YSOs in the southern part of the ring. This study shows that the main driver of the evolutionary stage is the density of the hosting condensation and not the (projected) distance to the ionizing star as expected earlier.

\section{Summary and conclusions}
\label{conc}
We used \herschel PACS and SPIRE images, complemented with existing data, to study the star formation observed towards the Galactic ionized region RCW~120.

\citet{zav10} presented the first results from \herschel, however this paper is an in-depth study under the HOBYS recipe which allow us to compare the results between different regions observed in this key program. Moreover, while the first \herschel results were focused on source 2, we produced the first reliable catalog of compact sources using \herschel data in this region.

The unprecedent coverage and sensitivity in the far infrared of the \herschel data allow us to derive, for the first time, the temperature and H$_2$ column density map for this region. The temperature ranges from 15~K to 24~K and the column density from $7\times 10^{21}$ cm$^{-2}$ up to $9\times 10^{23}$ cm$^{-2}$. The condensations defined by DEH09 at 870\mm corresponds to cold and dense regions where the majority of the sources are detected.

We also derive, for the first time, the envelope mass, envelope dust temperature and bolometric luminosity of compact sources detected there. The temperature ranges from 11.2~K to 34.1~K with a median of 19.1~K, from 1~$M_{\sun}$ to 174~$M_{\sun}$ with a median of 4~$M_{\sun}$ for the envelope mass and from 5~$L_{\sun}$ to 1163~$L_{\sun}$ with a median of 30~$L_{\sun}$. The volume density was computed by assuming a spherical source with the size defined at the reference wavelength (160\mm or 250\,$\mu$m) going from 2$\times$10$^{5}$~cm$^{-3}$ to 10$^{8}$~cm$^{-3}$. 

We use the physical parameters to discuss the star formation history in this region. We show that most of the compact sources (21 of the 35) are observed towards the PDR.

Thanks to \herschel data, we detected 21 sources, mostly in an early evolutionary stage, which were not detected and hence discussed in DEH09.

Using the L$_{\lambda\ge 350\mu m}$/L$_{\rm{bol}}$ criteria from \citet{bon10}, we classify the sources between Class~0, intermediate and Class~I. We found respectively 15, 16 and 6 sources in this classification.

We find that the projected distance to the ionizing source is not the parameter which controls the evolutionary stage of the sources, contrary to what was expected before, wrongly. In fact, the main driver for this is the density of the condensation where the source is located, whatever its distance to the ionizing sources. Consequently, there is no conflict between possible triggering and projected distance because the density plays a major role in the overall picture. Despite the fact that the southern layer of the region is compressed (Tremblin et al. 2014), \herschel data do not allow us to conclude on triggering. High resolution spectroscopic data are needed to determine the structure (possible fragmentation) of the cores and the evolutionary stage of the sources in these cores.

\begin{acknowledgements}
We thank the referee for his/her report which helps to improve the quality of the paper.
This work is based on observations obtained with \herschel-PACS and \herschel-SPIRE photometers. PACS has been developed by a consortium of institutes led by MPE (Germany) and including UVIE (Austria); KU Leuven, CSL, IMEC (Belgium); 
CEA, LAM (France); MPIA (Germany); INAF-IFSI/OAA/OAP/OAT, LENS, SISSA (Italy); IAC (Spain). This development has been supported by the funding agencies BMVIT (Austria), ESA-PRODEX (Belgium), CEA/CNES (France), DLR (Germany), ASI/INAF (Italy), and CICYT/MCYT (Spain). SPIRE has been developed by a consortium of institutes led
by Cardiff Univ. (UK) and including: Univ. Lethbridge (Canada);
NAOC (China); CEA, LAM (France); IFSI, Univ. Padua (Italy);
IAC (Spain); Stockholm Observatory (Sweden); Imperial College
London, RAL, UCL-MSSL, UKATC, Univ. Sussex (UK); and Caltech,
JPL, NHSC, Univ. Colorado (USA). This development has been
supported by national funding agencies: CSA (Canada); NAOC
(China); CEA, CNES, CNRS (France); ASI (Italy); MCINN (Spain);
SNSB (Sweden); STFC, UKSA (UK); and NASA (USA). 

This work is based on observations made with the \spitzer Space Telescope, which is operated by the Jet Propulsion Laboratory, California 
Institute of Technology, under contract with NASA. We have made use of the NASA/IPAC Infrared Science Archive to obtain data products from 
the 2MASS, \spitzer-GLIMPSE, and \spitzer-MIPSGAL surveys.

The Centre National d'Etudes Spatiales (CNES) is deeply acknowledged for the financial support.  

Part of this work was supported by the ANR ({\it Agence Nationale pour la Recherche}) project ``PROBeS'', number ANR-08-BLAN-0241. 
     
\end{acknowledgements}

\bibliographystyle{aa}
\bibliography{rcw120ps.bib}

\onecolumn
\begin{appendix}
\section{Herschel original fluxes for sources detected by getsources (see text)}
\begin{landscape}
\begin{longtable}{cccrrrrrrrrrrrrrrr}
\caption{Identification, position and fluxes for compact sources detected towards RCW~120}\label{comp}\\
\hline
Id & Ra & Dec & J & H & Ks & 3.60 & 4.50 & 5.80 & 8.00 & 24.0 & 70 & 100 & 160 & 250 & 350 & 500  \\
    & \multicolumn{2}{c}{J2000} & \multicolumn{8}{c}{mJy} & \multicolumn{6}{c}{Jy} \\
    \hline
\endfirsthead
\caption{continued.}\\
\hline
Id & Ra & Dec & J & H & K & 3.60 & 4.50 & 5.80 & 8.00 & 24.0 & 70 & 100 & 160 & 250 & 350 & 500  \\
    & \multicolumn{2}{c}{J2000} & \multicolumn{8}{c}{mJy} & \multicolumn{6}{c}{Jy} \\
    \hline
\endhead
\endfoot
  1&   258.04565&   -38.53315&  &  &  &        2.57&       19.12&       47.86&       69.24&     1758.09&       62.00&      112.80&      128.30&       83.46&       55.01&       23.46 \\
  2&   258.03625&   -38.51318&  &  &  &  &  &  &  &  &      116.70&      625.20&      904.20&      418.10&      240.80&      113.90 \\
  3&   258.13986&   -38.51368&  &  &        0.65&        5.52&       24.58&       63.16&      730.30&     1796.11&       41.61&       57.90&       87.31&       43.41&       13.89&        8.16 \\
  4&   257.96112&   -38.33202&  &  &        7.25&      137.90&      315.00&      554.40&      671.90&     2972.66&       13.63&       21.24&       23.04&       17.22&       12.14&       10.65 \\
  5&   258.14035&   -38.33111&  &        7.41&       55.92&      393.80&      558.15&      855.70&      951.10&     1896.43&        9.75&       12.01&       15.21&       17.30&       12.45&        7.75 \\
  6&   258.19193&   -38.42351&  &        3.53&       31.21&      255.00&      495.85&      819.60&     1048.00&     2138.66&        8.61&       12.58&       13.03&        4.69&        3.46&        3.52 \\
  7&   258.17612&   -38.33530&  &  &  &        0.56&        3.50&        2.79&  &      134.73&        7.05&       14.66&       15.45&       10.45&        7.79&        4.45 \\
  8&   258.14792&   -38.51212&  &  &  &  &  &  &  &  &        7.92&       21.46&       56.61&       36.86&       25.03&       11.53 \\
  9&   258.03720&   -38.52652&  &  &  &        2.56&        3.01&        7.56&        8.55&      113.64&        3.98&       12.02&       15.76&       87.22&       55.12&       19.56 \\
 10&   258.04517&   -38.51953&  &  &  &  &        1.35&  &  &  &        1.81&       11.17&       46.55&       73.89&       28.84&       18.83 \\ 14&   258.17627&   -38.44988&        1.38&        4.08&        8.14&       13.15&       22.63&       35.54&       25.73&      203.01&        7.34&       39.30&       58.34&       33.07&       17.25&       11.28 \\
 16&   258.13214&   -38.51672&  &  &  &  &  &  &  &  &       73.06&       93.17&       74.14&       18.33&        4.65&        2.76 \\
 19&   258.17145&   -38.45472&        9.76&       42.18&      104.90&      216.90&      277.90&      356.50&      384.40&     1176.02&        6.54&       21.61&       33.05&       11.79&        4.20&        0.18 \\
 24&   258.17780&   -38.44460&  &  &  &  &  &        0.00&        0.00&        0.00&       28.23&       49.36&       33.71&       11.55&       10.73&        0.51 \\
 26&   258.16742&   -38.34184&        3.25&        9.57&       19.89&       48.65&       87.12&      145.60&      225.70&      505.54&        2.11&        2.63&        2.67&        3.08&        3.40&        3.15 \\
 28&   258.16949&   -38.45095&        0.00&        0.00&        0.00&        0.00&        0.00&        0.00&        0.00&        0.00&       44.00&       89.36&       56.69&        8.75&        2.05&        0.10 \\
 33&   258.18164&   -38.39026&  &  &        7.08&        3.05&        8.34&        7.57&        9.08&      118.37&        1.77&        4.11&        7.59&        6.85&        4.33&        2.92 \\
 36&   258.10190&   -38.52818&  &  &  &  &  &  &  &  &        8.15&       15.50&       11.37&        5.40&        0.27&   \\
 39&   258.04541&   -38.52476&  &        2.53&        1.65&        3.89&        6.13&        7.91&       33.18&  &        0.59&        2.25&        0.90&       45.98&       14.59&        7.37 \\
 40&   258.10992&   -38.53234&  &  &  &  &  &  &  &  &        5.76&        5.94&       10.25&       10.82&        8.47&        6.09 \\
 44&   258.14203&   -38.34447&        1.34&        2.13&        2.73&        0.98&        3.01&        2.64&  &        1.93&        1.07&        2.44&        3.93&        3.92&        2.36&   \\
 48&   258.18912&   -38.38067&  &  &  &  &  &  &  &       10.53&        1.15&        2.87&        4.89&        2.34&        2.05&        0.01 \\ 49&   258.18124&   -38.37353&  &        1.44&        2.58&        5.59&       11.20&       22.07&       27.38&      137.68&        1.16&        2.18&        3.89&        1.47&        0.06&        1.23 \\
 50&   257.95599&   -38.40344&  &  &  &  &  &  &  &  &        0.89&        3.04&        5.73&        3.33&        2.45&        2.28 \\
 59&   258.08347&   -38.53136&  &  &  &  &  &  &  &  &        2.89&        8.75&       11.01&        7.26&        0.07&        0.62 \\
 63&   257.96445&   -38.53691&  &  &  &       20.50&       67.49&      125.00&      150.90&      438.16&        0.83&        1.02&        2.45&        3.16&        4.36&        4.96 \\
 82&   258.07132&   -38.53237&  &  &  &  &  &  &  &  &        1.44&        5.56&        8.89&        6.99&        0.10&   \\
 84&   258.17609&   -38.38145&  &  &  &  &  &  &  &  &        0.36&        2.38&        5.17&        2.38&        0.78&   \\
 94&   257.95505&   -38.33182&  &  &        2.54&        7.99&       15.94&       20.05&       21.25&     2477.74&        0.90&        2.14&        5.75&       29.01&       21.82&        1.04 \\
102&   258.16974&   -38.35928&  &        4.11&       12.21&       28.97&       27.71&       27.61&       18.23&        6.51&        2.03&        5.31&        8.22&        4.74&        4.10&        2.86 \\
123&   258.12061&   -38.40319&  &  &  &  &  &  &  &  &        4.28&        8.29&        7.94&        4.45&        1.97&        1.60 \\
126&   258.17963&   -38.48564&  &  &  &  &  &  &  &  &        6.47&       17.84&       17.31&        9.02&        0.20&        2.36 \\
155&   257.95459&   -38.40161&        0.81&        1.45&        2.49&       13.60&       21.54&       32.08&       46.89&      163.98&        0.38&        0.75&        2.00&        2.15&        2.17&        1.84 \\
175&   258.08353&   -38.56308&        0.52&        1.53&        2.13&        2.39&        3.06&        2.60&        1.06&        9.11&        0.35&        0.75&        4.57&        4.49&        5.64&        1.20 \\
179&   258.07333&   -38.58170&  &  &  &  &  &  &  &  &        0.15&        0.79&        3.44&        4.52&       11.27&        7.49 \\
\hline
 11&   258.04062&   -38.53902&  &  &  &  &  &  &  &  &        3.04&        6.87&        7.30&        9.03&  &       15.48 \\
 12&   257.98761&   -38.48531&  &  &  &  &  &  &  &      159.78&        4.14&        8.02&        9.48&        9.82&        6.69&        8.00 \\ 13&   258.03348&   -38.53858&  &  &  &        8.83&       45.50&      112.30&      180.50&      483.24&        3.82&        5.93&        9.21&       30.84&       50.55&   \\
 15&   258.04065&   -38.52089&  &  &        2.01&       19.71&       33.76&       42.35&       19.57&  &        2.56&  &  &  &       10.78&   \\
 17&   258.02078&   -38.51423&  &  &  &  &  &  &  &  &        3.44&        9.61&  &      171.00&       67.35&   \\
 20&   257.93423&   -38.32965&  &  &  &  &  &  &  &       69.07&        1.58&        2.52&        3.35&  &  &   \\
 23&   258.02631&   -38.51194&  &  &        1.67&        1.36&        4.90&        6.91&  &  &        1.49&        3.67&  &  &  &   \\
 27&   258.08841&   -38.30564&  &  &  &  &  &  &  &  &        2.25&        1.17&  &  &  &   \\
 32&   258.11938&   -38.51431&  &  &  &       28.17&       37.51&       54.58&       49.70&  &        2.71&        4.43&        1.13&  &  &   \\
 34&   258.29898&   -38.28543&  &  &  &        1.04&        1.01&  &        5.51&      408.27&        1.56&        0.96&  &  &  &   \\
 37&   258.02283&   -38.50959&  &  &  &  &        3.74&        7.20&       12.20&      238.86&        0.94&  &  &  &  &       35.88 \\
 38&   257.95184&   -38.32969&  &  &  &        1.72&        4.44&        5.37&        2.84&  &        1.47&        2.27&        4.52&        4.06&        8.59&       12.71 \\
 42&   258.19254&   -38.34798&  &  &  &        0.41&        1.35&  &  &       53.28&        1.29&        1.40&        2.15&  &  &   \\
 46&   257.95554&   -38.32800&  &  &  &        7.47&       22.37&       39.95&       37.70&      134.74&        1.04&        2.08&        2.65&  &  &   \\
 52&   258.09766&   -38.51765&  &  &  &        3.54&  &       13.45&       32.97&  &        3.11&        5.88&        6.99&  &  &   \\
 58&   258.04913&   -38.52470&  &  &  &        4.59&        3.01&       26.98&       84.79&  &        1.61&  &  &       40.04&       20.09&       14.33 \\
 66&   258.16336&   -38.35612&  &  &        2.62&       13.40&       26.86&       38.75&       50.58&      194.14&        0.77&        1.73&        4.06&  &  &   \\
 78&   258.09763&   -38.52643&  &  &  &  &  &  &  &  &        1.37&        1.76&  &  &  &   \\
 81&   258.13211&   -38.32971&  &  &  &        1.06&        2.52&        2.68&        1.42&       17.74&        0.55&        1.15&        2.17&  &  &   \\
 85&   257.94284&   -38.23342&  &  &  &        1.70&        1.16&        3.35&       11.08&       29.17&        0.44&        0.64&  &  &  &   \\
 86&   258.00015&   -38.55644&  &  &  &  &  &  &  &       62.96&        0.64&        0.92&  &  &  &   \\
 87&   258.12576&   -38.33563&  &  &  &        1.00&        2.54&        3.81&        3.74&       23.83&        0.60&        0.72&  &  &  &   \\
 91&   257.91385&   -38.45134&  &        0.98&        4.75&       21.70&       33.52&       37.45&       43.50&      139.88&        0.39&        0.85&        2.20&  &  &   \\
 95&   258.13080&   -38.31924&  &  &  &  &  &  &  &  &        0.40&        0.96&        2.40&        3.44&        6.59&        6.36 \\
100&   258.26227&   -38.35157&       17.98&       23.16&       22.80&       12.01&  &        6.78&        9.32&  &        2.63&        2.45&  &  &  &   \\
103&   258.17090&   -38.50493&  &  &  &  &  &  &  &  &        1.13&        2.08&  &  &  &   \\
106&   258.12173&   -38.53703&  &  &  &        2.54&        2.06&  &  &  &        2.60&        3.42&  &  &  &   \\
111&   258.21793&   -38.28977&  &  &  &  &  &  &  &     3016.04&        0.54&        0.30&  &  &  &   \\
112&   258.29178&   -38.46433&        0.56&        2.17&        3.01&        4.45&        5.42&        3.08&        5.30&  &        0.86&        1.58&        2.44&  &  &   \\
115&   258.26550&   -38.33118&       10.06&       18.53&       23.59&       16.53&       10.79&        8.72&        4.28&       15.13&        0.47&        0.56&  &  &  &   \\
116&   258.09653&   -38.54058&  &  &  &  &  &  &  &  &        2.17&        2.10&  &  &  &   \\
134&   258.19498&   -38.48449&  &  &  &  &  &  &  &  &        1.90&        3.15&  &  &  &   \\
135&   258.22525&   -38.36578&  &  &  &        4.27&       13.27&       24.62&       27.92&       27.79&        0.27&        0.50&  &  &  &   \\
137&   257.99088&   -38.28571&  &  &  &        1.69&        2.43&        0.99&  &       16.42&        0.29&        0.37&  &  &  &   \\
142&   258.12906&   -38.58107&  &  &  &        0.88&        0.68&  &  &  &        0.84&        1.48&  &  &  &   \\
145&   258.09253&   -38.52492&  &  &  &  &  &  &  &  &        1.02&        0.88&  &  &  &   \\
150&   258.08490&   -38.57631&  &  &  &        0.25&        0.45&  &  &  &        0.22&        1.36&        2.00&  &  &        9.32 \\
151&   258.00430&   -38.47361&  &  &  &  &  &  &  &  &        0.35&        0.37&  &  &  &   \\
152&   258.09113&   -38.43760&  &  &  &  &  &  &  &  &        3.19&        5.83&        6.89&  &  &   \\
157&   258.14062&   -38.40404&  &  &  &  &  &  &  &  &  &        3.50&  &  &  &   \\
158&   257.98721&   -38.32267&       13.25&       16.38&       14.59&        8.56&        8.62&  &  &  &        7.42&        7.64&  &  &  &   \\
160&   258.17786&   -38.50867&  &  &        1.19&        1.75&        1.26&  &  &  &        0.82&        1.71&  &  &  &   \\
167&   258.30264&   -38.51905&  &  &  &        0.78&        0.45&  &  &       17.07&        0.26&        0.25&  &  &  &   \\
172&   258.08502&   -38.49975&        9.06&        8.68&        6.64&        3.67&        2.63&        1.54&  &  &       20.20&       10.82&  &  &  &   \\
173&   258.17233&   -38.43617&        0.46&        8.63&       29.94&       41.70&       27.58&       27.66&  &  &        0.61&        0.86&  &  &  &   \\
180&   258.00781&   -38.49869&        0.42&        1.94&        3.75&        4.78&        2.77&        8.93&  &  &        0.37&  &  &  &  &   \\
183&   258.06735&   -38.40066&  &  &  &  &  &  &  &  &        1.77&        2.78&        3.10&  &  &   \\
184&   258.30685&   -38.27173&  &  &  &  &  &  &  &     1533.81&        0.22&        0.15&  &  &  &   \\
185&   258.10931&   -38.53597&  &        1.62&        3.74&        6.87&        5.38&  &  &  &  &        1.90&  &  &  &   \\
187&   258.09045&   -38.51449&  &  &  &        1.06&        0.68&  &  &  &        1.31&        1.25&  &  &  &   \\
189&   258.18784&   -38.45626&  &  &  &  &  &  &  &  &        0.27&        0.69&  &  &  &   \\
190&   257.98822&   -38.42171&  &        5.86&       66.73&  &  &     1191.00&     1186.00&      865.34&        0.39&        0.24&  &  &  &   \\
200&   257.86819&   -38.36474&        5.91&       76.97&      290.20&  &  &     1286.00&     1464.00&     1158.68&        0.23&  &  &  &  &   \\
202&   258.01770&   -38.35570&  &  &  &  &  &  &  &  &        0.20&        0.39&  &  &  &   \\
209&   258.08630&   -38.31583&      191.50&      380.80&      377.00&      218.90&  &       93.74&       59.72&  &        0.97&        0.61&  &  &  &   \\
210&   258.11935&   -38.59785&  &  &  &        0.42&        0.78&  &  &  &        0.29&        0.30&  &  &  &   \\
214&   258.00909&   -38.41885&  &        0.97&        1.36&        2.27&        1.99&        3.58&  &  &        0.75&        0.77&  &  &  &   \\
220&   258.09015&   -38.42498&  &  &        1.66&        3.13&        2.23&  &  &  &        0.81&        1.37&  &  &  &   \\
222&   258.16699&   -38.33837&  &  &  &  &  &  &  &  &  &        0.68&  &  &  &   \\
225&   258.07635&   -38.51399&  &  &  &  &  &  &  &  &  &        1.95&  &  &  &   \\
231&   258.12537&   -38.39469&  &  &  &  &  &  &  &  &        0.49&        0.84&  &  &  &   \\
234&   258.05734&   -38.54539&  &  &  &  &  &  &  &  &        0.53&        0.50&  &  &  &   \\
235&   258.08182&   -38.33235&  &  &  &  &  &  &  &  &        0.91&        2.63&  &  &  &   \\
241&   258.16605&   -38.51577&  &  &  &  &  &  &  &  &        0.82&        0.98&  &  &  &   \\
243&   258.01715&   -38.46565&        1.63&        2.09&        1.68&        1.04&        0.58&  &  &  &        0.45&        1.13&        1.29&  &  &   \\
244&   258.07840&   -38.38462&        3.06&        3.33&        2.30&        1.26&        0.78&  &  &  &        0.75&        1.61&  &  &  &   \\
245&   258.10086&   -38.35387&  &  &  &        0.70&        0.59&  &  &  &        0.84&        1.30&        1.70&  &  &   \\
248&   258.22702&   -38.48562&  &  &  &       12.46&        9.64&        8.35&  &  &        0.87&  &  &  &  &   \\
253&   258.05783&   -38.54014&  &        1.64&        3.84&        3.85&        2.85&        3.94&  &  &        0.28&  &  &  &  &   \\
259&   258.13406&   -38.38127&  &        1.85&        3.87&        4.80&        3.32&        3.68&  &  &        0.36&        0.62&  &  &  &   \\
260&   258.19949&   -38.34837&  &  &  &        0.26&        0.50&  &  &  &        0.14&        0.37&  &  &  &   \\
262&   257.97363&   -38.49245&  &  &  &        1.00&        1.01&  &  &  &        0.71&        1.37&        4.51&  &  &   \\
265&   258.18042&   -38.52432&  &  &  &        0.79&        0.97&  &  &  &        0.99&        1.39&  &  &  &   \\
267&   258.13535&   -38.46489&        4.42&       87.16&      395.90&  &  &     1853.00&     1500.00&     1070.69&        0.30&  &  &  &  &   \\
268&   258.11862&   -38.26870&  &        0.79&        1.11&        1.10&        0.85&  &  &  &        0.79&        1.23&  &  &  &   \\
271&   258.00577&   -38.49241&  &  &  &  &  &  &  &  &        0.25&  &  &  &  &   \\
274&   258.12790&   -38.32248&        3.68&        3.32&        2.37&        2.06&        4.21&        5.22&        1.67&  &  &        0.58&  &  &  &   \\
277&   258.22003&   -38.46579&  &  &  &        1.48&        1.31&  &  &  &        0.23&        0.93&  &  &  &   \\
279&   258.11310&   -38.36686&  &  &  &        1.73&        1.30&  &  &  &        0.20&        0.51&  &  &  &   \\
282&   258.19473&   -38.47963&  &  &        2.54&        4.57&        3.42&        6.82&  &  &        0.24&        0.54&  &  &  &   \\
292&   258.07513&   -38.57813&  &  &  &        2.80&        7.31&       10.89&       15.84&       63.85&        0.23&  &  &  &  &   \\
295&   258.05261&   -38.54072&  &  &  &  &  &  &  &  &        0.33&  &  &  &  &   \\
300&   258.09039&   -38.55998&       19.48&       16.87&       11.69&        7.03&        6.00&        6.22&        3.18&       34.85&        0.23&  &  &  &  &   \\
305&   258.18768&   -38.42732&  &  &  &        1.14&        1.17&  &  &  &        0.30&        0.35&  &  &  &   \\
307&   258.05716&   -38.39042&  &        0.82&        1.78&        2.55&        2.07&  &  &  &        0.23&        0.35&  &  &  &   \\
308&   257.87827&   -38.31401&        2.00&        8.39&       12.29&        9.13&        6.45&        5.01&        4.21&        4.74&        0.11&        0.29&  &  &  &   \\
329&   258.20914&   -38.41260&  &  &  &  &  &  &  &  &  &        7.00&  &  &  &   \\
\end{longtable}
 \end{landscape}
 
\section{Properties of tentative sources}\label{apendix:prop_tent}
\begin{longtable}{rlrrr}
\caption{Tentative-source properties} \label{prop_tent}   \\
\hline
      &    &                              &                             &      \\
Id      & T &   M$_{\rm{env}}$  &       L$_{\rm{bol}}$  & IR counterpart        \\
      & K       & $M_{\sun}$  &  $L_{\sun}$ &    \\
      \hline
      \endfirsthead
\caption{continued.}\\
\hline
   &    &                              &                             &      \\
Id      &T &    M$_{\rm{env}}$  &       L$_{\rm{bol}}$  & IR counterpart        \\
   & K  & $M_{\sun}$  &  $L_{\sun}$ &    \\
   \hline
\endhead
\endfoot
11&     17.2&   11.4&   25.3    &       No                              \\
12&     18.9&   9.1&    33.5    &       Yes                             \\
13&     16.7&   16.6&   31.2    &       Yes                             \\
15&     17.8&   186.9&  21.5    &       Yes                             \\
17&     17.0&   94.6&   28.3    &       No                              \\
20&     16.5&   6.6&    13.9    &       Yes                             \\
23&     17.6&   26.2&   13.1    &       Yes                             \\
27&     17.9&   7.3&    19.1    &       No                              \\
32&     22.3&   0.5&    22.7    &       Yes                             \\
34&     18.7&   4.3&    13.7    &       Yes                             \\
37&     17.8&   69.6&   8.5     &       Yes                             \\
38&     15.8&   11.2&   13      &       Yes                             \\
42&     16.1&   4.8&    11.5    &       Yes                             \\
46&     16.2&   5.7&    9.4     &       Yes                             \\
52&     23.2&   2.7&    25.8    &       Yes                             \\
58&     18.1&   98.9&   14.1    &       Yes                             \\
66&     16.6&   7.5&    7.1     &       Yes                             \\
78&     22.9&   1.9&    12.1    &       No                              \\
81&     15.9&   5.1&    5.3     &       Yes                             \\
85&     19.9&   1.8&    4.2     &       Yes                             \\
86&     16.9&   9.3&    6       &       Yes                             \\
87&     16.5&   8.9&    5.7     &       Yes                             \\
91&     17.6&   3.0&    3.8     &       Yes                             \\
95&     15.7&   6.1&    3.9     &       No                              \\
100&    18.1&   14.0&   22.1    &       Yes                             \\
103&    20.6&   4.5&    10.2    &       No                              \\
106&    20.9&   6.8&    21.8    &       Yes                             \\
111&    18.6&   1.4&    5.1     &       Yes                             \\
112&    17.9&   3.1&    7.9     &       Yes                             \\
115&    18.1&   3.3&    4.6     &       Yes                             \\
116&    22.4&   2.6&    18.5    &       No                              \\
134&    19.4&   10.9&   16.4    &       No                              \\
135&    16.8&   5.3&    2.7     &       Yes                             \\
137&    16.6&   4.3&    2.9     &       Yes                             \\
142&    17.4&   11.9&   7.8     &       Yes                             \\
145&    23.5&   0.8&    9.2     &       No                              \\
150&    16.1&   4.4&    2.3     &       Yes                             \\
151&    21.1&   0.7&    3.5     &       No                              \\
152&    20.1&   5.0&    26.4    &       No                              \\
157&    20.4&   8.2&    0.2     &       No                              \\
158&    18.9&   31.7&   57.3    &       Yes                             \\
160&    19.8&   5.1&    7.6     &       Yes                             \\
167&    18.1&   1.4&    2.6     &       Yes                             \\
172&    23.1&   11.1&   144.1   &       Yes                             \\
173&    20.0&   2.4&    5.8     &       Yes                             \\
180&    20.6&   5.5&    3.7     &       Yes                             \\
183&    20.1&   2.3&    15.4    &       No                              \\
184&    18.7&   0.7&    2.2     &       Yes                             \\
185&    21.1&   3.5&    0.2     &       Yes                             \\
187&    23.4&   1.2&    11.6    &       Yes                             \\
189&    19.1&   2.7&    2.8     &       No                              \\
190&    19.5&   0.8&    3.8     &       Yes                             \\
200&    19.8&   5.2&    2.4     &       Yes                             \\
202&    18.4&   2.0&    2.1     &       No                              \\
209&    18.5&   3.0&    8.8     &       Yes                             \\
210&    17.7&   2.1&    2.9     &       Yes                             \\
214&    21.4&   1.3&    7       &       Yes                             \\
220&    20.3&   3.4&    7.5     &       Yes                             \\
222&    15.9&   11.6&   0.2     &       No                              \\
225&    23.0&   2.1&    0.2     &       No                              \\
231&    19.4&   2.8&    4.7     &       No                              \\
234&    19.5&   1.6&    5.1     &       No                              \\
235&    19.0&   10.4&   8.3     &       No                              \\
241&    19.0&   3.8&    7.6     &       No                              \\
243&    21.0&   0.8&    4.4     &       Yes                             \\
244&    19.6&   5.1&    6.9     &       Yes                             \\
245&    20.1&   1.2&    7.8     &       Yes                             \\
248&    18.4&   43.3&   8       &       Yes                             \\
253&    19.6&   7.0&    2.8     &       Yes                             \\
259&    21.0&   1.2&    3.5     &       Yes                             \\
260&    16.1&   5.9&    1.5     &       Yes                             \\
262&    18.1&   5.4&    6.6     &       Yes                             \\
265&    18.0&   8.6&    9       &       Yes                             \\
267&    20.5&   4.7&    3       &       Yes                             \\
268&    18.8&   5.4&    7.3     &       Yes                             \\
271&    21.6&   2.4&    2.6     &       No                              \\
274&    15.9&   9.9&    0.2     &       Yes                             \\
277&    19.0&   3.7&    2.3     &       Yes                             \\
279&    20.4&   1.2&    2.1     &       Yes                             \\
282&    19.8&   1.6&    2.4     &       Yes                             \\
292&    16.1&   57.3&   2.3     &       Yes                             \\
295&    18.6&   14.7&   3.3     &       No                              \\
300&    18.2&   13.1&   2.3     &       Yes                             \\
305&    18.6&   1.6&    3       &       Yes                             \\
307&    19.8&   1.0&    2.3     &       Yes                             \\
308&    19.6&   0.9&    1.2     &       Yes                             \\
329&    18.5&   33.7&   0.2     &       No                              \\
\hline
\end{longtable}

\section{Image of the sources}
In this section, we present the three first sources of the final sample on the 2MASS, \spitzer GLIMPSE and MIPSGAL, \herschel and density maps (low and high resolution) together with the result of their SED fitting. The maps (1$\arcmin \times $1$\arcmin$ for IR maps and 2$\arcmin \times $2$\arcmin$ for \herschel maps) are centered on the coordinates given by {\it{getsources}} with the corresponding wavelength written in the upper-left part of the image.

\begin{figure}
 \centering
 \includegraphics[angle=0,width=130mm,height=220mm]{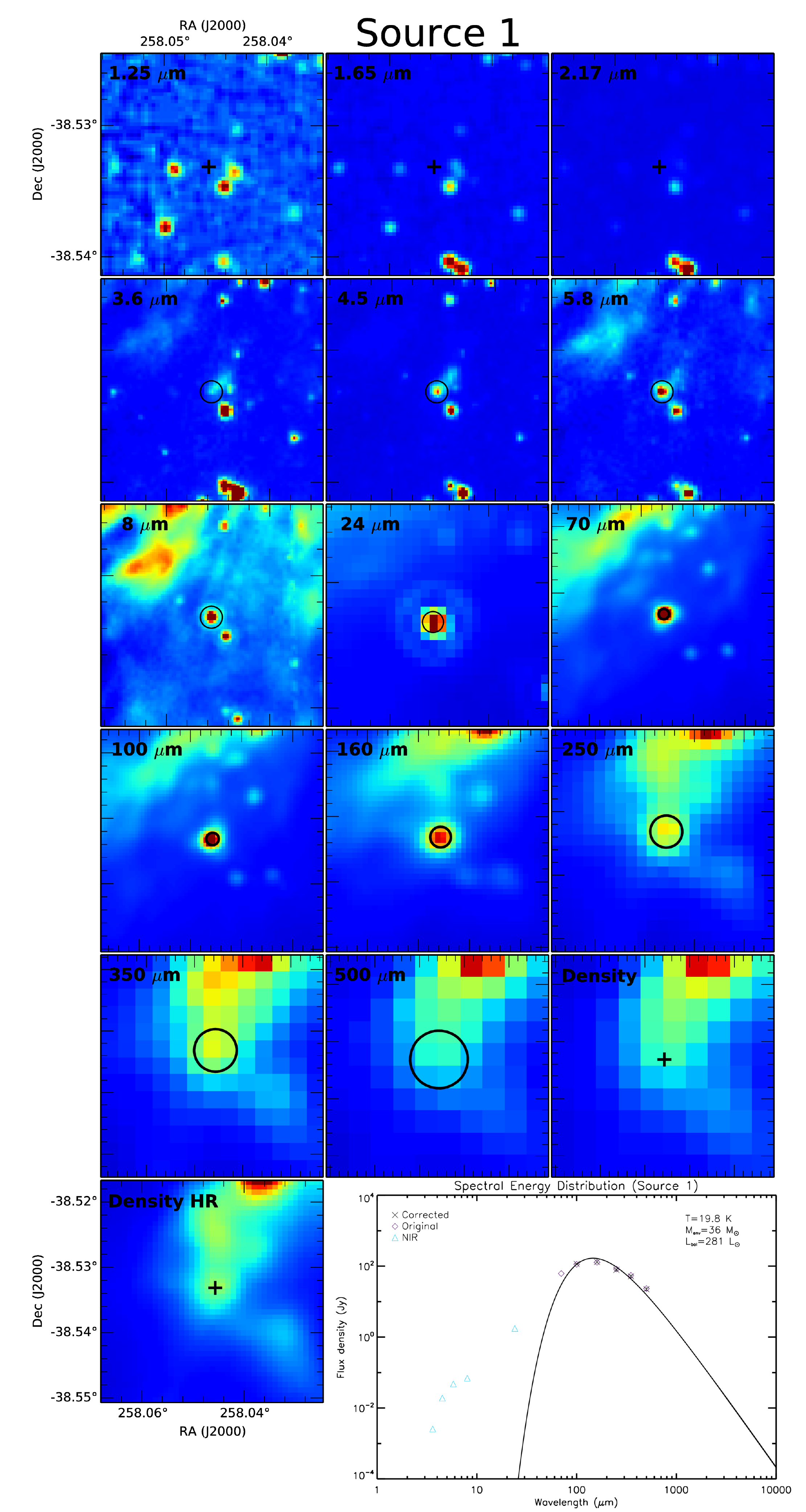}
 \caption{2MASS, GLIMPSE, MIPSGAL, \herschel, low-resolution and high-resolution images of source 1. If a counterpart is seen in the infrared catalogs, a black circle of 4$\arcsec$ radius is shown to indicate the location of this counterpart otherwise, the center (position of the source given by \textit{getsources}) is indicated by a cross. For \herschel images, the ellipses shown are the {\it{getsources}} parameters, A$_{\rm{FWHM}}$ and B$_{\rm{FWHM}}$. For the representation of the SED fitting, the original fluxes are represented by a magenta diamond, the corrected fluxes (flux scaling + color correction) at the wavelength used for the fitting are represented by a black cross, and the blue triangles represent the IR counterparts, if any. The identification number of the source is given in the title of the SED.}
 \label{Source1_m}
\end{figure}

\begin{figure*}
 \centering
 \includegraphics[angle=0,width=130mm,height=220mm]{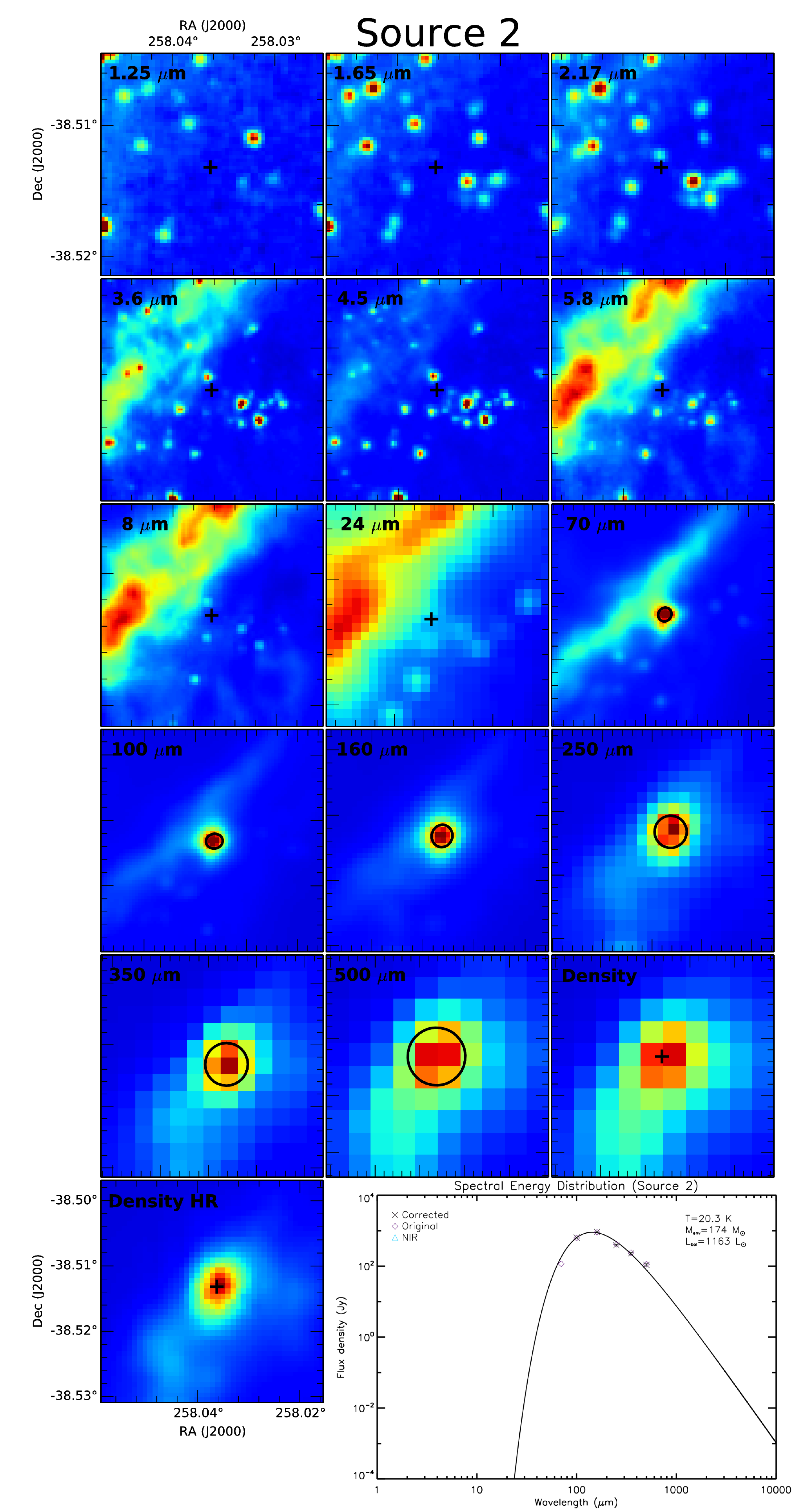}
  \caption{Same as Fig.~\ref{Source1_m} for source 2}
\end{figure*}

\begin{figure*}
 \centering
 \includegraphics[angle=0,width=130mm,height=220mm]{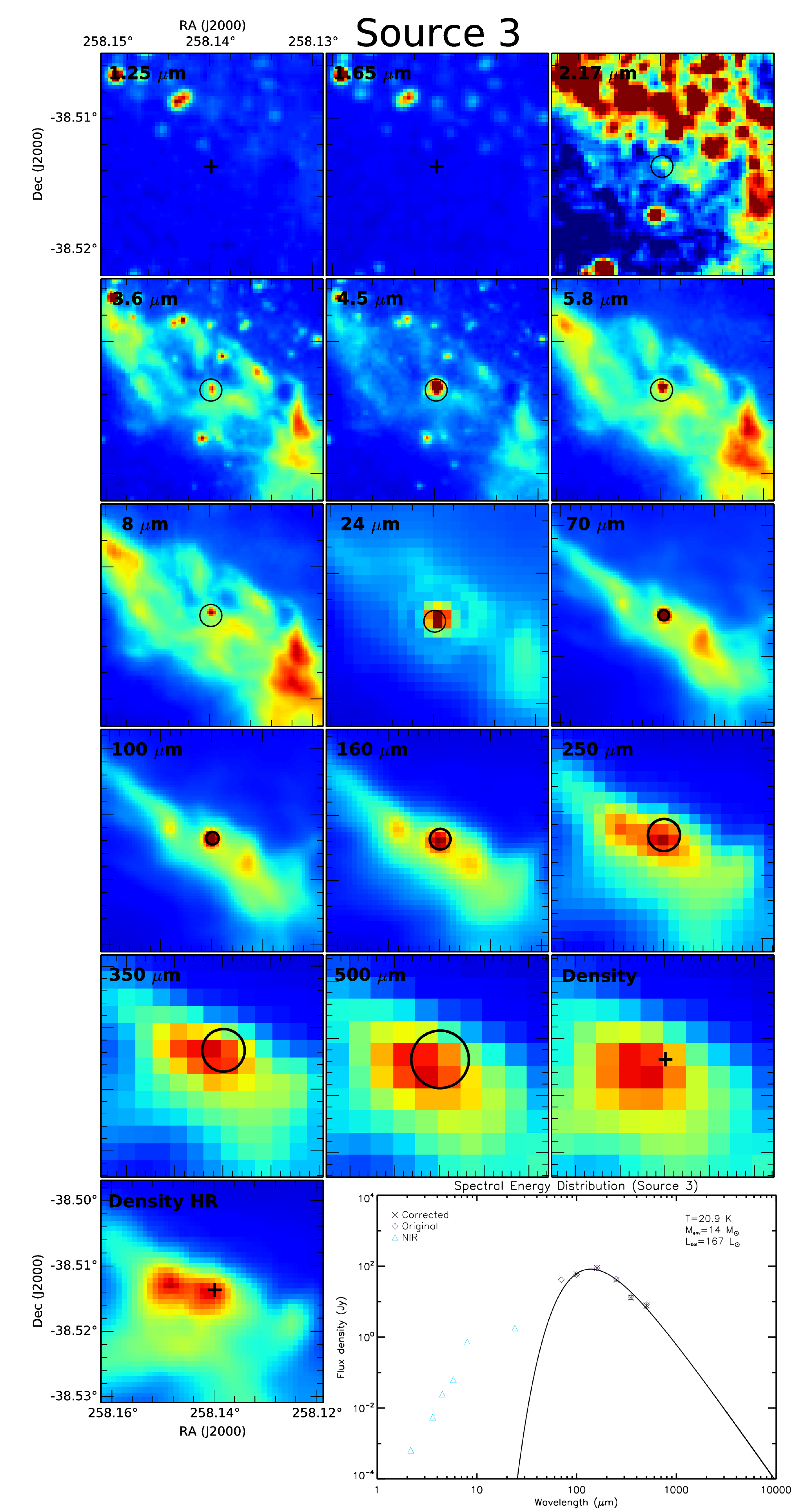}   
  \caption{Same as Fig.~\ref{Source1_m} for source 3}
\end{figure*}

\end{appendix}

\end{document}